\documentstyle[12pt,aaspp4,epsf]{article}

\received{28 June 1997; revised 13 May 1998}
\accepted{}
\journalid{337}{15 January 1989}
\articleid{11}{14}
\slugcomment{}

\def\,{\thinspace}
\def\ts{\thinspace}

\def\kms{km\thinspace s$^{-1}$}
\def\Lsun{L$_\odot$}
\def\Msun{M$_\odot$}

\def\Lco{{\hbox {$L_{\rm CO}$}}}
\def\Lhcn{{\hbox {$L_{\rm HCN}$}}}
\def\Lfir{{\hbox {$L_{\rm FIR}$}}}

\def\date{\number\day\space \ifcase\month\or
	January\or February\or March\or April\or May\or June\or 
	July\or August\or September\or October\or November\or December\fi
	\space\number \year}
\received{\date}

\begin{document}
\title{Rotating Nuclear Rings and Extreme Starbursts in Ultraluminous Galaxies}
\author{D. Downes}
\affil{Institut de Radio Astronomie Millim\'etrique, 38406 
 St.\ Martin d'H\`eres, France; downes@iram.fr}
\and\author{P. M. Solomon}
\affil{Department of Physics and Astronomy, State University of New York, 
 \break Stony Brook, NY 11794; psolomon@astro.sunysb.edu}
\received{10 July 1997 -- 1}
\accepted{12 June 1998}
\journalid{337}{10 November 1998}
\articleid{11}{14}
\slugcomment{}
\vskip 1cm
\centerline{to be published, {\bf Astrophysical Journal}, Nov. 10, 1998}

\vskip 1cm
\centerline{  Received \hskip 0.2cm \underline {28 June 1997}\hskip 0.4cm
 Accepted \hskip
0.2cm
\underline {12 June 1998}}
\bigskip
\vfil \eject
\begin{abstract}
New CO interferometer data show the molecular gas in IR ultraluminous
galaxies is in rotating nuclear disks or rings. The CO maps yield
disk radii, kinematic major axes, rotation speeds, enclosed dynamical
masses, and gas masses.  The CO brightness temperatures, the double
peaked CO line profiles, the limits on thermal continuum flux from
dust, and the constraint that the gas mass must be less than the
dynamical mass, all indicate the CO lines are subthermally excited and
moderately opaque ($\tau = 4$ to 10).  We fit kinematic models in
which most of the CO flux comes from a moderate-density, warm,
intercloud medium rather than self-gravitating clouds. Typical ring 
 radii are 300 to 800 parsecs.

 We derive gas
masses not from a standard CO-to-mass ratio, but from a model of
radiative transfer through subthermally excited CO in the molecular
disks.  This model yields gas masses of $\sim 5 \times 10^9 $\, \Msun  ,  $\sim
5$ times lower than the standard method, and a ratio $M_{\rm
gas}/L^\prime_{\rm CO}$
$\approx$ 0.8\,\Msun (K\,\kms\,pc$^2)^{-1}$.  In the nuclear disks, we derive a
ratio of gas  to dynamical mass, $M_{\rm gas}/M_{\rm dyn}\approx$ 1/6 and a
maximum ratio of gas to total mass surface density, $\mu/\mu_{\rm tot}$, of
1/3.

For the galaxies {\bf VII Zw 31}, {\bf Arp~193}, and IRAS {\bf
10565+2448}, the CO position-velocity diagrams provide good evidence
for rotating molecular rings with a central gap.

In addition to the rotating central rings or disks a new class of star
formation region is identified which we call an extreme starburst.  They have a
characteristic size of only 100 pc. with about  $10^9$ \Msun\ of gas and an IR
luminosity of $\approx 3 \times 10^{11}$ \Lsun\ from recently formed OB stars.
  Four extreme starbursts are identified in the 3 closest galaxies in the sample
including Arp~220, Arp~193 and Mrk~273. They are the most prodigious star
formation events in the local universe, each representing about 1000 times as
many OB stars as 30~Doradus. 

 In {\bf Mrk~231}, the CO(2--1) velocity diagram along
the line of nodes shows a $1''.2$-diameter inner disk and a 3$''$-diameter outer
disk.  The narrow CO linewidth, the single-peak line profile, the
equality of the major and minor axes, and the observed velocity
gradients all imply the molecular disk is nearly face-on, yielding low
optical and UV extinction to the AGN.  Such a geometry means that the
molecular disk cannot be heated by the AGN; the FIR luminosity of
Mrk~231 is powered by a starburst, not the AGN.

In {\bf Mrk~273}, the CO(1--0) maps show long streamers of radius
5~kpc(7$''$) with velocity gradients north-south, and a nuclear disk
of radius 400~pc ($0.6$'') with velocity gradients east-west.  The
nuclear disk contains a bright CO core of radius 120~pc (0.2$''$).
 
In {\bf Arp~220}, the CO and 1.3\,mm continuum maps show the two
``nuclei'' embedded in a central ring or disk at p.a. 50$^\circ$ and a
fainter structure extending $7''$ (3\,kpc) to the east, normal to the
nuclear disk.  Models of the CO and dust flux indicate the two
$K$-band sources contain high-density gas with $n$(H$_2$) = $2\times
10^4$\,cm$^{-3}$.  There is no evidence that these sources really are the
pre-merger nuclei.  They are more likely to be compact, extreme 
starburst regions
containing 10$^9$\,\Msun\ of dense molecular gas and new stars, but no
old stars.  Most of the HCN
emission arises in the two ``nuclei''.
The luminosity to mass ratios for the CO sources in
Arp~220 are compatible with the early phases of compact starbursts.
There is a large mass of molecular gas currently forming stars with
plenty of ionizing photons, and no obvious AGN.    The
{\it entire} bolometric luminosity of Arp~220 comes from starbursts,
not an AGN.

The CO maps show the gas in ultraluminous IR galaxies is in extended
disks that cannot intercept all the power of central AGNs, even if they
exist.  We conclude that in ultraluminous IR galaxies --- even in
Mrk~231 that hosts a quasar --- the far IR luminosity is powered by
extreme starbursts in the molecular rings or disks, not by dust-enshrouded
quasars.
 
\end{abstract}

\vfil
\eject

\section{INTRODUCTION}
The infrared ultraluminous galaxies are recent galaxy mergers in which
much of the gas in the former spiral disks has fallen into the center.
The gas re-establishes rotational support at a radius of a few hundred
parsecs.  This is the same size scale as the Narrow Line Region in
active galactic nuclei, and the rotating nuclear disks in some
elliptical galaxies.  During its infall from 5\,kpc to 0.5\,kpc, the
high-density gas forms many massive stars, as in the early starbursts
that formed the bulge stars in protogalaxies at high redshifts.  The
ultraluminous galaxies we observe in the local universe are
re-enacting this primeval process, with a large part of the central
mass in gas ready to turn into stars --- and ready to fall into a
black hole.

The large nuclear concentration of molecular gas in ultraluminous
galaxies has been detected in the millimeter lines of CO by many
groups during the past decade.  Scaling to the signal strengths from
Milky Way molecular clouds, however, soon led to a paradox for many of
the sources --- the estimated gas mass was equal to or larger than the
dynamical mass indicated by the linewidths.  For Arp~220, for example,
Scoville et al. (1991) found that nearly all of the mass in the
central few hundred parsecs was in the form of molecular gas.  To
resolve this dilemma, we showed that in the extreme environment in the
central 600\,pc of ultraluminous galaxies, much of the CO luminosity
must come from an intercloud medium that fills the whole volume,
rather than from clouds bound by self gravity, and therefore the CO
luminosity traces the geometric mean of the gas mass and the dynamical
mass, rather than just the gas mass (Downes, Solomon, \& Radford
1993).  This allows the nuclear disk gas to be overluminous in CO,
relative to self-gravitating clouds in galactic spiral arms.  The
important linewidth for this calculation is the dynamical linewidth of
the galaxy, not that of individual self-gravitating GMCs.  The
physical basis for estimating gas mass from the CO emission is
therefore different from that in Milky Way GMCs.  The Milky Way
conversion factor is relevant for an ensemble of GMCs in an ordinary
spiral galaxy, but not for the center of an ultraluminous galaxy.

In a subsequent survey paper on 37 ultraluminous galaxies observed
with the IRAM 30\,m telescope (Solomon et al. 1997), we derived the
molecular gas mass by several methods.  The minimum estimates of the
gas mass, for optically thin CO, did indeed reduce the gas mass to
much lower fractions of the dynamical mass.  The true gas mass must be
between the optically thin CO estimate and the dynamical mass, {\it
both} of which are less than the gas mass derived with the Milky Way
conversion factor for self-gravitating molecular clouds.

The significance of the lower molecular gas mass is twofold.  First,
it makes the gas mass smaller than the dynamical mass, as it must be.
Second, it means the total molecular gas mass in ultraluminous
galaxies is similar to, and not greater than, the molecular mass in
the disk of a gas-rich spiral.  It is thus not necessary to force into
the nuclear region all the gas in the galaxy --- H$_2$ and H~I.  The
nuclear starbursts can be fed by pre-existing molecular gas that falls
into the central few hundred parsecs from an original radius of a few
kiloparsecs, and not from the outer H~I disk.

These ideas show that the kinematics of the nuclear gas is critical to
understanding both the evolution of ultraluminous galaxies and the
molecular line formation.  The CO emissivity per unit gas mass surface
density is not constant, but varies with position in the nuclear disk.
A more correct treatment requires radiative transfer modeling, with
kinematic data from millimeter interferometers.

To carry out this next step in our study of molecular gas disks at the
centers of IR ultraluminous galaxies, we made high-resolution CO maps
with the IRAM interferometer.  We chose sources from our CO survey,
concentrating on the nearest ultraluminous galaxies in the northern
sky, for which we already had accurate measurements of the total flux,
linewidth, and radial velocity.  The new observations give accurate
positions, line and continuum fluxes at the interferometer resolution,
images, directions of the kinematic major axes, and radii of the
nuclear disks.
\footnote
{In this paper,  distances are for 
 $H_0 = 75$\,\kms \,Mpc$^{-1}$ and $q_0 = 0.5$.} 
These radii and the velocity extrema give the enclosed dynamical mass.
The velocity data allow modeling of the kinematics and the CO
brightness temperature versus radius.  For most sources, the CO data
indicate the gas kinetic temperature is higher than in typical
Galactic molecular clouds, but the CO is subthermally excited,
turbulent, and warm, quite unlike the molecular clouds in galaxy-scale
disks.  At the positions of maximum CO intensity in the disks, the CO
lines are moderately opaque, with optical depth $\tau$(1--0) $\approx
$ 4 to 10, but in other parts of the disk, the CO lines can be
optically thin.  The result is that the molecular gas mass is about a
factor of five lower than it would be if the CO emission came from
self-gravitating clouds.  The rest of the dynamical mass is accounted
for by young stars that provide the luminosity and older stars that
were already in the central bulge before the merger.

 We argue that the bright spots in the ultraluminous
galaxies are compact extreme starburst regions that occur in the dense gas
traced by HCN and CS emission.  They produce
 most of the HCN luminosity, and 1mm dust continuum while the more diffuse gas in the
nuclear disks dominates the CO luminosity. 
 These compact extreme starburst regions which we identify in Arp~193, Arp~220 and Mrk~273 are
the most  prodigious star formation events in the local universe. There are undoubtedly many more
too compact to be resolved. 

The plan of this paper is as follows.  Section~2 describes the
observations and summarizes the parameters obtained by direct
measurement.  Section~3 introduces our model of a turbulent rotating
disk and discusses the derivations of the rotation curve, disk radius,
disk height, turbulent velocity, gas mass, and dynamical mass obtained
by fitting the model disk to the data.  Sections~4 through 8 present
the results on the galaxies with CO(2--1) data, obtained with
sub-arcsecond resolution.  These galaxies are VII~Zw~31, Mrk~231,
Arp~193, Mrk~273, and Arp~220.  Section~9 gives the results on five
additional ultraluminous galaxies mapped in CO(1--0) only.  Section~10
discusses the source sizes, gas masses, the ratio of gas mass to
dynamical mass, and the mass of young stars needed to power the
starburst. 
 Section~11 contains our
conclusions.

\section{OBSERVATIONS AND RESULTS}
CO(1--0) and (2--1) were observed with the IRAM interferometer on
Plateau de Bure, France (Guilloteau et al. 1992).  Three to four
configurations of the four 15\,m antennas gave baselines from 24 to
410\,m for VII~Zw31, Mrk~231, Mrk~273, and Arp~220, and 24 to 288\,m
for the other sources.  The SIS mixers had receiver temperatures of 60
-- 80\,K, and operated single sideband at 3\,mm and double sideband at
1.3\,mm with system temperatures of 200 and 400\,K, respectively.  The
spectral correlator covered a 500\,MHz band at 2.5\,MHz resolution,
giving velocity ranges of 1300\,\kms\ at 7\,\kms\ resolution at 3\,mm,
and 700\,\kms\ at 3.4\,\kms\ resolution at 1.3\,mm.  For analysis, we
smoothed the data to 20 and 40\,\kms .  Amplitude and phase were
calibrated relative to quasars and the flux scale was adopted from
same-epoch measurements of quasars and planets with the interferometer
and the IRAM 30\,m telescope.  The beamwidths (FWHM) were 1$''$ to
3$''$ at CO(1--0) and $0''.5$ to 1$''$ at CO(2--1).   Table~1
summarizes the observing parameters.

Table~2 lists CO source positions and integrated CO line fluxes
from the interferometer.  We also list the CO line fluxes from the
30\,m telescope (Radford, Solomon, \& Downes 1991a; Solomon et
al. 1997).   Table~3 lists sizes from Gaussian fits to the CO
visibilities.  Sources observed in CO(1--0) and (2--1) had the same
sizes in both lines.  The visibility fits are independent of the
synthesized beam or the CLEAN algorithm, and there is no need to
deconvolve an apparent size on a map.  The phase calibration can
broaden a source due to the baseline error, varying as $2\pi \lambda
/D$ times the angle between the source and its phase calibrator
($\lambda =$ wavelength, $D=$ baseline).  For sources and calibrators
in this paper, the broadening is $< 0.1$ beamwidth.  After correcting
for atmospheric decorrelation, we obtained sizes $<0''.2$ for the
quasar calibrator sources.

\section{DISK MODELS}
\subsection{Geometry of Nuclear Disks from Kinematic Data}
CO surveys of our Galaxy show most of the gas in the inner $R<5$\,kpc
is in $\sim 6000$ giant molecular clouds (GMCs) of diameter $\sim
50$\,pc and H$_2$ density 150\,cm$^{-3}$.  In an ultraluminous IR
galaxy merger, these clouds fall into the central $R\sim 500$\,pc,
forming a disk of height $\sim 50$\,pc.  Because the GMCs' density is
too low to stabilize them against tidal shear at the merger's center,
and because the new volume is ten times smaller than the original
volume of all the GMCs in the galactic disk, the GMCs loose their
identity and blend into a continuous medium with a mean H$_2$ density
$\sim 10^3$\,cm$^{-3}$.  We therefore modeled the nuclear disks as a
continuous medium rather than an ensemble of individual clouds.

We used the CO spectral data to estimate how much of the gas was in a
high-density, inner ring or disk and how much in a lower-density,
outer disk.  To simulate the observed spectra and position-velocity
diagrams, we modified a model for rotating disks developed by
A.~Dutrey (see Dutrey, Guilloteau, \& Simon 1994).  This model has an
inner disk of high-density gas between radii $R_{\rm min}$ and $R_1$ =
$R_{\rm min}+\Delta R$, and an outer disk of lower-density gas between
radii $R_{\rm min}$ and $R_{\rm max}$.  The H$_2$ density between
$R_{\rm min}$ and $R_{\rm max}$ is
\begin{equation}
n(R) = 
n_0\, A\,\exp \bigg[ - 4 \ln 2 
\bigg({ {R-R_{\rm min}}\over {\Delta R}} \bigg)^2\ \bigg]  
		\ \ + \ \ n_0\, R^\alpha  \ \ \ \ ,
\end{equation}
where the inner disk is the Gaussian of width $\Delta R$ (FWHM), and
the outer disk is the power law.  We fit the outer disks with constant
density ($\alpha = 0$).  Our beams cannot distinguish between a ring
and a filled disk, so for most sources, we took $R_{\rm min}$ = 0.
Because of the radiative transfer through the rotating, inclined disk,
the predicted CO map looks like a ring, even if there is gas in to $R
= 0$.  For the galaxies VII~Zw~31, Arp~193, and IRAS 10565+2448, the
position-velocity diagrams are better fit with ring models, with
$R_{\rm min}$ = $R_0$, rather than filled disk models.   Table~4
gives the source geometry, derived as follows:

{\it Line of nodes:}
The visibility fits in individual spectral channels gave position
offsets vs.\ velocity, which showed the direction of the kinematic
major axis.  We checked the angle of the line of nodes from channel
maps and isovelocity contour maps. Errors are $\pm 10^\circ$.

{\it Rotation curve turnover radius $R_0$:}
In the data, the position-velocity diagrams along the kinematic major
axis often have two peaks near the source center.  In our models, twin
peaks occur near the points where the rotation curve turns over and
becomes flat.  The observed offsets of the two intensity peaks on the
major axis were tried as first guesses for the rotation curve turnover
radii $R_0$.  Model fits converged close to these values.  In angular
units, the uncertainties are $\sim\pm 0''.1$.  In our models, $R_0$ is
the projected radius with the largest gas column density at the same
line-of-sight velocity, and the highest flux.  For most of the
sources, the solid angle inside this zone is too small to affect the
maps, even if the gas is really in a filled disk rather than a ring.

{\it Inner disk half-intensity radius, $R_1$ = $R_{\min}+\Delta R$:}
In our models, most of the CO flux comes from the high-density inner
disk.  The extent of the central double peak in the position-velocity
diagrams constrained the radial width $\Delta R$, which we varied
until the model position-velocity diagrams matched the data. The
uncertainty in $\Delta R$ is $\sim\pm 0''.2$

The {\it outer radius, $R_{\max}$,} of the low-density disk was taken 
to be the observed CO maximum extent on the line of nodes.

{\it Disk thickness $H$:}
Along the $z$-axis perpendicular to the equatorial plane, the gas
volume density $n(z)$ near $z=0$ was approximated by a gaussian with
full width to half-maximum $H$.  For a disk with a flat rotation
curve, the disk's thickness can be estimated with the Mestel formula
 (Mestel 1963; see also Binney \& Tremaine 1987): 
\begin{equation}
H(R)  = 1.4 \ \sigma (R) \bigg( {R\over V_{\rm rot}} \bigg)  \ 
\bigg[1+ {\rho_{\rm gas}\over \rho_1 } \bigg(1- {H \over 2R}\bigg)\bigg]^{-0.5}
\ \ \ ,
\end{equation}
where $\sigma$ is the 1-D velocity dispersion (assumed 
independent of $z$) and $\rho_1(R)$ is the total mass density (gas plus stars)
in an equivalent sphere with a flat rotation curve.
This equivalent density, in \Msun pc$^{-3}$, is related to 
the rotation velocity, in \kms , 
through the usual integral for dynamical mass, 
$\rho_1(R) = 18.5 (V_{\rm rot}/R)^2$.
We dropped the second-order term and estimated the disk thickness as 
\begin{equation}
H \approx  1.4 \ { \sigma (R) \bigg( {R\over V_{\rm rot}} \bigg) } 
\bigg[1+ {\rho_{\rm gas}\over 18.5}\bigg( {R\over V_{\rm rot}} \bigg)^2 \ 
\bigg]^{-0.5}
 \ \ \  ,
\end{equation}
where $H$ and $R$ are in pc, $V_{\rm rot}$ is in \kms , $\rho_{\rm
gas}$ is in \Msun pc$^{-3}$.  Because the velocity dispersion
decreases with distance from the center, our model disk thickness
tends to a constant value.  In reality, the molecular disks are warped
and twisted, and their shape depends on the merger's infall history.

\subsection{Rotation Curves}
We combined our density models with rotation curves that  
reproduced the observed channel maps, position-velocity diagrams, 
spectral profiles, interferometer visibilities, and line intensities.  
We took the rotation velocity to be
\begin{equation}
V_{\rm rot}(R) = V_0\, \bigg({R \over R_0 } \bigg)^\beta \ \ \ \ ,
\end{equation}
with $\beta = 1$ for $R<R_0$ and $\beta= 0$ for $R_0 \leq R \leq R_{\rm max}$,
that is, the curve rises from the center and flattens
 after $R_0$.  Most of the derived values of $R_0$ are  
smaller than our beam, so more elaborate rotation curves are
not justified for now.

For the {\it inclination, $i$}, we started from the major/minor axis
ratio of the integrated CO source, with $i=\cos^{-1}$(minor/major),
and then varied the inclination to match the position-velocity data
and to ensure the derived gas mass was less than the dynamical mass.
The nuclear rings or disks may not be circular; they may be
elliptical, analogous to $x_2$ orbits at the center of a barred
galaxy.  The inclination $i$ listed in  Table~5 should be
interpreted in the sense $\sin i = \sin I \cos \phi$, where $I$ is the
true inclination of the disk relative to face-on, and $\phi$ is the
azimuth of the major axis of the ellipse, if there is one.  With our
current beams, we cannot tell the difference between circular and
elliptical orbits.

For the {\it rotation velocity, $V_0$}, we first guessed the apparent
speed, $V_0 \sin i$, to be half the velocity range of the twin CO
peaks on the major axis.  The best fits were close to these values,
giving 200 to 300\,\kms\ (corrected for inclination) on the flat part
of the curve (Table~5).  This is an equivalent circular velocity
if the orbits are elliptical.

The {\it turbulent velocity, $\Delta V$,} of the gas was taken from
the model fit, not from the observed line profiles, which include both
turbulent broadening and the rotational velocity gradient in the
beam. Rotation alone cannot explain the observed profiles.  We need a
local, turbulent line broadening as well. The models actually
constrain this parameter rather well.  We assumed the local line
broadening had the form
\begin{equation}
f(V-V_0) = 
\exp \bigg[ -\bigg({ {V-V_0}\over {\Delta V}} \bigg)^2\ \bigg] 
\end{equation}
where $\Delta V$ is the local line halfwidth to the $1/e$ level
($\Delta V$ = $0.6 \times$ FWHM = 1.4$\sigma$, where $\sigma$ is the
r.m.s.\ velocity dispersion along the line of sight).  The radiative
transfer model included the local linewidth at each point and the
rotational velocity versus radius.  We iterated the model fits until
the local turbulence and the rotational velocity gradient ---
convolved with our beam --- matched the observed position-velocity
diagrams and the observed line profiles.  Turbulent velocities $\Delta
V$ that fit the data were 30 to 140\,\kms , with some of the
position-velocity diagrams indicating lower turbulence at greater
radii from the nucleus. At the rotation curve turnover point $R_0$,
the values of $\Delta V$ correspond to local FWHM linewidths of 70 to
230\,\kms , so the nuclear disks are highly turbulent.

\subsection{CO line parameters}
After deriving sizes and velocities from the data, we adjusted 
our models to match the observed CO line intensities. As a  
first guess, we set the gas kinetic temperature equal to the 
dust temperature derived from blackbody fits to the IRAS fluxes.   
At the gas densities in our models, collisions 
cannot raise the CO excitation temperature to the gas kinetic 
temperature, nor populate the CO levels thermally.  
Instead, there are different excitation temperatures for each
transition.  The line brightness temperatures, excitation temperatures,
and gas kinetic temperatures are in the sense $T_b<T_{\rm ex}<T_{\rm kin}$,
with $T_{\rm ex}(2-1) < T_{\rm ex}(1-0)$.
We therefore assumed the CO level populations 
were determined by the local gas kinetic temperature, density, and 
velocity gradient, and we calculated, with an escape probability program, 
CO column densities per unit velocity width 
for a local velocity gradient of 1\,\kms\,pc$^{-1}$ and a 
[CO/H$_2$] abundance  of $8\times 10^{-5}$, both typical of 
Milky Way molecular clouds.  
For these values, the escape probability method yields 
the following non-LTE excitation temperatures $T_{\rm ex}$: 
\begin{equation}
T_{\rm ex}(1 - 0) = 0.186 \ n^{0.622} \ T^{0.240}_{\rm kin}
\end{equation}
\begin{equation}
T_{\rm ex}(2 - 1) = 0.117 \ n^{0.678} \ T^{0.247}_{\rm kin} \ \ \ , 
\end{equation}
where $n$ is the H$_2$ number density in cm$^{-3}$, and 
$T_{\rm kin}$ is the gas kinetic temperature in K. 
These formulae are valid for 
$50<T_{\rm kin}<150$\,K and $100<n<3000$\,cm$^{-3}$.  

We then calculated the radiative transfer on the lines of sight
through the disk (eqs.\ 7 \& 8 of Dutrey et al. 1994), convolved the
computed map with our beam, and adjusted the parameters to match the
observed brightness temperatures.  Highest CO opacity occurs at the
rotation curve turnover radius, $R_0$, which yields the largest column
density at the same line-of-sight velocity. Table~6 gives the
best-fit model CO(1--0) excitation temperatures and gas densities, and
 Tables~7 and 8 give the model CO(1--0) and (2--1) line opacities
and Rayleigh-Jeans brightness temperatures.  We also ran models for
higher density gas, with excitation temperatures closer to the kinetic
temperatures, but these models gave too much CO and dust flux, and did
not reproduce the position velocity diagrams.

 At the rotation curve turnover radius $R_0$, our models yield
CO(1--0) opacities of 2 to 8.  These mean opacities apply to smoothly
distributed gas.  The model of subthermal excitation yields the
observed CO(2--1)/(1--0) ratios of 0.6 to 1.0 (Radford et al. 1991a).
In the lower-density, outer disks detected in these sources, the
volume-averaged densities in the model are too low to yield detectable
CO lines.  For these outer disks, our model must be corrected with an
area filling factor for molecular clouds, as in CO maps of disks of
normal galaxies.  For the mean values for the outer disks in 
Table~6, we adopted a volume filling factor of 0.1, an area filling
factor of 0.3, and local gas densities 10 times higher than the
volume-averaged values listed in the Table.

\subsection{Gas Mass and Dynamical Mass}
We used the CO parameters to estimate the gas mass and the dynamical
mass {\it vs.} radius (Table~9).  After we found a density law
that reproduced the observed brightness temperatures, we integrated
over the source to get the gas mass.  These masses are for a
[CO/H$_2$] abundance of $8\times 10^{-5}$, as in molecular clouds in
our Galaxy.  For the sources observed here, it is difficult to lower
the model CO abundance below the Milky Way value. Keeping the same CO
line intensity at a lower CO abundance would force us to make the gas
mass bigger than the dynamical mass, invalidating the model.

The dynamical masses were taken to be $R\,V_{\rm rot}^2/G$, where
$V_{\rm rot}$ is the rotation speed listed in Table~5.  Most of
the sources have flat rotation curves, so the dynamical mass increases
linearly with radius.   Table~9 lists the ratio of gas mass to
dynamical mass in the inner, high-density disk alone (to the radius
$R_1$ in Table~4) and in the inner and outer disks together (to the
radius $R_{\rm max}$ in Table~4).  Relative to our assumed Hubble
constant, the gas mass scales as $H_0^{-2}$, the dynamical mass as
$H_0^{-1}$ and their ratio as $H_0^{-1}$.

We also list in Table~9 the maximum ratio of gas to total (gas +
stars) surface density in our models.  We calculated the ratio of gas
to total mass surface density, $\mu/\mu_{\rm tot}$, in 15 bins in
radius, and found maximum values of $\sim $1/3 to 1/4, midway between
the rotation curve turnover radius $R_0$ and the half-intensity radius
$R_1$.  The models in this paper are for the distributed gas, but
there is also denser gas in star-forming cores that give rise to
HCN and CS lines.  Most of the CO flux comes from the distributed
medium, but our gas masses may have to be corrected upward to allow
for the cores, depending on future interferometer results in the
dense-gas tracer lines.  We return to this point in the Discussion section.

\subsection{Dust Continuum Flux}
We used the model gas mass to predict the dust flux, from 
\begin{equation}
 S(\nu_{\rm obs}) = (1+z) \kappa(\nu_r)M D^{-2}_L B(\nu_r, T_d)  \ \ \ \ \ , 
\end{equation}
where $S$ is the dust continuum flux density, 
$\nu_{\rm obs}$ and $\nu_r$ are the observed and rest-frame frequencies,
$M$ is the {\it gas} mass, $D_L$ is the luminosity
distance, $B$ is the Planck function, and $T_d$ is the dust temperature. 
We took the dust temperatures from our fits to 
the IRAS fluxes (Solomon et al. 1997) --- the same
as the gas temperatures in Table~6 --- and   
we used a dust mass absorption coefficient 
$\kappa (\nu_r ) = 0.1 \times \nu^n$,
where $\nu_r$ is in THz, $\kappa$ is in 
cm$^2$\,gm$^{-1}$ of interstellar matter, and we took the index $n =1.5$.  
At 230\,GHz, this absorption coefficient is  
$\kappa$ $=$ 0.011\,cm$^2$ gm$^{-1}$ of interstellar matter, or, for
a gas-to-dust mass ratio of 100, $\kappa_d$ $=$ 1.1\,cm$^2$ gm$^{-1}$ 
of dust, as in previous estimates for dense molecular clouds
(see  Kr\"ugel \& Siebenmorgen 1994, their Fig.~12). 
The predicted thermal fluxes from dust are then 
\begin{equation}
 S(\nu_{\rm obs}) = 6.4\times 10^{-7} (1+z) M \ D^{-2}_L \ T_d \  
\nu^{2+n}_r  \ \ \ \ \ , 
\end{equation}
where flux density $S$ is in Jy,  gas mass $M$ is in \Msun , 
luminosity distance $D_L$ is in Mpc, dust temperature $T_d$ is in K,
and rest frequency $\nu_r$ is in THz.
The predicted thermal dust fluxes at 2.6 and 1.3\,mm are compared  
 with the observed fluxes in Table~10.
  
Four sources, VII~Zw~31, Mrk~231, Mrk~272, and Arp~220, were observed
in both the CO(1--0) and (2--1) lines, and on the longest baselines of
410\,m.  Because we have better angular resolution for these sources,
we discuss them in more detail than our other sources.

\section{ {\bf  VII Zwicky 31} }
This source, catalogued by Zwicky (1971), was identified by Fairclough
(1986) as an ultraluminous IR galaxy.  Its optical surface brightness
profiles resemble those of elliptical galaxies and its optical line
ratios fit a starburst rather than an AGN (Djorgovski, de Carvalho, \&
Thompson 1990).  Sage \& Solomon (1987) found this source had one of
the highest known CO luminosities and deduced that the molecular gas
was a large part of the dynamical mass.  Scoville et al. (1989) showed
the CO source had a radius $< 3.8\,$kpc.

The CO maps of the galaxy VII Zw 31 are the best evidence in 
our sample for a rotating ring.  Models that best fit the data are
those with a minimum ring radius $R_{\rm min}$ in eq.(1) equal to
$R_0$ = 290\,pc.  Filled-disk models, with $R_{\rm min}$ = 0, give
poorer fits, with much less contrast.  In 40\,\kms -wide channel maps,
the CO peak migrates from south to north with increasing velocity
(Fig.~1).  The source is resolved east-west, perpendicular to
the kinematic major axis.  This indicates the source is inclined, but
more face-on than edge-on.  Our modeling gives a good fit to the data
for a ring inclined at 20$^\circ$ to face-on.   Figure~2 shows
the maps of integrated intensity and isovelocity contours across the
source in both CO lines.  The total CO(1--0) flux measured with the
interferometer is equal to the flux obtained at the 30\,m telescope
(Radford et al. 1991a).  At CO(2--1), the source is partially resolved
out, and the flux from the interferometer is only half the single-dish
flux.  In channels off the CO line, the continuum flux is $<2$\,mJy at
109\,GHz and $<10$\,mJy at 221\,GHz.  This limit on any thermal dust
flux is consistent with the gas mass of $1.1\times 10^{10}$ deduced
from the CO luminosity, for a dust temperature $<50$\,K.

The CO velocity contours in Fig.~2 indicate the north-south
velocity gradient.  In the position-velocity diagrams (Fig.~3)
there is a 200\,\kms\ velocity shift over 4$''$ north-south.  The
model in the Tables reproduces well the observed diagrams, the CO
profiles across the source, the channel maps, and the observed line
intensities.  The CO spectra (Fig.~4) have twin peaks separated
by $\pm 70$\,\kms .  The blue and redshifted CO peaks are respectively
$0''.65$ south and north of the source centroid.  The line profiles
are remarkably symmetric along the north-south kinematic major axis.
The CO line of nodes differs from the northwest-southeast dust lane
found by Djorgovski et al.\ (1990) on a 10$''$ scale by subtracting a
model from their optical images.

\section{ {\bf  Markarian 231 } }
The Seyfert~I galaxy Markarian 231 has an IR luminosity of $3.3\times
 10^{12}$\,\Lsun\ (e.g., Sanders et al. 1987).  Most of the power is
 emitted at the center of a major galaxy merger which has tidal tails
 extending over 75\,kpc (Cutri, Rieke, \& Lebofsky 1984; Hutchings \&
 Neff 1987; Sanders et al. 1987).  Optical spectroscopy indicates
 stars formed in the past 1~Gyr over a widespread region
 (10--15\,kpc).  In this large extranuclear region the H$\alpha$
 emission comes from shocks rather than stellar photoionization (e.g.,
 Hamilton \& Keel 1987; Lipari, Colina, \& Macchetto 1994).  The best
 previous CO study is that by Bryant \& Scoville (1996), who found an
 east-west velocity gradient in the CO source.  Our higher-sensitivity
 results have a factor of two better resolution than their study, and
 not only confirm the velocity gradient, but also distinguish the
 inner and outer disks in the position-velocity diagrams.

\subsection{CO in Mrk 231's molecular disk}
In contrast to the large optical extent, the CO source is very
compact, on a scale twenty times smaller in radius.  In 40\,\kms\
channel maps in CO(1--0) and (2--1), the line peak shifts from west to
east with increasing velocity (Fig.~5).  The gradient is clearly
shown on the CO isovelocity maps (Fig.~6).  The CO(2--1)
intensity contours (Fig.~6) are symmetric and only slightly
broader than the beam.  The CO(2--1) position-velocity diagram along
the line of nodes (Fig.~7) shows an inner nuclear disk of
diameter $1''.2$ (radius 460\,pc) and a 3$''$ outer disk with a lower
velocity gradient.  The asymmetric CO(2--1) line profiles $0''.4$ from
the center of the galaxy (Fig.~8) are a characteristic signature
of a rotating nuclear disk.  The maximum linewidth in both CO lines is
190\,\kms\ FWHM, which is narrow for ultraluminous galaxies.

The CO data imply the molecular disk is face-on, for three reasons.
{\it 1)} The CO major and minor axes are nearly equal, indicating an
inclination $i \leq 20^\circ$.  {\it 2)} To reproduce the
single-peaked CO profile, narrow CO linewidth, and observed velocity
gradients, our model disks must be within $20^\circ$ of face-on.  For
$i \geq 20^\circ$ to face-on, the gas mass would exceed the dynamical
mass.  {\it 3)} The CO does not absorb the nuclear continuum source at
any velocity.  This agrees with the CO disk being face-on, leaving the
Seyfert~I nucleus unobscured.  For $i= 10^\circ$ to face-on, the true
rotation velocity would be 345\,\kms , the same as we deduce for
Arp~220.  If the disk were even more face-on, the observed velocity
gradient would imply a rotation speed $> 400$\,\kms , higher than in
all the other galaxies in our sample, which seems unlikely.  We adopt
$i= 10^\circ$ and $V_{\rm rot}$ = 345\,\kms , although the kinematic
data alone would also allow a rotation velocity of 250\,\kms .

In CO, we do not detect the gas emitting in the 1667\,MHz OH megamaser
with 760\,\kms\ linewidth (Staveley-Smith et al. 1987).  nor do we
detect anything at the IR source $3''.5$ south of the nucleus, to a
limit of 5\,mJy in 40\,\kms\ channels (see also Bryant \& Scoville
1996).  Armus et al. (1994) interpreted the southern IR source as the
nucleus of the merger partner, but recent HST images show this feature
to be a dense arc of star-forming knots (Surace et al. 1998).  In any
case, it is very weak in both CO and the mm continuum.

In the center of Mrk~231, the AGN is detected by centimeter wavelength
 VLBI as a variable, nonthermal radio continuum source with a size $<
 1$~pc (Preuss \& Fosbury 1983; Neff \& Ulvestad 1988; Lonsdale,
 Smith, \& Lonsdale 1993; Taylor et al. 1994). New VLBA images at
 1.4\,GHz resolve this nonthermal source into a nuclear core with
 pc-scale lobes.  Carilli, Wrobel \& Ulvestad (1998) subtracted their
 VLBA map from a VLA map and found an $0''.4$ nonthermal nuclear disk
 source that emits half the nuclear flux, the rest coming from the
 core and the pc-scale lobes.

At millimeter wavelengths, outside the CO line channels, we detect an
unresolved (size $<0''.1$) nonthermal, mm continuum source at the
nucleus, which is probably the nonthermal VLBA core.  The mm continuum
source and the CO centroid both coincide with the 8.44\,GHz continuum
source (Condon et al.\ 1991).

\subsection{Mrk~231's Luminosity: 2/3 Starburst, 1/3 AGN} 
 The CO kinematic data for the central 1 kpc of Mrk~231 and our model
of a face-on disk show the far-IR luminosity comes from a starburst,
not a black hole accretion disk.  We assume the central UV-to-FIR
continuum has three components: the visible and UV continuum is
blackbody flux from the AGN accretion disk, the 2$\mu$m peak is
blackbody flux from a dusty AGN torus, and the far-IR peak is
blackbody flux from the molecular disk.  The extinction to the
accretion disk is low, $A_V$ = 2\,mag, because both the dusty torus
and the larger-scale molecular disk are face-on.

\noindent
{\it The accretion disk's luminosity} may be derived from the UV and
visible fluxes and the extinction to the Seyfert~I nucleus.  The UV
fluxes measured by IUE and HST are $2\times 10^{-15}$\,erg s$^{-1}$
cm$^{-2}$ \AA$^{-1}$ (Schmidt \& Miller 1985; Hutchings \& Neff 1987;
Smith et al. 1995). A blackbody with this flux, corrected for a UV
extinction of 5\,mag (Smith et al. 1995) and a distance of 170\,Mpc,
has a UV luminosity of $4\times 10^{11}$\,\Lsun .  For an accretion
disk temperature of 20000 to 30000\,K, as in the models of Malkan \&
Sargent (1982) and Sanders et al. (1989), the blackbody peaks at 2000
to 1000\,\AA, and has a radius of $1\times 10^{-3}$\,pc.  This result
is consistent with the optical continuum, which has $A_V$ = 2\,mag and
an intrinsic magnitude $M_V = -25.1$ (luminosity $5\times
10^{11}$\,\Lsun ; Boksenberg et al. 1977).

\noindent
{\it The dusty torus' luminosity} may be derived from the near-IR
continuum. As in other Seyfert~I galaxies, the 2$\mu$m bump is well
fit by a 1470\,K blackbody, close to the dust sublimation point
(Kobayashi et al. 1993).  The flux at the 2$\mu$m bump, $9.3\times
10^{-11}$\,erg~s$^{-1}$ cm$^{-2} \mu$m$^{-1}$, corrected for an IR
extinction of 0.2 to 0.6\,mag (Krabbe et al. 1997), implies a
blackbody luminosity of (3 to 4) $\times 10^{11}$\,\Lsun , and a
radius of 0.18\,pc.  The compact 10 to 25\,$\mu$m source in Mrk~231
(Matthews et al. 1987; Keto et al. 1992; Miles et al. 1996) arises
farther out in the dusty torus, where the dust temperature is
$\sim$200\,K, at a radius of $\sim$10\,pc from the AGN.  This dusty
torus is heated by the AGN.  If its thickness is equal to its radius,
then the torus absorbs half the AGN's power, so the AGN's total
luminosity is 6 to $9\times 10^{11}$\,\Lsun , the same as estimated
from the optical and UV flux.

\noindent
{\it The molecular disk's FIR luminosity} may be derived from the IRAS
fluxes, and is $2\times 10^{12}$\,\Lsun\ (e.g., Solomon et al. 1997).
The molecular disk has a dust optical depth close to unity at
100\,$\mu$m and causes the deep 10$\mu$m silicate absorption (Allen
1976; Rieke 1976; Roche, Aitken, \& Whitmore 1983).  The blackbody
radius calculated from the FIR luminosity is 200\,pc, half the
measured size of the molecular disk.  The molecular disk receives some
heat from the dusty torus.  With the thickness-to-radius ratio in our
CO model, it would intercept half of the power, or $2\times
10^{11}$\,\Lsun\ of re-processed radiation from the AGN.  Since this
is only 10\% of the FIR luminosity, most of the FIR luminosity of
Mrk~231 must come from the starburst in the molecular ring or disk.  A
similar model was used by Rigopoulou, Lawrence, \& Rowan-Robinson
(1996) to fit the FIR continuum spectrum.

Our CO model fit gives a gas mass of $3.1\times 10^9$\,\Msun\ ---
equal to the mass of all the molecular gas in the Milky Way --- in a
high-density disk between $R=76$ and 850\,pc.  There is another
$0.9\times 10^9$\,\Msun\ of gas in a lower density outer disk
extending to a radius of 1.7\,kpc.  The high-density molecular disk is
the region observed in H$_2$ at 2\,$\mu$m by Krabbe et al. (1997).
The brightest H$_2$ line is 200\,pc from the nucleus, in the molecular
disk, so the low 2\,$\mu$m extinction on the line of sight to the
nucleus itself does not apply to the H$_2$ flux.

Our conclusion that most of Mrk~231's FIR luminosity comes from a
starburst can be checked for consistency (Table~11).  We assume
the new stars in the burst, $M_{\rm new\star}$, yield $L_{\rm
FIR}/M_{\rm new\star}$ $\approx$ 500\,\Lsun \, \Msun$^{-1}$, a
luminosity ratio that can be attained in a fast starburst (Leitherer
\& Heckman 1995).
Dividing this ratio into the FIR luminosity yields 
$M_{\rm new\star}$, from which we then calculate 
$M_{\rm old\star}= M_{\rm dyn}-M_{\rm gas}-M_{\rm new\star}$,
  where $M_{\rm old\star}$ is the mass of old stars 
 in the nuclear bulge before the merger.
  For the region $R \leq 460$\,pc in Mrk~231, 
the rotation velocity inferred from our 
CO data  yields a dynamical mass
  $M_{\rm dyn}$($<460$\,pc) = $12.7\times 10^9$\,\Msun\ , and
  our CO model yields $M_{\rm gas}=$  
$1.8\times 10^9$\,\Msun .
The consistency check then yields
$M_{\rm old\star}= 7.1 \times 10^9$\,\Msun , 
which is about the same as the
  mass of old stars in a similar radius in the
  Milky Way (e.g., Oort 1977).  Because of our assumed $L/M$,
the derived stellar masses in the starburst and the bulge
are uncertain by at least a factor of two.  Our point is simply that
with a starburst powering the FIR luminosity, 
all the gas and stellar masses are quite plausible.
In summary, we think Mrk~231's black hole accretion disk emits 
$\sim 1\times 10^{12}$\,\Lsun , or $\sim 30$\% 
of the total bolometric luminosity.  Most of the luminosity,
 $\sim 2\times 10^{12}$\,\Lsun , comes from 
the starburst in the molecular ring or disk.  

\section{ {\bf Arp 193} }
The ultraluminous merger Arp~193 shows two tidal tails in the visible
and the near IR (Smith et al. 1995, 1996).  The IR images show a
single, elongated nucleus, but cm-radio continuum maps show two
sources separated by 1$''$.  Strong lines of Br$\gamma$ and H$_2$ are
seen in $K$ band (Goldader et al. 1995).  The near-IR continuum is
dominated by young supergiants formed in the merger-induced starburst.
Previous CO line ratio studies indicated sub-thermal CO excitation and
moderate densities (Radford et al. 1991).  Our detection of HCN in
Arp~193 indicated, however, that large amounts of dense gas were also
present (Solomon, Downes, \& Radford 1992).

 The CO(2--1) emission is nearly unresolved
in individual 20\,\kms\ channels, with temperatures up to 8\,K (Fig.~9).
The CO intensity and isovelocity contour maps (Fig.~10) reveal a 
rotating disk with
a line of nodes at the same orientation as the optical isophotes in
the center of the merger.   The
position-velocity diagrams and CO spectra (Figs.~11 and 12) 
along the line of nodes have the characteristic signature of a
rotating ring, with two peaks near the rotation curve turnover radius,
as in VII~Zw~31 and IRAS 10565+24.  Models that best fit the data are
those with a minimum ring radius $R_{\rm min}$ in eq.(1) equal to
$R_0$ = 220\,pc.  As in VII~Zw~31, filled-disk models, with $R_{\rm
min}$ = 0, give poorer fits, with much less contrast.  The centroid of
the CO source coincides with the main radio continuum peak on the 8.4~GHz
map by Condon et al. (1991).  
The strongest CO peak is  1$''$ to the southeast of the centroid, 
and peaks at $-$120\,\kms\
(relative to 225.282\,GHz). It is nearly unresolved on our maps, coincides with 
a secondary radio continuum peak.  This
hot, compact, southeast CO core in Arp~193 is responsible for the large
difference in the CO(2--1)/(1--0) ratio in the blue and redshifted
sides of the line profiles we observed at the 30\,m telescope (Radford
et al. 1991) . This region is hotter and denser than the disk and has 
properties similar to a huge
molecular cloud core.

This compact southeast core is one of several sources 
we identify in this study as {\it Extreme
Starburst Regions}  (see Section 8 on Arp~220 for a detailed discussion).  
Using the observed CO
luminosity, linewidth, radius, and temperature,  
we can estimate the gas mass, dynamical mass and
far IR luminosity of the core. The results, summarized in Table 12, 
indicate an object with a gas mass
$\sim 6 \times 10^{8}$\Msun , a luminosity of  $\sim 2\times 10^{11}$\,\Lsun , 
and about $\sim 1\times 10^{9}$\,\Msun\  of newly formed  stars. 

\section{ {\bf  Markarian 273 } }
The galaxy Markarian 273 has at least one Seyfert~2 nucleus with
optical linewidths of 700\,\kms\ (e.g., Sargent 1972; Koski 1978).
There is evidence for two Seyfert 2 nuclei (Asatrian, Petrosian, \&
B\"orngen 1990), and two nuclei are seen on IR images (Armus et
al. 1992; Majewski et al. 1993; Knapen et al. 1997).  Centimeter radio
maps show a $0''.3$ northwest peak coinciding with the CO, and a
weaker $0''.2$ southeast peak with no CO or IR counterpart (Ulvestad
\& Wilson 1984; Schmelz, Baan, \& Haschick 1988; Sopp \& Alexander
1991; Condon et al. 1991).  There is also an 18\,cm VLBI source of
16\,mJy (Lonsdale et al. 1993; Smith, Lonsdale, \& Lonsdale 1998a).
The best previous CO study is by Yun \& Scoville (1995), who observed
the source with a 2$''$ beam and found a component extended
north-south as well as an unresolved nuclear component.  Our
higher-resolution CO maps also show the extended structure, but now
resolve the nuclear source into a nuclear disk oriented east-west, and
a very compact core embedded in the nuclear disk.

\subsection {Extended molecular gas and nuclear disk in Mrk~273}
The extended gas is best traced in the CO(1--0) channel maps,
blueshifted to the south and redshifted to the north (Fig.~13a).
The CO at $+$250 to $+$350\,\kms\ runs in an arc from 3$''$ to 7$''$
(2 to 5\,kpc) north of the nucleus (Fig.~14). This extended gas
also appears in the maps of CO(1--0) integrated intensity and
isovelocity contours (Fig.~15a).  The CO(1--0) lines are weaker
and narrower in the northern arc than in the nuclear disk (Fig.~16).  
The arc has 4\% of the CO(1--0) flux of Mrk~273, so if gas
mass is proportional to CO flux, then the northern arc has a gas mass
of $1\times 10^8$\,\Msun .  Figure~17a is the CO
position-velocity cut in declination, through the nucleus.  The
narrow-line emission extends 7$''$ (5\,kpc) south at 0\,\kms\, and
7$''$ north at $+$300\,\kms .  It is tempting to speculate that the
streamers are bringing molecular gas into the center. The north-south
extent of the streamers on our CO map agrees with that on the map by
Yun \& Scoville (1995).  While those authors deduced a line of nodes
at p.a. 30$^\circ$, our isovelocity contour maps show there are two
perpendicular kinematic systems.  In the extended gas the velocity
gradient is north-south, and in the nuclear disk it is east-west.
  
The $2''$ nuclear disk with its compact core are best seen in the
CO(2--1) channel maps, made with a $0''.6$ beam (Fig.~13b).  The
nuclear disk kinematics are shown in the CO(2--1) map (Fig.~15b), 
where the isovelocity contours trace the east-west
velocity gradient of the nuclear disk over a 1\,kpc diameter.  
Figure~17b is the CO position-velocity diagram of the nuclear disk in
right ascension.  The broad nuclear disk line has a full width to half
maximum of 380\,\kms , and its velocity centroid changes from $-300$
to $+200$\,\kms\ over $0''.5$ from west to east.  This is a projected
velocity gradient of 1.5\,\kms\,pc$^{-1}$, the same as deduced by
Schmelz, Baan, \& Haschick (1988) from H~I absorption.  Table~11
lists the stellar and gas masses that would be needed in a starburst
in the Mrk~273 molecular disk if the $L_{\rm FIR}/M_{\rm new\star}$
ratio is the same as we adopted in our starburst model of Mrk~231.

\subsection{An Extreme Starburst Region in Mrk~273}
The high-resolution CO(2--1) maps show a remarkable 
molecular-line source in the Mrk~273 nuclear disk --- a bright,  
 $0.35''\times <0.2''$ CO core (Fig.~15b), that 
resembles the west nucleus of Arp~220. 
This is the most luminous extreme starburst region   
in our sample of 10 galaxies.
It has an IR luminosity of about $6 \times 10^{11}$\,\Lsun\ , 
generated in a region with a radius of only 120\,pc and 
 a current molecular mass of  $1 \times 10^9$\,\Msun\  
 (Table~12). 
To put this in perspective, 
the entire molecular core has a radius about 5 times that of
an IR luminous Milky Way GMC  (for example, W51) 
but with about 3,000 times the molecular mass and
$\approx \, 10^5$ times the IR luminosity from OB stars. 

This core has a broad CO line 
with a zero-intensity width of  1060\,\kms , the same
as the OH megamaser  (Staveley-Smith et al. 1987).  
 It coincides with the northwest extended continuum peak  
at 8.44\,GHz, which has a size $0''.32 \times 0''.18$
(Condon et al. 1991), the same as the compact CO.  
The radio spectrum is nonthermal, with a spectral index of $-0.6$.
The thermal dust emission dominates the spectrum above 350\,GHz
(Chini et al. 1989; Rigopoulou et al. 1996).
At the CO core, we detect continuum fluxes of
 11$\pm 2$\,mJy at 111\,GHz and $8\pm 2$\,mJy at 225\,GHz. 
Extrapolation of the synchrotron and thermal dust spectra 
indicates the dust contributes 50\% of the flux at 1.3\,mm 
and $<10$\% at 3\,mm. The extended nonthermal continuum emission coincident with a high
mass of dust and gas leaves little doubt that this region is powered by star formation.

\section{{\bf Arp 220} }
The center of the ultraluminous galaxy Arp~220 has 
two radio continuum and two IR sources 
$1''.0$ apart that are interpreted as merger nuclei.  
These are not exactly the same objects in the radio and the IR, because
both of the east and west IR nuclei are four times larger ($0''.8$) 
than the radio sources ($0''.2$) (Condon et al. 1991;
Graham et al. 1990; Majewski et al. 1993; Miles et al. 1996; Scoville
et al. 1998).   The $K$-band continuum is starlight associated with the nuclei,
and is best seen on the HST NICMOS images (Scoville et al. 1998). 
These images resolve the eastern IR nucleus into a strong NE and weaker SE
component, the latter of which coincides with the eastern radio continuum
component and the eastern OH masers (Diamond et al. 1989).
The two radio sources are extended and nonthermal, and are produced by 
supernovae in the most active star-forming regions.
The 1.3 and 1.6\,$\mu$m [Fe II] lines in the west source indicate iron 
evaporated from dust in shocks (Armus et al. 1995a; 
van der Werf \& Israel 1998).  
 The best previous CO study is that
by Scoville, Yun, \& Bryant (1997), who derived the kinematics of
the nuclear disk with a resolution of 1$''$.  

\subsection{CO and dust emission from the two nuclei of Arp 220}
Our CO(2--1) and 1.3\,mm continuum maps (Fig.~18a) show 
two compact sources embedded in more extended emission.
The continuum fluxes at 1.3\,mm are due to dust, 
because they are well above the extrapolated radio synchrotron spectra 
(Fig.~20).
The different appearance of the dust continuum and the CO maps is 
 a temperature effect. At 1.3\,mm the dust is optically thin
and its flux varies as the column density and the dust temperature, while
the CO flux is partly opaque, with flux varying with temperature at the 
surface, and partly optically thin, with flux 
varying inversely with CO excitation temperature.
 The Arp~220 molecular disk contributes strongly to the CO maps, 
while the warmer compact peaks dominate the dust continuum maps.
  From the 1.3\,mm dust flux alone, and for dust at 
100~K and solar metallicity, we derive gas masses for Arp~220-west and -east  
of 0.6 and  $1.1\times 10^9$\,\Msun .

In {\it Arp~220-west}, the $0''.3$ (100\,pc) mm continuum dust source coincides 
within $0''.1$ of the cm-continuum west peak, and 
the $K$-band west peak (Scoville et al. 1998). 
In the  CO(2--1) channel maps (Fig.~19b), Arp~220-west appears 
as a strong source in the range $-$310 to $+$210\,\kms ,  peaking at 
$-$110\,\kms\ ($cz_{\rm lsr} =$ 5340\,\kms ), with an 
{\it observed} CO(2--1) brightness temperature of 27\,K, 
of which 10\,K is from the more extended Arp~220 disk.
The positive-velocity emission near Arp~220-west is centered at $+$80\,\kms ,
and appears in both the CO(2--1) channel maps and in a 
position-velocity cut at p.a. 300$^\circ$ (Fig.~22b).  This positive-velocity
emission has about 25\% of the CO integrated intensity of the negative-velocity 
emission of Arp~220-west. 

The dust continuum is centered between the 
negative and positive velocity emission, suggesting that Arp~220-west is
a composite structure, with a kinematic axis nearly perpendicular to that
of Arp~220's main disk.  
At the negative velocity west core, the emission extends 
asymmetrically from  $-$50 \,  to $-$ 400 \, \kms\ (Fig.~22b).
 At the positive velocity core, it runs from $+$50 \,  to $+$350
\, \kms.  These extreme, asymmetric line profiles, 
confined to a small region, suggest possible radial
flows along a {\it bar--like} structure, or immense high-velocity molecular 
{\it outflows} similar
to those observed in high mass star formation regions.   
Molecular outflows  are also the likely sites
of the strong H$_2$ lines  (Sturm et al. 1996) .

In {\it Arp~220-east}, the CO covers
a wide velocity range.  The negative-velocity gas at $-$230 to $-$30\,\kms\ is   
separated from Arp~220-west by  $0.''85$ at p.a. 110$^\circ$ --- the
same angle as the cm-radio continuum sources, the 18\,cm OH masers, 
and the $K$-band SE peak (Scoville et al. 1998).
The center velocity of this gas is $cz_{\rm lsr} =$ 5330\,\kms , 
close to that of the OH megamasers and 
H$_2$CO emission at the cm-radio east source (Baan \& Haschick 1995; 
Lonsdale et al. 1998).   There is also
more extended, positive-velocity gas ($0''.9$ FWHM) at 130 to 290\,\kms .  
It is $1''.3$ from Arp~220-west at p.a. 85$^\circ$ --- the same 
orientation as the $K$-band NE source (Scoville et al. 1998).
This positive-velocity gas in Arp~220-east, 
at $\sim +$200\,\kms\ ($cz_{\rm lsr} =$ 5650\,\kms ), has 
the same velocity offset from Arp~220-west as the ionized gas seen 
in Br$\gamma$ and Pa$\beta$ (Larkin et al. 1995), 
and the weak OH masers at 1612~MHz (Baan \& Haschick 1987).
Its {\it observed} maximum CO(2--1) brightness temperature 
 is 19\,K, of which 4\,K is from the extended Arp~220 disk.
The 1.3\,mm continuum from dust in Arp~220-east has a diameter of $0''.6$ (205\,pc) 
and is displaced by $0''.3$ from the cm-radio east source.  It 
falls in between the $K$-band NE and SE peaks on the images by 
Scoville et al. (1998), and in between the positive and negative-velocity CO.
The kinematic data suggest that Arp~220-east has a steady progression 
of velocity between the CO peaks at 5330\,\kms\ and 5650\,\kms , along 
the same position angle (p.a. 50$^\circ$), and in the same sense of rotation
as the main, larger-scale molecular disk of Arp~220 (Fig.~22c).

\subsection{CO in the Arp 220 molecular disk}
The molecular disk is best seen as an extended source in the CO(1--0) maps  
(Fig.~18b). The CO(2--1) maps also show this same extended source,
when the compact peaks are subtracted.   
  The disk has CO halfwidths of $2''.0\times 1''.6$ at p.a. 50$^\circ$,  
 with a line of nodes at this same position angle,   
 blueshifted in the southwest and redshifted in
the northeast, as seen in the CO spectra along 
the major axis (Fig.~21). 
Our model fits yield a rotation curve turnover radius of 200\,pc, 
a disk outer radius of 480\,pc, and 
a disk rotation velocity of 330\,\kms\ on the flat part of the rotation curve.
The CO emission centroid of the
molecular disk is $\sim 0''.3$ east of the western nucleus.
 These parameters reproduce the
position-velocity diagrams along the kinematic major axis (Fig.~22),
and are similar to the values derived by Scoville et al. (1997) by a 
different algorithm.  On a larger scale, CO can be traced in the 
8$''$ diameter {\it outer disk} at the same position angle. The inner and 
outer disks both contain shocked gas, detected in the near-IR vibrational 
lines of H$_2$ (van der Werf 1996).

The CO(1--0) maps also show an eastern streamer extending 7$''$ (3\,kpc) 
perpendicular to the disk, in the range $+$45 to $+$245\,\kms ,  
curving in to the east nucleus 
with increasing velocity offset (Fig.~19a).  This east streamer 
is most intense at 85 to 165\,\kms\ (Fig.~18b).
It may be material that is still falling into the center.  
Surprisingly, this CO streamer appears on the HST 
$V$-band image by Shaya et al. (1994).  
The CO inner and outer disks coincide with the 
prominent optical dust lane at 50$^\circ$ (Fig.~23a), while  
the CO east streamer coincides with the 
perpendicular dust lane in the optical image (Fig.~23b).

Our model fits to the CO position-velocity diagrams of the Arp~220 disk 
(Fig.~22) yield a much higher dynamical mass
than estimates from the IR (e.g., Doyon et al. 1994;
Shier, Rieke, \& Rieke 1994, 1996; Larkin et al. 1995). 
From the masses in Table~9, 
we estimate $M(<350$\,pc) $=8.8\times 10^9$\,\Msun  ,
and $M(<600$\,pc) $=1.5\times 10^{10}$\,\Msun  , three to four times
higher than the IR estimates. The millimeter CO data and the 
2.3\,$\mu$m CO bandhead absorption data yield different masses 
because the two IR nuclei do not lie on the line of nodes traced by the 
CO maps and because the near IR starlight is obscured and 
does not trace the full extent of the molecular disk.

The gas mass derived from both the dust and the CO implies a
column density of $1\times 10^{24}$\,cm$^{-2}$ and $A_V \sim 1000$\,mag
through the disk, consistent with the dust having an opacity 
of unity at 180\,$\mu$m (Emerson et al. 1984; Scoville et al. 1991).  
The high opacity 
in the far IR may partly explain why the C$^+$ 158\,$\mu$m line is
 weak in Arp~220 (Fischer et al. 1998.).
The ratios of fine structure lines observed by the $ISO$ satellite
  yield an equivalent screen $A_V \sim 45$\,mag (Genzel et al. 1998).
As is well known, a screen model gives only an extreme lower limit.
The corresponding extinction when the emitting gas and dust are
completely mixed is $A_V \sim 1000$\,mag, the same as we deduce from
the column density derived from the CO maps.

Figure~24 shows our model of the molecular disk and 
the two ``nuclei".  
The CO disk is inclined  40$^\circ$ from face-on, Arp~220-west
has a radius of 68\,pc, and Arp~220-east has a radius of 110\,pc.
The disk thickness is 90\,pc, which means the path to the near surface
of the east and west K-band sources is 0 to 20\,pc,
which is why the
two nuclei are visible at all at $K$ band.  That is, although
the visible extinction through the entire molecular disk is 1000\,mag, the 
visible extinction to the near surfaces of the east and west sources 
is only about 50\,mag, because of the shorter path. 
Our diagram is similar to that of Scoville et al. (1997; their Fig.~7),
except our presentation is a view from the pole of the disk. 
In our model, the near side is north, 
the far side is south.  This agrees with the optical colors, which are  
bluer on the north side of the dust lane (Shaya et al. 1994).
 Since the CO east nucleus is south of the major axis, it is on 
the far side.  The CO disk or ring includes both nuclei
but the nuclei are oriented east-west, and hence the difference in
their line of sight velocities does not give the
full rotation speed of the disk. 

Our data and model have some important consequences for the interpretation
of the Arp~220 nuclear sources.  

1) The rotation curve of the CO disk indicates a dynamical mass of 
$12 \times 10^9$\,\Msun\ interior to 480\,pc, which corresponds to
the central bulge mass of a large spiral like the Milky Way.

2) The gas mass in each of the two  extreme starburst ``nuclei'' is only $6\times
10^8$\,\Msun . Their individual luminosities are $\sim 3\times 10^{11}$\,\Lsun .
(About half of the Arp~220 FIR luminosity comes from the molecular disk,
not the two nuclei).
With an L/M ratio of 1000, corresponding to a super-starburst in its 
initial phase, the mass of new stars in each nucleus 
would be only $3\times 10^8$\,\Msun ,
sufficient to explain all the $K$-band continuum luminosity.  
The velocity dispersion of the CO in each of the two nuclei implies
a dynamical mass of $\sim 1\times 10^9$\,\Msun , about the same as our
estimate of the sum of the gas and new stars.  

3) Hence, there is no room left over for old stars in the two ``nuclei" 
--- they cannot be the relicts of the old nuclei of the pre-merger 
galaxies.   Furthermore, there is no observational evidence --- radio,
infrared, or optical --- that they contain old stars.

4) In any case, the masses of the two ``nuclei'' are negligible in comparison
with the mass that controls the motion of the molecular disk.
Furthermore, 
the two ``nuclei" of Arp~220 have radial velocities indicating that
they take part in the general disk rotation, i.e., they share the 
general rotation in the potential of the old bulge, and are 
dominated by {\it its} gravity, not their own.

5) In our interpretation, the two ``nuclei" of Arp~220 are not the
pre-merger nuclei at all.  We think they are just large ensembles of
pre-existing GMCs that have been strongly compressed on their infall
into the old bulge potential as a result of the merger. 
The masses of these large, compressed gas structures can reach
10$^9$\,\Msun , and the strong compression, in about one orbital period,
leads to the extreme starbursts, with luminosities of a few times 
10$^{11}$\,\Lsun ,
from each compact 100-pc structure.  These extreme starbursts power
the ultraluminous galaxies.  We are not seeing the birth of quasars.

\subsection{Arguments for a starburst}

What produces Arp~220's luminosity? We think it's stars, for three reasons: 
there is no obvious AGN, there are enough ionizing photons for a starburst,
and there is enough dense molecular gas to make stars.

\noindent
1. {\it Arp~220 has no AGN lines.}  The mid-IR lines of [O~IV] and [Ne~V] 
are {\it not} detected (Sturm et al. 1996; Lutz et al. 1996).
The Br$\alpha$ width of 1300\,\kms\ reported by Depoy, Becklin, \&
Geballe (1987) as evidence for an AGN was not confirmed in later measurements 
(Goldader et al. 1995; Larkin et al. 1995).

\noindent
2. {\it Arp~220 has no AGN in the radio continuum.} 
New VLBA maps (Smith et al. 1998b) indicate all the VLBI point sources are 
young radio supernovae in a dense medium, with radio luminosities 
comparable to SN1986J in NGC\,891.

\noindent
3. {\it The radio nuclei are extended starburst regions.}
At centimeter wavelengths, their diameters are: west, $0''.21\times 0''.14$ 
($72\times 48$\,pc) and  east, $0''.32\times 0''.19$ ($110\times 64$\,pc)
(Condon et al. 1991). The source extent, the low brightness at 5\,GHz, and
the high FIR-to-radio flux ratio all show the
nonthermal radio continuum is starburst-dominated, not AGN-dominated
(Sopp \& Alexander 1991; Condon et al. 1991; Baan \& Haschick 1995).

\noindent
4.  {\it Arp~220 is not a ``warm" ultraluminous galaxy.}
IR ultraluminous galaxies with an AGN, like Mrk~231, Mrk 1014, and 
08572+3915, have warm mid-IR colors ($f_{25}/f_{60} > 0.2$) and a 
bright, symmetric nucleus  resembling a reddened QSO (Surace et al. 1998). 
Unlike these sources, Arp~220 has strong far-IR and weaker mid-IR flux  
($f_{25}/f_{60} = 0.08$).  It has been argued that a QSO is
hidden in Arp~220, but the silicate absorption depth at 10\,$\mu$m in 
Arp~220 (e.g., Smith, Aitken, \& Roche 1989; Dudley \& Wynn-Williams 1997)
has been greatly overestimated, due to the strong PAH lines 
directly adjacent to the silicate feature (Genzel et al. 1998), so
 a deeply embedded AGN is not required.

\noindent
5. {\it The diffuse OH megamasers do not indicate black holes.} 
A new VLBI study (Lonsdale et al. 1998)
shows Arp~220 has two OH megamaser components in each nucleus, 
one diffuse and the other
compact.  The {\it diffuse} megamaser component arises in the 
extended starburst medium, and can be explained by IR pumping via photons 
absorbed in the 35 and 53\,$\mu$m lines of OH.  The observed 9-Jy 
depth of the 35\,$\mu$m OH line can only be explained by absorption against an 
{\it extended} source (Skinner et al. 1997).   The {\it compact} OH megamasers,  
probably collisionally pumped shocks, are currently the only evidence for  
possible AGNs in Arp~220 (Lonsdale et al. 1998), but it is hard to
understand why such AGNs would 
have no associated radio continuum core-jet sources. 

\subsection{The ratio of bolometric flux to ionizing flux resembles
that in Sgr~B2.}
From the observed Br$\gamma$ flux, Armus et al. (1995b) 
and Shier, Rieke, \& Rieke (1996) estimated 
that $\leq 10$\% of Arp~220's luminosity came from a starburst. 
Armus et al.\ noted, however, if they had underestimated the 
$K$ band extinction, then a starburst could power the source. 
  The higher near-IR extinction is indeed 
 required by the high column densities found from the CO maps.
From the radio free-free continuum,
 Scoville et al. (1991; 1997) claimed there were not enough 
ionizing photons for a starburst to power Arp~220.
We think this expectation was wrong, because galactic star-forming regions 
like W49, W51, and Sgr~B2 have FIR luminosities
greatly exceeding the Lyman luminosities derived from their radio fluxes.  
 In compact HII regions like W3(OH), or in our galactic center, the FIR excess 
is a factor of $\sim 20$ (e.g., Zylka et al. 1995).

In H~II regions with densities $>10^3$\,cm$^{-3}$, 
most of the Lyman photons heat dust rather
than ionizing the gas (Jennings 1975;
Panagia 1977; Fazio 1978; Mezger 1985), so
the Lyman photon rate derived from the radio flux is only a lower limit.
We may estimate this lower limit to Arp~220's Lyman continuum photon flux
from the continuum at 113\,GHz, which includes 15\,mJy from the 
synchrotron spectra and 13\,mJy from dust (Fig.~20).  
This leaves 13\,mJy as optically thin free-free flux, and implies a 
 Lyman continuum photon rate  $> 1\times 10^{55}$ s$^{-1}$.  
Similar lower limits come from the H92$\alpha$ radio recombination line 
(Zhao et al. 1996), and the low-frequency turnover of the radio  
spectra of the two nuclei (Fig.~20), due to 
free-free absorption in ionized gas with an emission measure of 
10$^8$\,cm$^{-6}$\,pc 
and a mass of $5\times 10^{7}$\,\Msun\ (Sopp \& Alexander 1991).

For comparison, in the galactic giant H~II region and 
molecular cloud Sgr~B2, the FIR power is 10 times the ionizing luminosity 
derived from the radio (Gatley et al. 1978).  Sgr~B2 resembles 
Arp~220's molecular gas in column density, strong molecular lines, 
and far-IR opacity, 
but no one has ever claimed Sgr~B2's FIR excess proves it is powered by a 
black hole.   
Arp~220 and Sgr~B2 both have the same ratio of 
FIR flux (in W m$^{-2}$) to radio free-free flux 
(in Jy), and hence the same FIR excess.  So our lower
limit on Arp~220's ionizing flux, scaled by the Sgr~B2 template, 
yields exactly  Arp~220's  FIR luminosity of $1.2\times 10^{12}$\,\Lsun ,
and is indeed compatible with a starburst.
Similar conclusions are reached by Genzel et al. (1998) from
$ISO$ line data on the ionizing flux.

\subsection{Arp~220 has plenty of gas for a starburst.}
Our models yield gas masses of 
$0.6\times 10^9$\,\Msun\ for Arp~220-west, 
 $1\times 10^9$\,\Msun\ for Arp~220-east, and
$3\times 10^9$\,\Msun \ for the disk to $R_2=1.2$\,kpc (Table~9).
The total gas mass is $5\times 10^9$\,\Msun , about 6 times lower than 
previous estimates with a standard ratio of gas mass to CO luminosity.
Our value agrees with the gas mass of $3.5\times 10^9$\,\Msun\ 
obtained by Sturm et al. (1996) from the H$_2$ lines at 6.9 and 17\,$\mu$m,
although this latter value depends sensitively on the assumed
temperature.
Our method differs from that of 
 Scoville, Yun, \& Bryant (1997), who assumed all
 the dynamical mass is in gas, rather than stars, thereby  
forcing their model CO disk to be only 16\,pc thick at $R=$250\,pc, 
and to have a {\it mean} density of $1.5\times 10^4$\,cm$^{-3}$.
Because we do not assume all the mass is gas,
we derive a disk thickness of 80\,pc and a 
lower mean density than Scoville et al. 
Our higher-resolution data indicate the gas density is highest   
 at the two nuclei, not at the center of the molecular disk as in
the model of Scoville et al., and our value for the total gas mass is therefore a factor
of two lower than their estimate.

Even the lower molecular gas mass, however, is about the same as in 
the entire disk of a gas-rich spiral!  
It is enough gas for a huge nuclear starburst.
In the Arp~220 inner disk, 
the gas surface density is 100 times the
 peak value in the Milky Way 4-kpc molecular ring 
and more than 20 times the surface density 
in the inner few hundred pc of our Galaxy.
To estimate the {\it stellar} mass in the nuclear bulge, we 
adopted a global luminosity to starburst mass ratio in the Arp~220 disk 
 of  300\,\Lsun \,\Msun$^{-1}$, based on 
starburst $L/M$ ratios from the models of Leitherer \& Heckman (1995).
This gives a estimate of 
the mass in newly-formed stars from the
current merger-induced starbursts.  Subtracting this mass and the gas mass
from the dynamical mass then gives an rough estimate of 
the mass in old, pre-merger stars.  
The resulting bulge mass is $8\times 10^9$\,\Msun\ within $R= 480$\,pc  
(Table~11),  about the same as in a large galaxy like our own 
(e.g., Oort 1977).  The starburst
interpretation thus implies that old stars are a 
significant part of the dynamical mass.

\subsection{Arp~220-west has 25\% of the total 
luminosity.}
  The CO lines, the 1.3\,mm dust flux, and the cm-radio continuum 
all suggest the currently most active starburst is 
Arp~220-west. Its radius of 68\,pc
and dust temperature of 75 to 80\,K imply a blackbody luminosity of 
$3\times 10^{11}$\,\Lsun , or about 25\% of 
Arp~220's total IR luminosity of $1.3\times 10^{12}$\,\Lsun , as expected 
from the $R^2 T_d^4$ ratio of the west and disk sources.  
This estimate also agrees with the fraction of the 
mid-IR flux absorbed by the 35\,$\mu$m OH line at 5320\,\kms\  
(Skinner et al. 1997), the velocity of Arp~220 west.  
If the Lyman continuum scales 
as FIR luminosity, then Arp~220-west has 1000 times the ionizing flux of 
30~Doradus, which is 10$^{52}$ photons s$^{-1}$ (Kennicut \& 
Chu 1994), from  2400 OB stars in a region of comparable size   
(Parker 1993). Because Arp~220-west is small and the dynamical time scale is 
short, its starburst must be younger (age  $5\times 10^6$\,yr) than
that in the larger Arp~220 disk. The initial phase of a compact starburst 
can be highly luminous, reaching $L/M$ ratios of 
1000 -- 3000\,\Lsun \,\Msun$^{-1}$  (see models in Leitherer \& Heckman 1995).
Our estimate of the total mass (from the CO linewidth) 
and the gas mass (from the CO luminosity), suggests that in 
 Arp~220-west,  50 to 60\% of the original gas mass has already turned 
 into the new stars in the current, extreme starburst 
(Table~12).

\section{SOURCES OBSERVED IN CO(1--0) ONLY}
We observed these sources with lower resolution and
  thus our kinematic models are less precise.  
All these sources show the same phenomena as the previous group of sources, 
however, with velocity gradients 
and line profiles that suggest rotating rings or disks.

\subsection{ {\bf 00057$+$4021} }
The galaxy IRAS 00057$+$4021 has not been well studied, 
despite its high luminosity
of $L_{\rm IR} = 4\times 10^{11}$\,\Lsun , and its OH megamaser 
(Kaz\`es, Mirabel, \& Combes 1988).  
The $R$-band image by Armus, Heckman, \& Miley (1987)
shows a 60$''$ (48\,kpc)  southeast-northwest disk with 
a tidal tail stretching a further 45$''$ (36\,kpc) southeast.  
In contrast to the optical light, the CO source is quite compact. 
Most of the molecular gas is in a $1''.1\times <0''.6$ core source.
The CO has wider lines (140\,\kms \ FWHM) and 
is more intense on the southeast side of the line of nodes
than on the northwest (Fig.~25). 
The continuum flux from the CO source is 
$<$10\,mJy at 110\,GHz.  This limit on the thermal dust flux is 
consistent with the gas mass of $9\times 10^8$\,\Msun\
deduced from our model fits to the CO.

In the position-velocity diagram along the line of nodes at p.a. 135$^\circ$ 
the CO line has 
 a velocity range of 300\,\kms\ over a diameter of 3$''$ (Fig.~25).   
The CO lines are blue-shifted in the northwest 
and redshifted in the southeast. In strips parallel to the kinematic major
axis the CO spectra show the double-peak behavior of a rotating
disk or ring.  To the southeast, the CO profile is steep 
toward the red, and gently sloping toward the blue.  To the northwest,
the profile is reversed.  The nuclear velocity gradient in CO has the same  
major axis as the larger-scale optical isophotes.
The position-velocity data and the spectra both indicate 
another, lower-intensity (0.5~K) disk
with a broad linewidth (200\,\kms ) in the same region as the brighter,
high-density disk.

\subsection{ {\bf 02483$+$4302} }
The galaxy IRAS 02483+4302 
is a merger with a tidal tail extending 90$''$ to the west.
The merger has two nuclei separated by $3''.8$  
east-west.  In optical lines, nucleus $A$ 
(west) has a Seyfert~2 spectrum, nucleus $B$ (east) has a LINER spectrum.  
The optically more intense nucleus $A$ appears to belong to an
elliptical galaxy plowing through the disk of a former spiral containing 
nucleus $B$ (Kollatschny et al. 1991).  
The optical continuum of nucleus $B$ comes partly
from hot stars (Womble et al. 1990). 

The CO (Table~2) coincides with nucleus $B$, 
supporting the idea that this was originally a gas-rich spiral.  
Our CO position also agrees within 0.5$''$ of the 13\,mJy source seen at
5~GHz (Crawford et al. 1996).  
At 109\,GHz, the continuum flux from the CO source is 
$<10$\,mJy. This limit on the thermal dust flux is 
consistent with the gas mass deduced from the CO.
In velocity channels 40\,\kms\ wide, the CO source is very compact.
Figure~26 shows the maps of CO integrated intensity and 
isovelocity contours across the source.
The CO has a north-south velocity gradient, with  
a kinematic major axis perpendicular to the optical tidal
tail extending west from the merger nuclei.

In strips parallel to the kinematic major
axis the CO spectra show the symmetric behavior of a rotating
disk or ring.  In the south, the CO profile is steep
toward the blue and gently sloping toward the red.  In the north,
the profile is reversed.  
The position-velocity diagram along the line of nodes shows
a gradient of 200\,\kms\ over $0''.9$ in declination (Fig.~26).  
The narrow CO linewidth and the low optical extinction to the  
nucleus  both suggest the molecular disk around 
nucleus $B$ is face-on. 

The CO source, at $z=0.05144,$ ($cz_{\rm lsr} = 15420$\,\kms ) 
and the quasar Q0248+430  at $z=1.311$ are separated on the sky
by $16''.9$ (the galaxy-quasar separation is incorrectly listed as 
$3''.5$ by Burbidge (1996) and by Hoyle \& Burbidge (1996)).
The tidal tail that extends $>80$\,kpc from the two merger nuclei 
crosses the quasar at a projected distance of 15.4\,kpc from the CO source. 
At this position, low-density atomic gas in the tidal 
tail is seen in absorption against the quasar in the 
Na~I~D and Ca~II~H and K lines at $z=0.0515$ and 0.0523, 
with linewidths $< 150$\,\kms\ (Womble et al. 1990).
There are two other absorption systems, in Mg~II, at $z = 0.394$ and 0.451,
from the halos of other, more distant galaxies on 
the line of sight to the quasar (Womble et al. 1990; 
Sargent \& Steidel 1990; Borgeest et al. 1991).        
After correcting for the primary beam, we derived the 
quasar's continuum flux to be 190~mJy at 110~GHz in June-October 1994.  
 Toward the quasar, our data at 20\,\kms\ resolution  
do not show any CO(1--0) absorption within
 $\pm 600$\,\kms\ of the center frequency (Table~1),
to a limit of 30\,mJy, or 15\% of the quasar's millimeter continuum flux, 
  This range includes the $z=0.0523$
($cz = 15300$\,\kms ) redshift seen in absorption against the quasar 
in  Na~I and Ca~II.   
If the absorbing atomic gas extends  0.3 to 3\,kpc
on the line of sight, the width of the tidal tail on optical images, 
then the column densities derived from the Na~I and Ca~II 
lines by Womble et al. (1990) imply H~I densities of 1 -- 10\,cm$^{-3}$, 
 too low for any gas to be in molecular form.

\subsection{ {\bf  10565$+$2448} }
The main optical and near IR peak of IRAS 10565+2448 has strong Br~$\gamma$
lines (Goldader et al. 1995) and  H~II region-type line ratios (Armus, 
Heckman, \& Miley, 1989, 1990; Veilleux et al.\ 1995).
Murphy et al. (1996) suggest this merger is a triple galaxy system.
On their $r$ band image there is a secondary source on a tidal tail,
26$''$ (20\,kpc) northeast of the main peak, at the same redshift. 
A third object with unknown redshift is 7.9$''$ (6.2\,kpc) southeast 
of the main peak.  In our 500\,MHz band 
we detect no CO toward the other two $r$-band sources, 
to a limit of 20\,mJy~beam$^{-1}$ in 20\,\kms\  
channels.  At centimeter wavelengths, there is extended emission and an 
unresolved, nonthermal core (Condon et al. 1991; Crawford et al. 1996).  

Figure~27 shows the maps of CO integrated intensity, isovelocity 
contours, and CO linewidth. The peak of the $2''.3\times 1''.7$ CO source 
coincides with the compact cm-radio continuum core.
Along the line of nodes, the CO line profiles are very symmetric,  
with well-defined blueshifted peaks in the east and redshifted peaks in the 
west (Fig.~28).  Together with VII~Zw~31 and Arp~193, the galaxy 
10565$+$2448 is one of the best examples of a rotating 
ring in our sample. 
In the position-velocity diagram (Fig.~29), there are two 
prominent peaks, and a
velocity shift of 180\,\kms\ over the central $1''$ in R.A. 
 Models that best fit the data are
those with a minimum ring radius $R_{\rm min}$ in eq.(1) equal to 
$R_0$ = 230\,pc.   
As with VII~Zw~31 and Arp~193, 
filled-disk models, with $R_{\rm min}$ = 0, give
poorer fits, with much less contrast.   
The CO ring must be nearly face-on, because of the small 
separation (80\,\kms ) of the twin peaks, 
the narrow CO linewidth (140\,\kms\ FWHP), 
and the small extent of the integrated CO intensity on the 
kinematic major axis. 
Unlike Arp\,220 or VII\,Zw 31, there is not much CO in an outer
 disk beyond the nuclear ring of 10565+2448.  

\subsection{ {\bf 17208$-$0014} }
The ultraluminous galaxy 17208$-$0014 
has $L_{\rm IR} =2.2\times 10^{12}$\,\Lsun\ and
an optical and near-IR line spectrum 
that indicates H~II region-type excitation and high reddening
(Martin et al. 1989; Kim et al. 1995; Goldader et al. 1995; 
Veilleux et al.\ 1995). 
Images at $r$-band (6550\,\AA) show two tidal tails
from a merger (Melnick \& Mirabel 1990; Murphy et al.
1996).  The innermost parts of the two tails, extending north and east to 
radii $>20$ kpc are nicely shown in the $r$-band image by Sanders \& Kim 
(in Solomon et al. 1997). At $2''.9$ (2.2\,kpc) southeast of the nucleus,
the $r$ band and $I$ band (0.82\,$\mu$m) images  show 
 a secondary peak on a tidal tail.
The $K$-band (2.2\,$\mu$m) surface brightness has an $r^{1/4}$
profile out to a radius of 10\,kpc, as in elliptical galaxies, probably
due to the merger.  
The nucleus has a size of $1''.8 \times 1''.4$ in $K$ band and appears 
single, which suggests a completed merger 
(Zenner \& Lenzen 1993; Murphy et al. 1996). 
The nucleus also has one of the strongest known OH megamasers, 
with 10$^3$\,\Lsun \ in the 1667\,MHz OH line, 
and a radio continuum source of size $0''.32 \times 0''.26$ 
(220$\times 270$\,pc; Martin et al. 1989).

Figure~30 shows the maps of CO(1--0) integrated intensity and 
isovelocity contours.  The CO source has a 
size of $1''.8 \times 1''.6$,  smaller than
measured by Planesas, Mirabel, \& Sanders (1991), and about the same size
 as the $K$~band source.  The centroid of the CO
source (Table~2) agrees to within $0''.2$ with the  
cm-radio continuum source (Martin et al. 1989).
The isovelocity contours and the CO spectra show the molecular gas is 
blueshifted in the northwest and redshifted in the southeast.
The position-velocity diagram (Fig.~30) 
indicates a strong velocity gradient 
with a change  of 400\,\kms\ over $1''.5$ at p.a. 120$^\circ$.  
We interpret this angle 
as the line of nodes of the nuclear disk.  The CO spectrum at the 
peak of the source (Fig.~30) has a linewidth of 
375\,\kms\ FWHM, and 700\,\kms \ to zero intensity.  At the CO
peak, there appears to be a weak continuum at the 5\,mJy
level.

\subsection{ {\bf 23365$+$3604} }
On $R$-band images, the ultraluminous galaxy 23365$+$3604
has two tidal tails, but only a single, blue nucleus
(Klaas \& Els\"asser 1991; Murphy et al. 1996).   
The optical spectrum (e.g., Veilleux et al.\ 1995)
is of the LINER type, with line ratios 
implying shock velocities of 80 -- 90\,\kms\ and pre-shock electron
densities of 1 -- 10\,cm$^{-3}$ in tenuous gas, as in old
supernova remnants (Klaas \& Els\"asser 1991).  In the nucleus,
The [O III] line has a width of 260\,\kms\ FWHM (Kim et al. 1995).
The source has strong Br$\gamma$ and H$_2$ lines at 2\,$\mu$m 
(Goldader et al. 1995). In the 6\,cm radio continuum, there is a secondary
source extending 1$''$ south of the nucleus (Crawford et al. 1996).
 
In our 40\,\kms\ channel maps, the CO source is compact in all channels
over a range of 320\,\kms .  
The CO source has a size of $1''.0\times 0''.9$, with a peak 
2$''$ south of the optical position listed by Klaas \& Els\"asser (1993).  
The gradient in the isovelocity contours is at p.a. 135$^\circ$, and 
the CO spectra along this line of nodes
show asymmetric profiles characteristic of a rotating ring or disk. 
Along this kinematic major axis, the position-velocity diagram  
shows a shift of 300\,\kms \ over 4$''$ (Fig.~31).

\section {DISCUSSION}
\subsection{Source sizes, gas masses from CO imaging, and the ratios 
  $M_{\rm gas}$/$L^\prime_{\rm CO}$ 
and $M_{\rm gas}$/$M_{\rm dyn}$
}
The main results of this study are the CO source 
sizes --- the halfwidths of the integrated CO emission (Table~3), 
and the inner and outer disk radii from the model fits to the 
position-velocity diagrams and the channel maps (Table~4).  
The measured radii of the integrated CO correspond to the 
half-power radii $R_1$ derived from the kinematic data.
We had previously noted that for ultraluminous galaxies the CO(1--0)  
flux (Jy\,\kms ) was $\sim 4$ times the 100\,$\mu$m flux (Jy), as expected 
from a blackbody model for the far IR (Downes, Solomon, \& Radford 1993).  
We derived the dust temperature from the FIR fluxes, and 
took this to be the intrinsic CO brightness temperature.  From the CO 
luminosity, we then predicted the mean radius, $R_{\rm CO}$, for a spherical 
CO source (Solomon et al. 1997). Interestingly, the sizes measured with 
the interferometer are within a factor of two of the blackbody sizes.  
The radii $R_0$ and $R_1$ from our fits to the 
kinematic data (Table~4) bracket our single-dish estimate:
$R_0 < R_{\rm CO} < R_1$.  The CO radii measured with the interferometer 
now allow us to better estimate the 
gas and dynamical masses (Table~9). 

In the centers of ultraluminous
galaxies, we find $M_{\rm gas}$/$L^\prime_{\rm CO} \approx 0.8$
\Msun\,(K\,\kms\,pc$^2$)$^{-1}$, about 5 times lower than the standard value
 for self-gravitating molecular clouds.
We had anticipated this result from single-dish data (Downes et al. 1993; 
Solomon et al. 1997), and it is now confirmed by the 
interferometer maps.  
We had earlier used the CO linewidths and radii from our blackbody model 
to obtain dynamical mass, which in turn led us to revise downward the 
gas masses from  single-dish CO luminosities.  We estimated
a minimum gas mass by assuming optically thin CO.
These estimates, which we labeled $M_{\rm thin}$ (Solomon et al. 1997),
are within a factor 1.5 of the gas masses 
derived from our model fits to the interferometer data.
Table~9 also lists the ratio of gas mass to dynamical mass,
which is $\sim 1/6$ in the inner, high-density disk, 
and 1/10 in the outer, low-density disks. 

 We showed earlier (Downes et al. 1993) that 
the $M_{\rm gas}$/$M_{\rm dyn}$ and 
$M_{\rm gas}$/$L^\prime_{\rm CO}$ ratios are related by
\begin{equation}
	\bigg({ M_{\rm gas} \over M_{\rm dyn} }\bigg) =  
	\bigg( {M_{\rm gas} \over L^\prime_{\rm CO}} \bigg)^2
 	{1 \over \alpha^2 }  \ \ \ ,
\end{equation}
where, letting  $f\equiv M_{\rm gas}/M_{\rm dyn}$, we have    
\begin{equation}
	M_{\rm gas} = f^{0.5} \alpha L^\prime_{CO} \ \ \ .
\end{equation}
The factor $\alpha = C n^{0.5}/T_b$, where $n$ is the mean H$_2$ number 
density over the whole volume and $T_b$ is the CO brightness temperature.
In the units used here, the constant $C=2.6$ for a sphere
and $\sim $1.0 for a flared disk with thickness/radius ratio of
0.15, as in some of the disks in our sample.   
These equations depend only on gravity, and are independent of
whether the gas is self-gravitating or not, or whether the CO lines are 
optically thick or thin.

The derived values for  $M_{\rm gas}$/$M_{\rm dyn}$ and 
$M_{\rm gas}$/$L^\prime_{\rm CO}$ are compatible with starbursts.
Assume a starburst powers an ultraluminous IR galaxy, with 
$L/M_{\rm new\star}$ = 300\,\Lsun \Msun$^{-1}$ for the new stars in
the burst, all the power emerging in the IR. 
Such galaxies typically have  
$L_{\rm IR}/L^\prime_{\rm CO}$ = 200\,\Lsun (K\,\kms\ pc$^2$)$^{-1}$.
Combining the two luminosity ratios gives 
\begin{equation}
M_{\rm new\star} = (2/3) L^\prime_{\rm CO} \ \ \ ,
\end{equation}
or $\sim 3\times 10^9$\,\Msun\ as the mass of new stars 
in the burst.  Most galaxies prior to the merger would have had 
a rotation velocity $\sim $250\,\kms\ at a radius of 500\,pc, which 
implies an old stellar bulge about twice as massive as the newly-formed
population.  Since $M_{\rm dyn}$ = $M_{\rm gas}$ + $M_{\rm new\star}$ 
+ $M_{\rm old\star}$, we have from the previous equation that
\begin{equation}
	M_{\rm dyn} = M_{\rm gas} + 2 L^\prime_{\rm CO}.
\end{equation}
Solving for the molecular gas mass yields
\begin{equation}
	{ M_{\rm gas} \over L^\prime_{\rm CO} } = (1+\alpha^2)^{1/2} - 1 \ \ \ .
\end{equation}
This equation can be generalized to arbitrary mass ratios of the old
and new stellar populations.  For the Arp~220 disk, with a {\it mean}
H$_2$ density $\sim 500$\,cm$^{-3}$ and $<T_b>$ = 20\,K, these
equations predict $\alpha\sim 1$, $M_{\rm gas}/M_{\rm dyn}$ = 0.2, and
$M_{\rm gas}/L^\prime_{\rm CO}$ = 0.4\,\Msun\,(K\,\kms\,pc$^2)^{-1}$,
close to the values derived from our model fits to the interferometer
data.  In other words, the starburst scenario is compatible with the
derived ratios.

\subsection{Dense gas traced by HCN and extreme starburst regions}
There is another important gas component that is better traced by
emission from high dipole moment molecules like HCN, rather than CO.
The HCN emission requires densities $n$(H$_2$)$ >10^4$\,cm$^{-3}$.  In
giant molecular clouds in the disks of the Milky Way and normal
spirals, the average HCN emission is weak with a typical ratio of
\Lhcn/\Lco $= 1/20$ to 1/40 (Gao \& Solomon 1998). In the Milky Way,
HCN emission is strong only in cloud cores that form high-mass stars.
We showed previously that ultraluminous galaxies have abnormally high
HCN luminosities (Solomon, Downes, \& Radford 1992), with \Lhcn/\Lco
$= 1/4$ to 1/8, indicating a higher fraction of the total molecular gas 
is in
dense, $ >10^4$\,cm$^{-3}$, star-forming cores than in normal
galaxies.  The densities we derive for the smoothly distributed gas in
the rotating disks (see Table 6) are typically 10 times lower than that 
needed for HCN excitation, except
for the extreme starburst regions Arp220-west, Arp220-east and the
disk of Mrk231.  The denser gas responsible for the HCN emission (and
CS emission) is thus not adequately accounted for in most sources by
the ``diffuse" CO emitting component that fills the disk.

In Arp~220, the HCN lines we observed have the same velocities as 
the east and west ``nuclei'', 
so it seems that the east and west ``nuclei'' alone account for most of
the HCN emission.  These dense, compact sources have a hydrogen column
density of $0.6 \times 10^{25}$\,cm$^{-2}$ and mean density of 20,000
cm$^{-3}$, enough to thermalize the lower rotational levels of HCN by
a combination of collisions and radiative trapping.  Within the
Arp~220 east and west sources, the HCN emission may thus have the same
intrinsic brightness temperature as the CO(2--1) emission, namely,
50\,K (Table~8).  Using the sizes and linewidths from Tables~4 and 5,
we estimate the HCN luminosity of these two regions alone to be 
\Lhcn $\approx 7 \times 10^8 $ K\,\kms\,pc$^2$, which is 3/4
of the observed total (Solomon, Downes, \& Radford 1992). 
These two regions thus emit only 1/4
of the CO luminosity but most of the HCN luminosity. 
Their total gas mass of 1.7 $\times 10^9$\,\Msun\  is
already accounted for in the mass budget in Table~9.  
Thus for Arp~220, only a small increase in 
 the  mass budget would be enough to account for the HCN emission. 

 While some of the HCN emission undoubtedly comes 
 from dense star forming molecular cores  embedded 
 in the (relatively) diffuse CO disks of ultraluminous galaxies,  
 it is likely that high density, extreme starburst regions  
 similar to Arp~220 east and west exist in 
 most of the other ultraluminous galaxies in our sample, and are the 
real sources of most of the HCN emission. 
 Arp~220 is the closest
ultraluminous galaxy and has the best resolved structure. 
The second closest
galaxy, Arp~193  (see section 6 and Figs.~10 and 11), 
also shows evidence of an extreme starburst in the
southeast, that appears similar
to the strong sources in Arp~220.  
We suggest that the high-density, extreme starburst regions are the source of
much of the HCN luminosity but only a fraction of the CO luminosity.  
Most are beyond detection at our
current resolution in CO emission since their size of $\approx$100 pc 
would be $< 0''.14$
 for all but two of the galaxies (see Table~3).  
These objects may then add about $1 \times 10^9$\,\Msun , 
or 25\%,  to the total gas mass in galaxies with very high HCN emission.  
This is less than we estimated in our 1992 paper. 
In line with this re-interpretation, we note that 
the galaxy VII~Zw~31 has much weaker HCN emission than most of the others 
in our sample. It also has
the largest CO disk, a lower ratio of \Lfir/\Lco\ 
(less star formation per solar mass of gas), and may not
have any extreme starburst regions like Arp~220 east or west.

\subsection{Stability of the molecular disks}
The standard parameter for characterizing 
the stability against local, axisymmetric perturbations
of a disk that is supported
by differential rotation and random motion is  
\begin{equation}
	Q= {{\sigma_v \kappa}\over {\pi G \Sigma} } \ \ \ \ ,
\end{equation}
where $\sigma_v$ is the one-dimensional random 
velocity dispersion, $\kappa$ is the local epicyclic 
frequency, and $\Sigma$ is the mass surface density.
(Safronov 1960; Toomre 1964; Goldreich \& Lynden-Bell 1965; 
for a recent review of the criterion applied to gravitationally coupled
stars and gas in a disk, see Jog 1996).
If $Q$ is $<1$, the structure is unstable, and
large massive star clusters may form.  
We used our interferometer data on the sizes, turbulent velocities, 
rotation velocities, and mass surface densities to estimate the 
stability of the nuclear disks of the ultraluminous galaxies.
In our model rotation curves, the epicyclic frequencies are 
$\kappa \approx 2V/R =$ constant on the rising part of the rotation curve,
and $\kappa \approx \sqrt 2 V/R$ on the flat part of the rotation curve.
The objects in our sample 
have high gas mass fractions and 
typically have $Q_{\rm gas}\leq 2$, but $Q_{s+g} <1$ for stars plus gas 
(the regime in Fig.~1e of Jog 1996).
For Arp~220, we obtain $Q_{\rm gas}$ = 2.2 at $R=480$\,pc for the gas alone, 
but 
$Q\sim 1$ for the gas plus stars.  
For VII~Zw~31, at a radius of $R=1100$\,pc, 
we obtain $Q$ = 1.1 and 0.9 for these two masses.
These values suggest the molecular disks  
of Arp~220 and VII~Zw~31 are globally unstable against 
axisymmetric perturbations 
and will form massive star clusters.

Empirically, this instability in the central disks appears to 
produce one or more 
large clumps of dense molecular gas 
--- the compact, nearly unresolved peaks on the CO maps.  
Table~12 lists some of these 
compact regions of dense molecular gas, 
 which
we identify with extreme starbursts. 
Their typical radius of 70 to 100\,pc may be the scale on which 
 the two-component (stars + gas) system becomes unstable.  Their mean 
H$_2$ density can reach  
$2\times 10^4$\,cm$^{-3}$, and their mass appears to be  
$\sim 1\times 10^9$\,\Msun\ of gas
initially, which then forms 
 $\sim 10^6$ OB-type stars over 10$^7$\,yr,
or $\sim 1000$ times the number of OB stars in 30~Doradus. The total
luminosity of one of these extreme starburst regions is $3\times 10^{11}$
to $1\times 10^{12}$\,\Lsun .  The central disk of an ultraluminous IR
galaxy typically contains two or three such extreme starburst 
regions at any given time.
It is mainly these regions that provide the input 
power to the ultraluminous galaxies.
They heat the dust in the central disks to typical temperatures of 75\,K,
(the blackbody fits to the colors measured by IRAS).  Since the dust 
 is opaque at 100\,$\mu$m, the disk radiates as a black body.  The
typical disk radius is 200 to 300\,pc, and the Stefan-Boltzmann formula
for the disk as a whole yields a luminosity of the order of 10$^{12}$\,\Lsun ,
the typical output power of the ultraluminous galaxies.  If these
starbursts occur in the old, pre-merger nuclei, they change them 
considerably, creating a new cusp with a much greater density of   
stars.  If the extreme starbursts occur slightly outside of the old 
pre-merger nuclei, they will form new nuclei -- new cusps of high 
stellar density. 
\subsection{Why stars outshine black holes}
The CO disks' radii of $\sim$500\,pc and 
 rotation speeds of $\sim$300\,\kms\ yield orbital periods of
10\,Myr.  If their FIR luminosity of $\sim 10^{12}$\,\Lsun\  comes 
from new stars, then our estimates of
 $L_{\rm FIR}/M_{\rm new\star}$ = 300\,\Lsun \Msun$^{-1}$
imply star forming rates of 50\,\Msun\,yr$^{-1}$.  The mass of gas plus 
new stars is $\sim 6\times 10^9$\,\Msun , so 
 in 10 rotations of the molecular disks, half the gas turns into new stars.

If part of the $\sim 10^{12}$\,\Lsun\ luminosity came from a black hole 
accretion disk radiating at 
$L\sim 0.1\ \dot m c^2$, the accretion rate would be 
$\sim 1$\,\Msun\,yr$^{-1}$,  so in 10 molecular disk rotations 
the black hole could only accrete $\sim 1$\% of the molecular gas.
The other 99\% of the gas would continue to form stars over this long period,
outshining the black hole.

Alternatively, if the gas fell in faster and created 
a $6\times 10^9$\,\Msun\ black hole in one or two orbital periods, then 
the accretion rate would be 60\,\Msun\,yr$^{-1}$, and a 
standard Shakura \& Sunyaev (1973) optically thick 
accretion disk, radiating at one-tenth the Eddington luminosity, would have a    
luminosity shooting up to 10$^{15}$\,\Lsun\ --- 1000 times
more luminous than the IR ultraluminous galaxies and 
quasars in the local universe.
If the gas fell in rapidly at an  accretion rate of 60\,\Msun\,yr$^{-1}$, 
but the 
luminosity stayed at 10$^{12}$\,\Lsun , then the accretion disk's
radiative efficiency would be only 
$L = 10^{-4}\ \dot m c^2$  --- not any more efficient than producing 
energy from starbursts.  Although it is unlikely that such a massive,
high-density, accretion flow would be advection dominated (e.g., Rees 1982),
if the flow were in this regime, then almost by definition, the accretion
disk would not be a luminous object.

The model of rapid accretion to a black hole thus poses more problems 
than it solves.  With a very high accretion rate of 60\,\Msun\,yr$^{-1}$,
why would the IR ultraluminous galaxies only have an output of 
10$^{12}$\,\Lsun\ if their power source were a black hole?  Why aren't 
they as luminous as the powerful quasars at high redshifts,
since there is no shortage of fuel?  Inversely, if the power source were
a black hole accreting at a modest rate of 1\,\Msun\,yr$^{-1}$,
it would last for $6\times 10^9$\, years and 
there would be many more ultraluminous
galaxies than are observed.   Since the source statistics indicate
the ultraluminous
phase in mergers lasts for $\sim 10^8$ years, the more likely answer is
that the gas is used up in forming stars -- not quasars.

In summary, even in 10 rotations of the molecular disk, a black hole
radiating 10$^{12}$\,\Lsun\ at high efficiency
could accrete only 1\% of the molecular disk's mass.
So even though new stars convert matter into energy less 
efficiently than a standard accretion disk, 
they make up for this in ultraluminous
IR galaxies by occurring in burst, 
in a 2000 times larger volume, with a 2000 times
larger total mass than is available within the Bondi radius $G\,M_{\rm bh}/V^2$
of a supermassive black hole.   The molecular disks probably
survive for at least 10 rotations.  During this time, at most
10$^7$\,\Msun\ is accreted to a black hole.  The rest of the molecular 
gas forms stars.

\section{CONCLUSIONS}
\noindent
{\it 1.) Rotating disks:} At sub-arcsecond resolution,  
our CO maps of ultraluminous galaxies 
show rotating disks of molecular gas that has been 
driven into the centers of the mergers.  
The maps of Mrk~273 and Arp~220 also show large-scale 
streamers or tidal tails roughly perpendicular to the nuclear disks.

\noindent
{\it 2.) CO only moderately opaque:} 
Model fits show that the observed double-peaked CO spectra, and the 
peak-to-center contrast in the twin-peaked patterns in 
CO position-velocity diagrams, are produced by radiative transfer  
through the rotating disks, in CO that is only  moderately opaque. 
 The CO(1--0) opacities are 4 to 10 at the peaks, and lower elsewhere in
the disks.

\noindent
{\it 3.) High turbulence in the molecular disks:}  In the Arp~220
disk, the 1-D velocity dispersion of the molecular gas is 
$\sigma = 100$\,\kms\ 
(FWHM 230\,\kms ). This high turbulence is one of the reasons why the 
CO is less opaque than in quiescent Milky Way molecular clouds. 
The turbulence also determines the disk   
thickness, and the heat input to the gas.
The turbulent heat input, $M_{\rm gas} \Delta V^3/R$,  is
$\sim 10^9$\,\Lsun , the same as the output from 
the 17\,$\mu$m S(1) line of H$_2$, a principal cooling line (Sturm et al. 
1996).  The line luminosity is that expected from $10^9$\,\Msun\ of gas
at $\sim 100$\,K.

\noindent
{\it 4.) Most of the CO luminosity comes from relatively ``low-density" gas:}  
Except for the compact cores 
in Mrk~231, Mrk~273, and  Arp~220, the molecular disks have 
true CO brightness temperatures  $\sim 20$\,K.   This is 
rather lower than  the dust and gas kinetic temperatures, which are 
 $\sim$65 to 100\,K.  
This implies that the density over most of the molecular disk is relatively  
low (300 to 2000\,cm$^{-3}$), giving subthermal excitation.
The CO(2--1)/(1--0) ratios are also consistent with  
 the CO being subthermally excited.

\noindent
{\it 5.) The CO lines in the molecular disks 
come from a continuous medium, not from 
self-gravitating (``virialized") clouds:}   
At the densities of 10$^3$\,cm$^{-3}$ 
needed to explain the observed flux, low-density molecular clouds 
would be unstable
against tidal shear in the rotating disks.  Because we also detect  
high density tracers like HCN and CS, we estimate 
$\sim 10$\% of the gas is in highly opaque, 
high-density (10$^5$\,cm$^{-3}$), self-gravitating 
clouds that are stable against tidal forces.   
In the outer disks, however, the CO emission 
may come from ``normal" molecular clouds.   In this outer 
disk gas, the volume filling factor would be $\sim 0.1$, the area filling 
factor 0.3, as in normal galactic disks.

\noindent
{\it 6.) Low dust flux consistent with lower molecular mass:} 
Only in Arp~220 is there significant thermal continuum flux from dust
 at 1.3\,mm.
The continuum in Mrk~231 at 3 and 1.3\,mm is nonthermal.  
At 3\,mm, no thermal continuum is detected from dust, 
to limits of $\sim $2 to 10\,mJy, in accord with the gas masses 
obtained from CO, corresponding to a CO luminosity to gas mass conversion
factor about 5 times lower than in self-gravitating clouds in the Milky Way.

\noindent
{\it 7.)  Still lots of gas, nevertheless:}
In spite of the ``low" gas densities and CO line opacities, 
the derived gas mass is high; the mass of  $\approx 5 \times 10^9 $\, \Msun\ is
equal to the mass of molecular clouds in a large gas rich spiral galaxy.   
Within
the molecular disks, the ratio of gas mass to the enclosed dynamical mass is
$M_{\rm gas}/M_{\rm dyn}$ = 1/6. The ratio of gas to total mass surface
density, $\mu/\mu_{\rm tot}$,  reaches a maximum value of 1/3 within the
molecular disks.   The ratio $M_{\rm gas}/L^\prime_{\rm CO}$ of gas mass to CO
luminosity is about one-fifth of its value in self-gravitating molecular clouds.
For the galaxies in this sample, typical ratios are 
$M_{\rm gas}/L^\prime_{\rm CO}$ = 
0.8\,\Msun (K\,\kms\,pc$^2)^{-1}$.  

\noindent
{\it 8.)  Extreme Starburst Regions:}  
  Four extreme starbursts are identified in the 3 closest galaxies in the sample
including Arp~220, Arp~193 and Mrk~273. They are the most prodigious star
formation events in the local universe, each representing about 1000 times as
many OB stars as 30~Doradus. They have a
characteristic size of only 100 pc, with about  $10^9$ \Msun\ of gas and an IR
luminosity of $\approx 3 \times 10^{11}$ \Lsun\ from recently formed OB stars.

Arp~220 has 2 extreme starbursts.
The integrated CO and 1.3\,mm continuum maps of Arp~220
show two compact peaks.
The west nucleus, at $cz_{\rm lsr}$ = 5340\,\kms , is the same as the
cm-radio and IR west peak.
the CO east gas at 5330\,\kms , is associated with the cm-radio east
peak and its OH megamasers.
The CO east gas at 5650\,\kms\ is associated with the ionized gas in 
the $K$-band east nucleus.
The IR luminosity of the compact peaks in Arp~220 can be 
explained by extreme starbursts in their early phases. 
 The cm-radio continuum, the CO intensity,  and the 1.3\,mm dust flux 
all suggest the currently most active starburst is the west peak.
Arp~220 west also shows a complex velocity
structure which may indicate a bar in formation or a huge molecular outflow.
 The mass of
these compact regions is dominated by molecular gas and young stars, not by a
bulge population.

\noindent
{\it 9.) Regions of dense molecular gas are regions of extreme starbursts:}
We suggest that the HCN emission, 
which is very strong in most ultraluminous galaxies, 
originates in the high-density, extreme starburst regions 
similar to Arp~220 east and west. 
 These regions
are larger than ordinary GMCs, but are
 filled with molecular gas at a density usually found only in small cloud cores.
They do not produce most
of the CO luminosity, but they do emit most of the HCN luminosity. This explains
the high HCN luminosity and directly relates the HCN emission to star
formation, as we suggested previously (Solomon et al. 1992).

\noindent
{\it 10.) Ultraluminous galaxies are powered by starbursts, not AGNs:}
The CO data show the gas in ultraluminous
 galaxies is in extended disks that cannot intercept all the power of
 central AGNs, if they exist.  As a rule of thumb, if you can see the
AGN in the UV/visible, then it is not heavily absorbed, and 
cannot be responsible for the far-IR/sub-mm luminosity.  
Furthermore,
if the rotating molecular disks were very thin (30\,pc) and extended
all the way in to central AGNs, then their dust would be very hot, emitting
most of their power at 10 to 20\,$\mu$m.  In fact, most of the dust (and 
gas) observed in the ultraluminous IR galaxies is cool (70\,K), 
emitting at 60 to 100\,$\mu$m --- the
usual temperature of dust in molecular clouds heated by starbursts. 
We conclude that in 
ultraluminous galaxies --- even in Mrk~231 that hosts a
quasar ---   the far IR luminosity is powered by extreme 
starbursts in the molecular disks, not by dust-enshrouded quasars.

\acknowledgments
We thank the telescope operators at Plateau de Bure and the
IRAM staff astronomers for their help in taking the data, 
A.\ Dutrey for use of her disk modelling program, C.M.\ Walmsley
for the escape probability program,  and S.\ Guilloteau, R.\ Lucas, 
and J.\ Wink for help in data reduction. 
We also thank H.\ Ungerechts for re-measuring some 
CO fluxes at the IRAM 30\,m telescope, and P.\ van der Werf, R.\ Genzel, and
N.\ Scoville for helpful talks.  
We thank the referee for many helpful comments. P.M.S.\ is grateful for a 
Research Award from the Alexander von Humboldt Foundation and for 
a senior scientific fellowship from the North Atlantic Treaty 
Organization.

\clearpage

%-----------------------------------------------------

\newpage
%-------------------------------------------------------------------

%\begin{figure}
%\plotone{sgi9259.ps}
%\caption{We use \LaTeX\ {\tt figure} environment syntax in this caption.}
%\end{figure}

%FIGURE CAPTIONS

\clearpage
%FIGURE 1 ---------------------------------------------------------
\begin{figure}
\epsscale{0.5}
%\epsfsize=0.5\hsize
\plotone{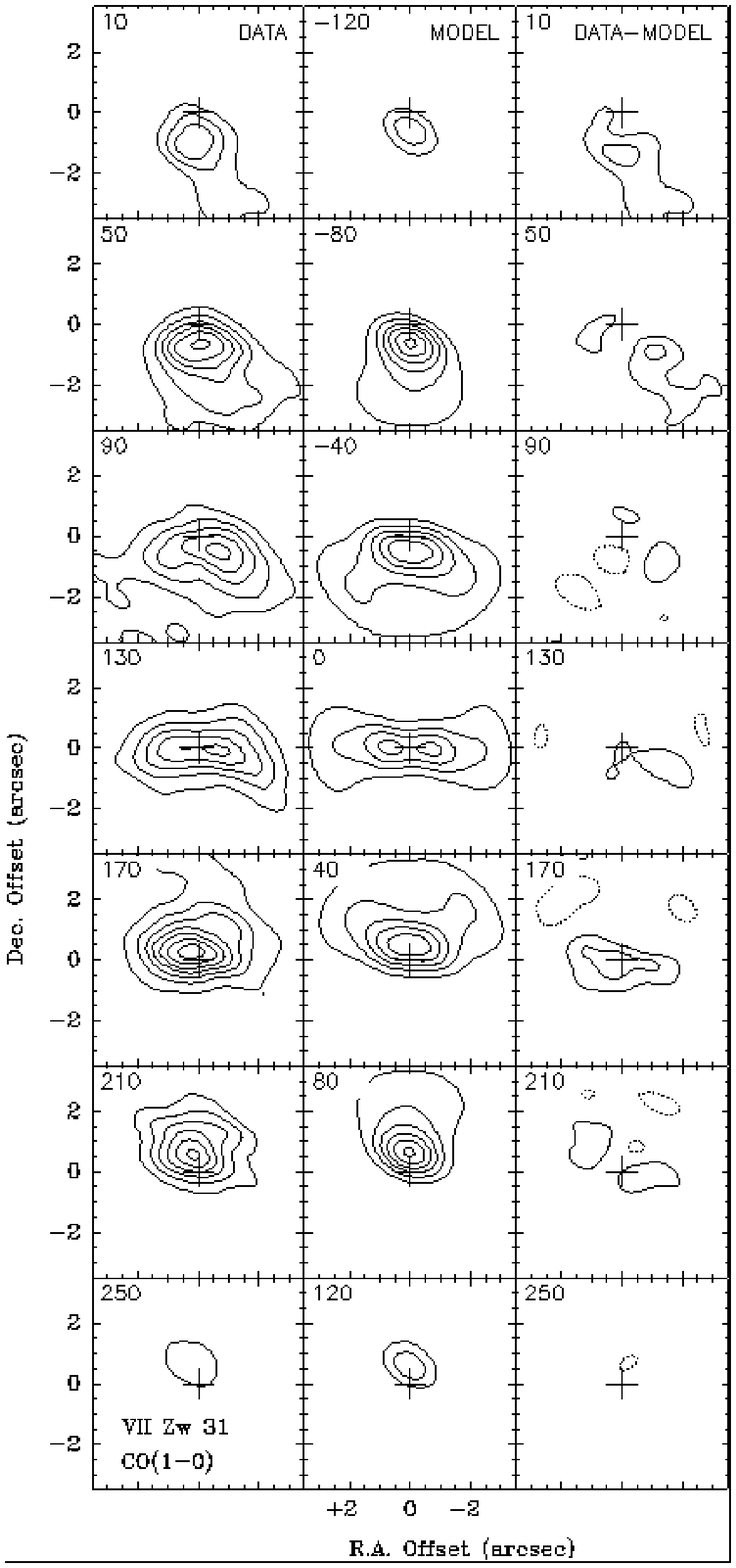}
\caption[{\bf VII Zw 31 :} Channel maps, model, and difference]
{ {\bf VII Zw 31 :}  
{\it left:}
CO(1--0) in 40\,\kms\ channels.
Radial velocities (\kms , upper left of each box) are relative to 
109.380\,GHz ($cz_{\rm lsr} =$ 16147\,\kms ).  
Contour unit: 11\,mJy beam$^{-1}$,
 with $T_b/S = 88$\,K/Jy. Beam $= 1''.4\times 0''.8$. 
The cross at (0,0) is 
05$^{\rm h}$16$^{\rm m}$46.$^{\rm s}$51,
79$^\circ$40$'$12.$''8$ (J2000).
%\noindent
{\it middle:} Maps from the model, with 
same contours and resolution as the data.
%\noindent
{\it right:}  Data minus model, with same contour unit. 
}\end{figure}

\clearpage
%FIGURE 2 ---------------------------------------------------------
\begin{figure}
\plotfiddle{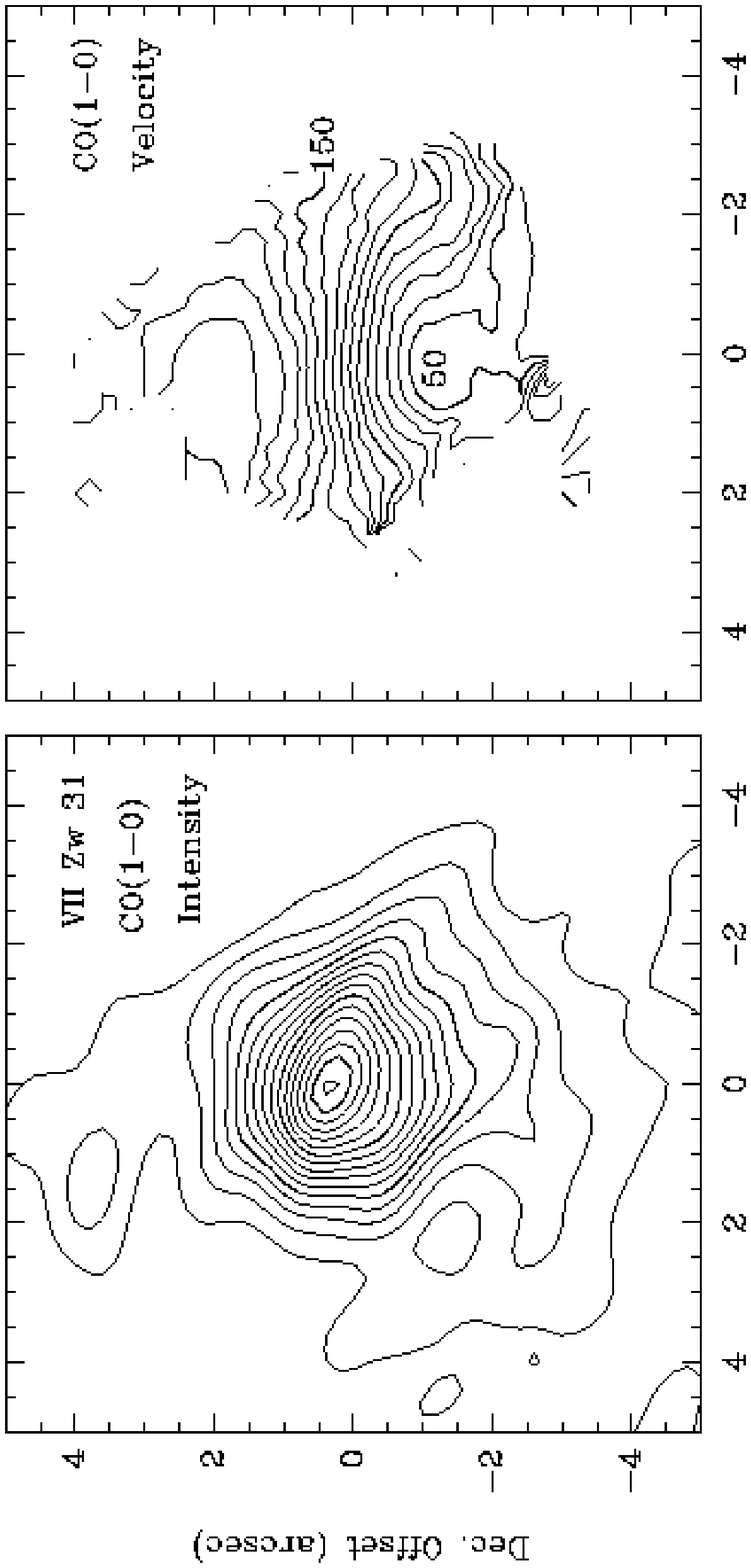}{\hsize}{-90}{65}{65}{-270}{610}
\plotfiddle{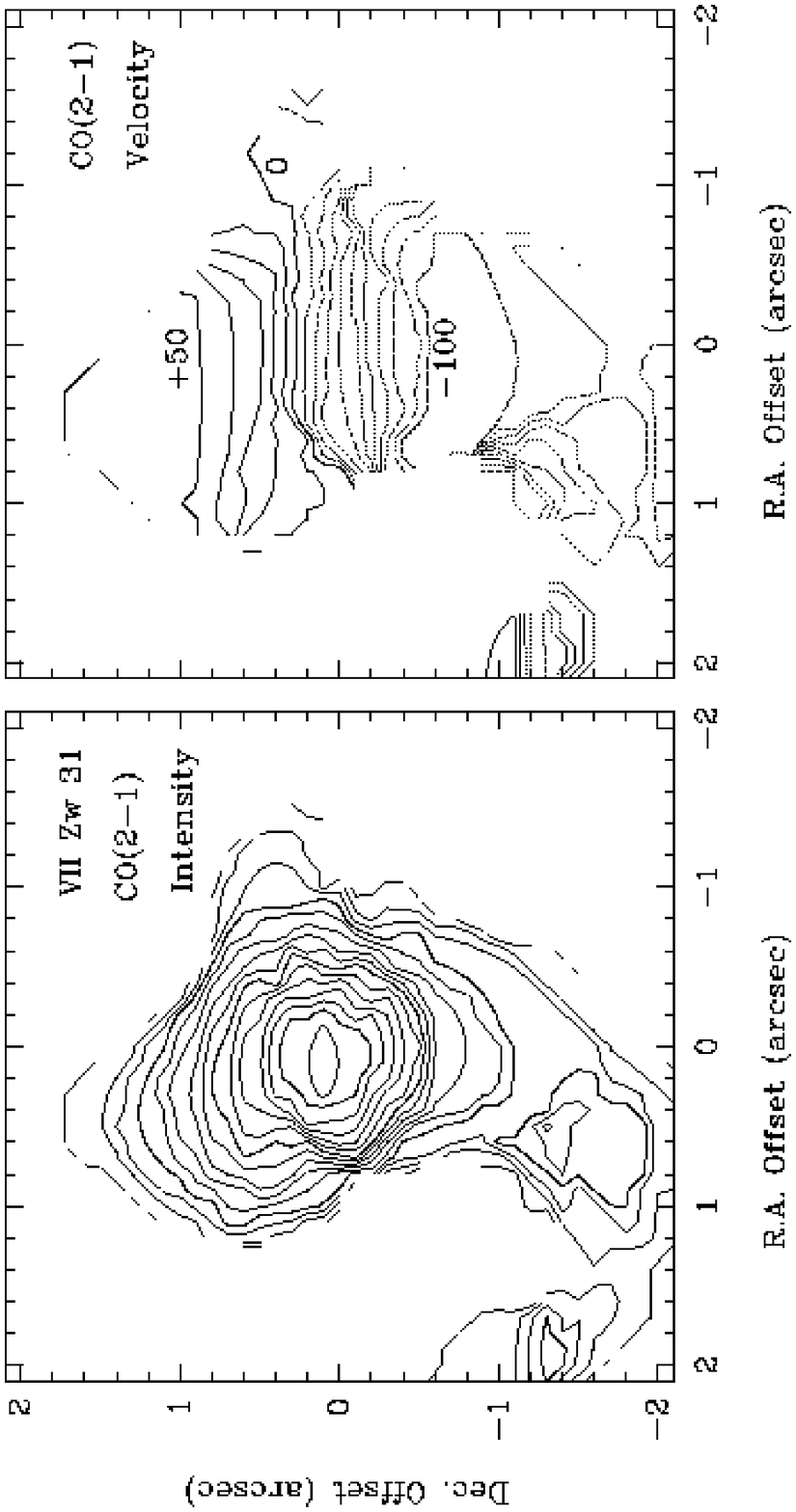}{\hsize}{-90}{65}{65}{-270}{860}
\end{figure}
\begin{figure}
\vspace{-5cm}
\caption[{\bf VII Zw 31 :} CO(1--0) and (2--1) Moment maps ]
{ {\bf VII Zw 31 :}
{\it  upper left:}
CO(1--0) integrated over ($-$40, $+$260\,\kms ).
Beam $= 1''.4\times 0''.8$. Contour step 1.2\,Jy beam$^{-1}$ \kms , 
with $T_b/S = 88$\,K/Jy.
\noindent
{\it  upper right:} Velocity contours in steps of 10\,\kms\ 
relative to 109.380\,GHz
($cz_{\rm lsr} =$ 16147\,\kms ).  Labels are in \kms .
\noindent
{\it  lower left:}
CO(2--1) integrated over ($-$140, $+$80\,\kms ).
Beam $1''.1\times 0''.8$. Contours: 
2 to 16 by 1, in units of 1.6\,Jy beam$^{-1}$ \kms , with $T_b/S = 30$\,K/Jy.
% The (0,0) position is 05$^{\rm h}$16$^{\rm m}$46.$^{\rm s}$51,
% 79$^\circ$40$'$12.$''5$ (J2000).
\noindent
{\it  lower right:} Velocity contours in steps of 10\,\kms\ 
relative to 218.650\,GHz
($cz_{\rm lsr} =$ 16300\,\kms ).  Labels are in \kms .
Center position in all maps is 
05$^{\rm h}$16$^{\rm m}$46.$^{\rm s}$51,
79$^\circ$40$'$12.$''5$ (J2000).

}\end{figure}
%\nopagebreak

\clearpage
%FIGURE 3 ---------------------------------------------------------
\begin{figure}
\plotfiddle{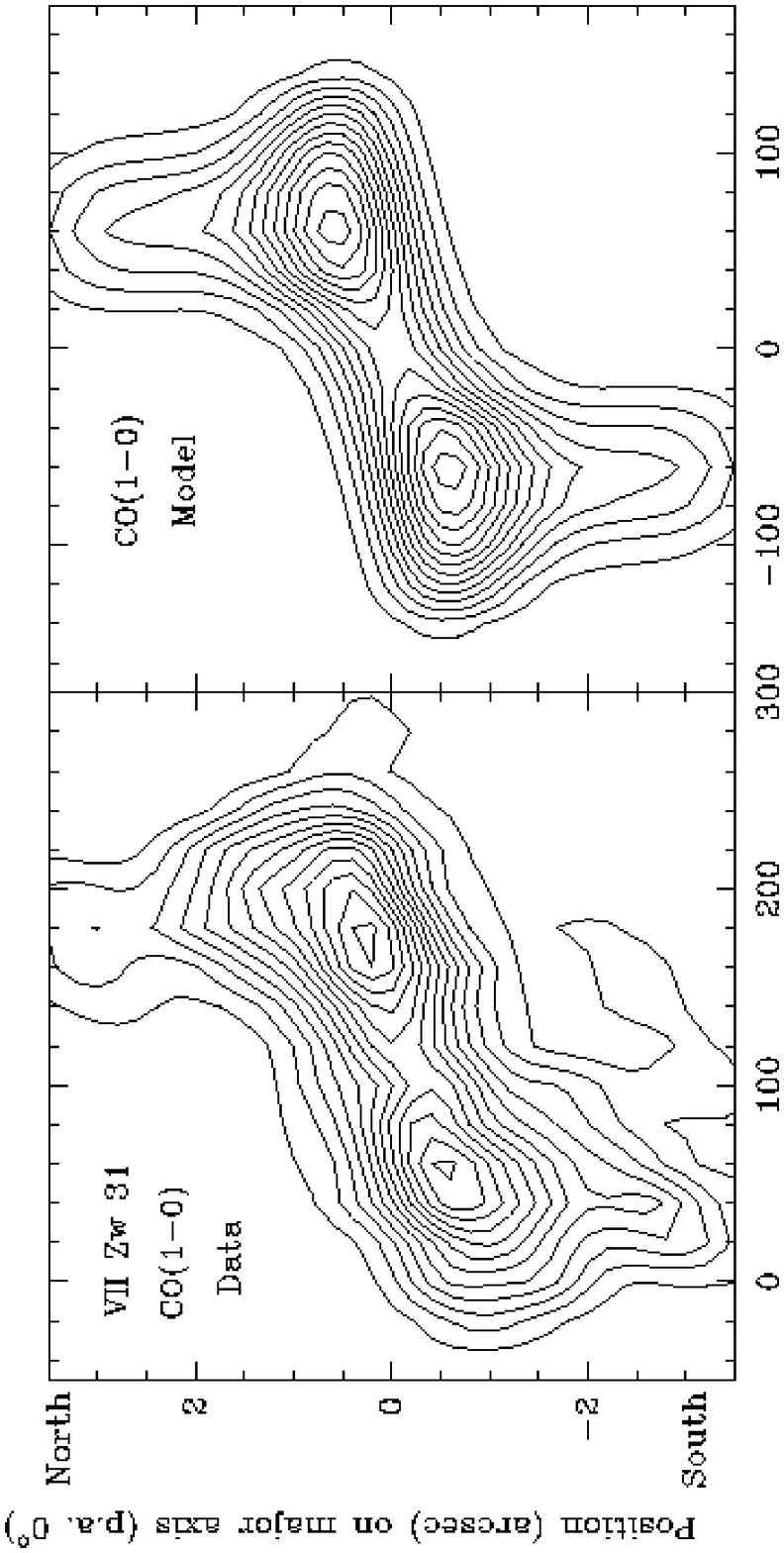}{\hsize}{-90}{65}{65}{-270}{610}
\plotfiddle{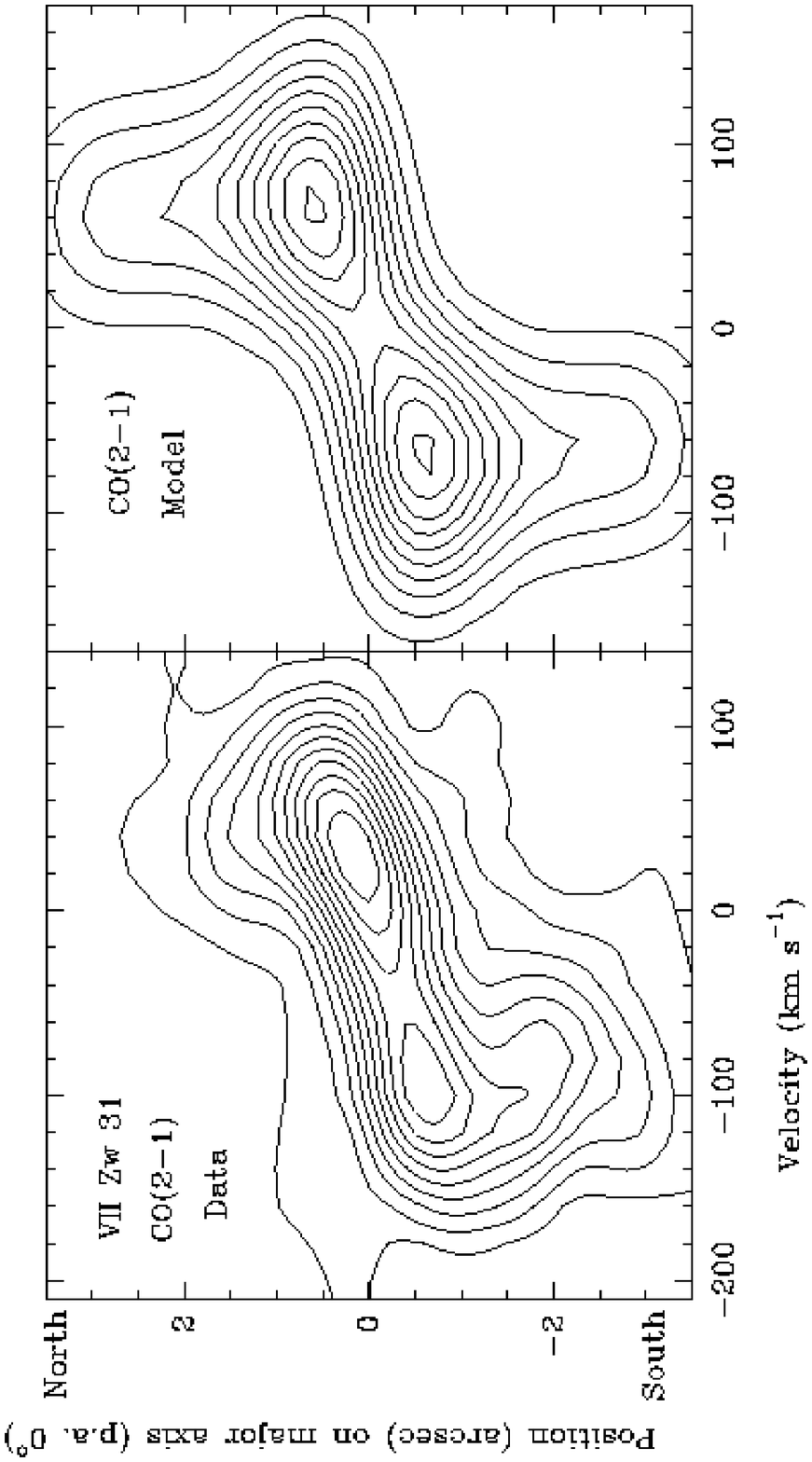}{\hsize}{-90}{65}{65}{-270}{860}
\end{figure}

\begin{figure}
\vspace{-5cm}
\caption[{\bf VII Zw 31 :} CO(1--0) and (2--1) Position-velocity plots: data
\& model]
{{\bf VII Zw 31 :} Position-velocity cuts in declination, with 
zero position at 05$^{\rm h}$16$^{\rm m}$46.$^{\rm s}$51,
79$^\circ$40$'$12.$''5$ (J2000). 
{\it upper left:} CO(1--0) in  $1''.4\times 0''.8$ beam. Contour step:  
6\,mJy beam$^{-1}$, with $T_b/S = 88$\,K/Jy.
Velocity is relative to 109.380\,GHz ($cz_{\rm lsr} =$ 16147\,\kms ).  
{\it upper right:} Predicted CO(1--0)  
 from the model, with same contours and resolution  as for the data.
%\noindent
{\it lower left:} CO(2--1) in $1''.1\times 0''.8$ beam. 
%Contour levels are 5 to 55, in steps of 
Contours: 8 to 90, in steps of 
% corr. fac. =1.63
%3\,mJy beam$^{-1}$, with $T_b/S = 30$\,K/Jy.
4.9\,mJy beam$^{-1}$, with $T_b/S = 30$\,K/Jy.
%  
% The (0,0) position is 
% 05$^{\rm h}$16$^{\rm m}$46.$^{\rm s}$51,
% 79$^\circ$40$'$12.$''8$ (J2000). 
Velocity is relative to 218.650\,GHz
($cz_{\rm lsr} =$ 16300\,\kms ). 
{\it lower right:} Predicted CO(2--1)
from the model, with same contours and resolution 
as for the data.

}\end{figure}

\clearpage
%FIGURE 4 ---------------------------------------------------------
\begin{figure}
\epsscale{0.5}
\plotfiddle{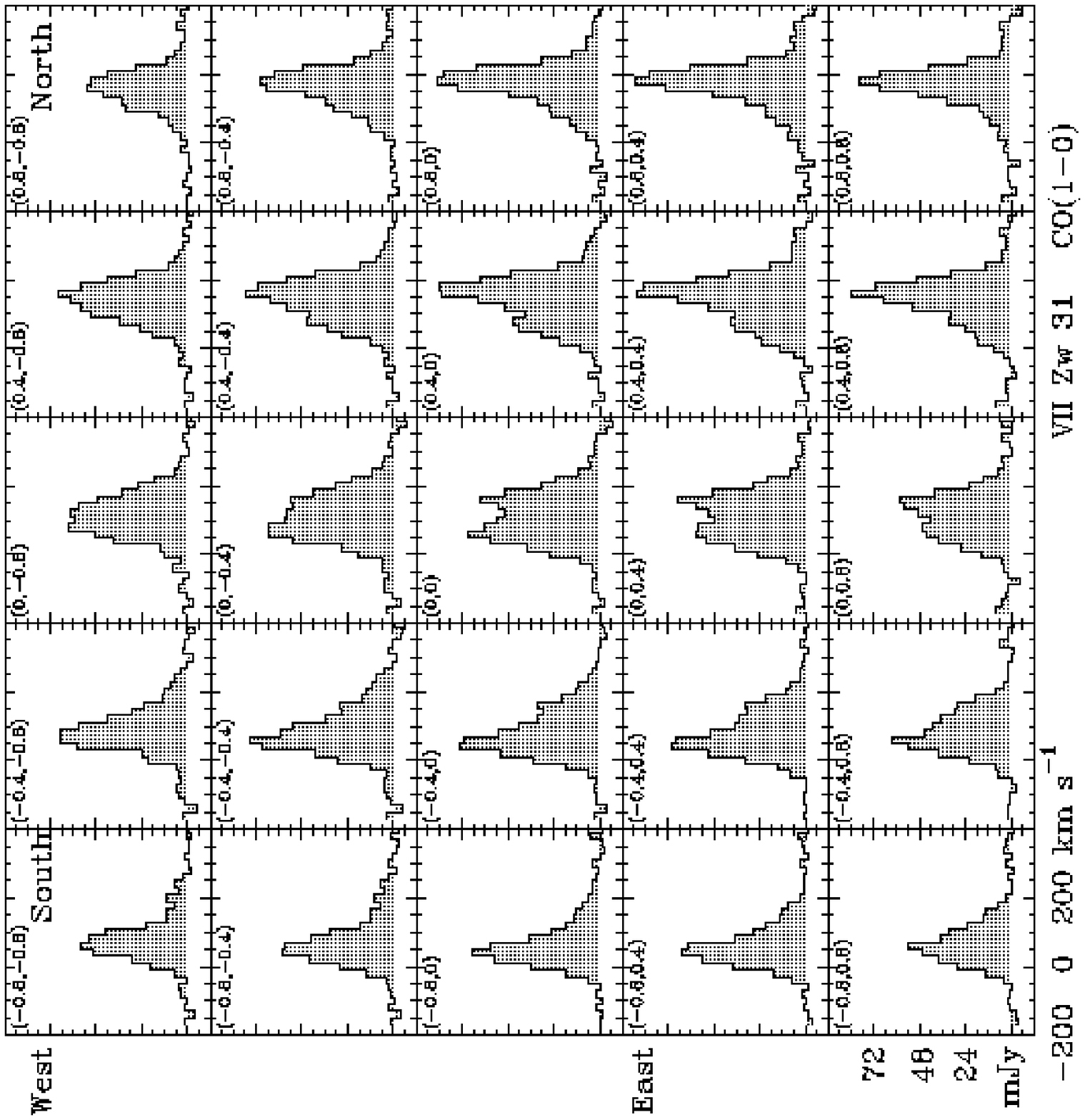}{\hsize}{-90}{56}{56}{-230}{500}
\plotfiddle{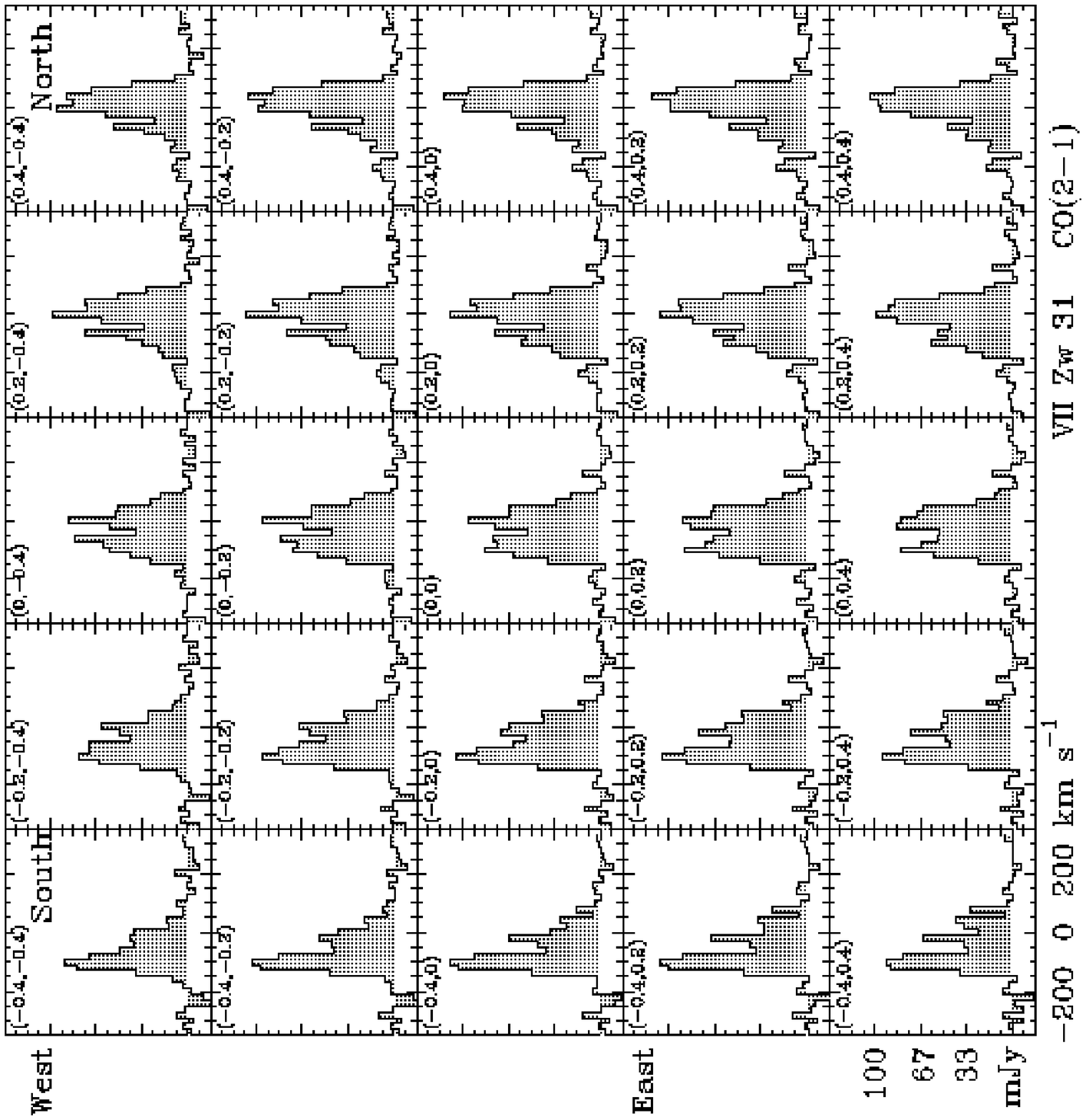}{\hsize}{-90}{56}{56}{-230}{656}
\end{figure}
\begin{figure}
\caption[{\bf VII Zw 31 :} CO(1--0) and (2--1) spectra.]
{{\bf VII Zw 31 :} 
{\it (a), upper:} CO(1--0) spectra vs.\    
%% with large ticks at  0, 24, 48, and 72\,mJy  beam$^{-1}$;  
radial velocity relative to 109.380\,GHz
($cz_{\rm lsr} =$ 16147\,\kms ).
In the upper left of each box are offsets (arcsec) 
on the kinematic major and minor axes.
Beam $=  1''.4\times 0''.8$, with $T_b/S =  88$\,K/Jy.

{\it (b), lower:} CO(2--1) spectra vs.\ 
%% with large ticks 
%  at  0, 20, 40, 60, and 80\,mJy  beam$^{-1}$;  --- uncorrected
%% at  0, 33, 66, 100, and 133\,mJy  beam$^{-1}$; 
%corr fac.=1.63
radial velocity relative to 218.650\,GHz ($cz_{\rm lsr} =$ 16300\,\kms ). 
%% with large ticks at $-200$,  0, and $+$200\,\kms\ (left to right).
In the upper left of each box are offsets (arcsec) 
on the kinematic major and minor axes.
Beam $1''.1\times 0''.8$.  
The (0,0) position in both diagrams is 
05$^{\rm h}$16$^{\rm m}$46.$^{\rm s}$51,
79$^\circ$40$'$12.$''5$ (J2000). 
}\end{figure}

\clearpage
%FIGURE 5 --------------------------------------------------------- 
\begin{figure}
\plottwo{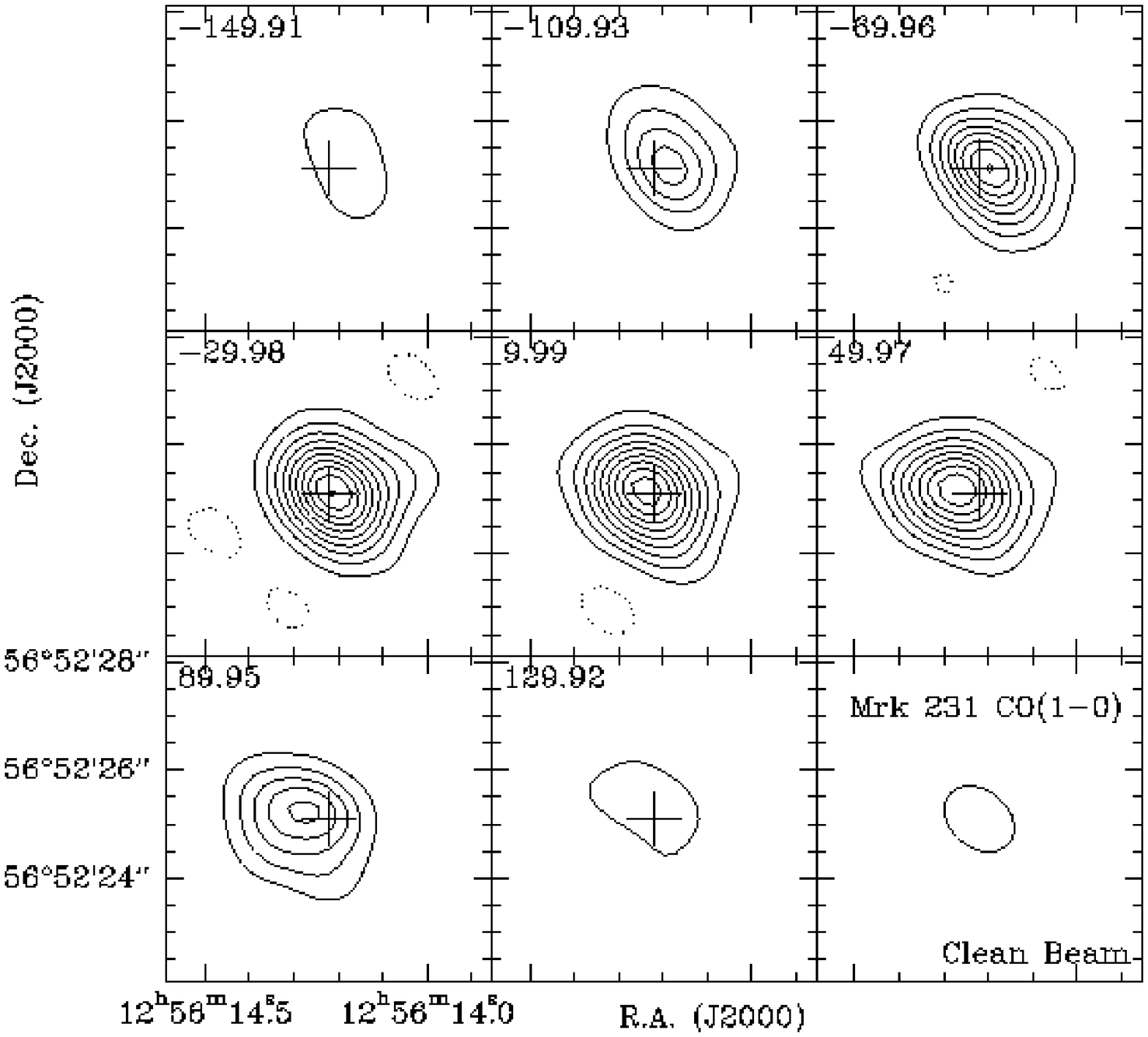}{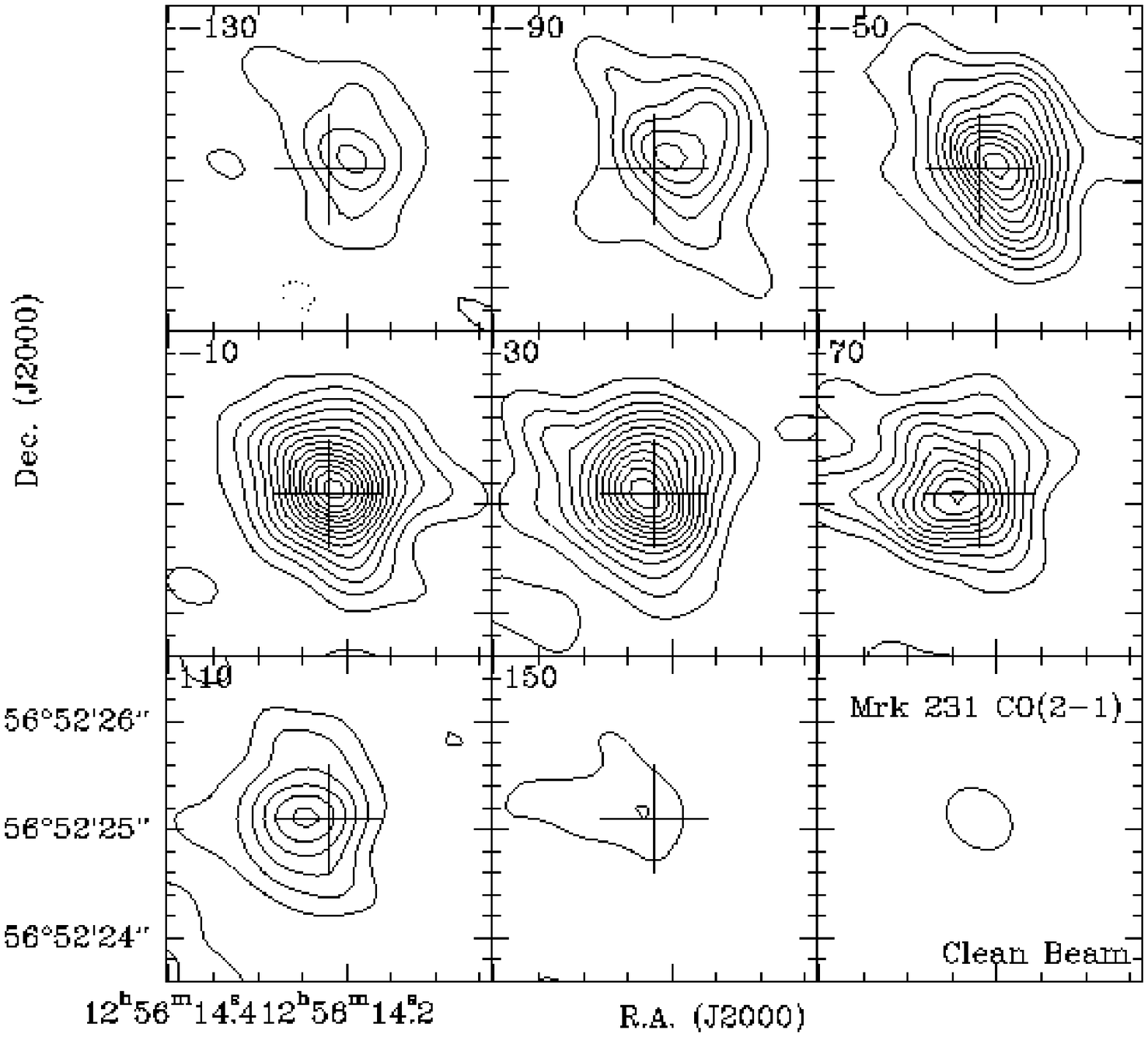} 
\caption[{\bf Mrk 231 :} CO(1--0) and (2--1) channel maps]
{ {\bf Mrk 231 :} 
{\it (a), left:}  CO(1--0) in 40\,\kms\ channels.
A 63\,mJy continuum point source has been subtracted 
at the CO centroid (cross). 
Radial velocities (\kms , upper left of each box) are relative to 
110.602\,GHz ($cz_{\rm lsr} =$ 12656\,\kms ).
Contour unit: 20\,mJy beam$^{-1}$. 
Beam $=  1''.3\times 1''.1$  (lower right) with $T_b/S = 70$\,K/Jy.

{\it (b), right:}  CO(2--1) in 40\,\kms\ channels.
A 36\,mJy continuum point source has been subtracted 
at the CO centroid (cross).
Radial velocities (\kms , upper left of each box) are relative to 
221.204\,GHz ($cz_{\rm lsr} =$ 12650\,\kms ).
Contour unit: 25\,mJy beam$^{-1}$.
Beam $=  0''.7\times 0''.5$  (lower right) with $T_b/S = 75$\,K/Jy.
}\end{figure}

\clearpage
%FIGURE 6a ---------------------------------------------------------
\begin{figure}
%%% \plotfiddle{f06.ps}{\hsize}{-90}{65}{65}{-220}{520}
\plotfiddle{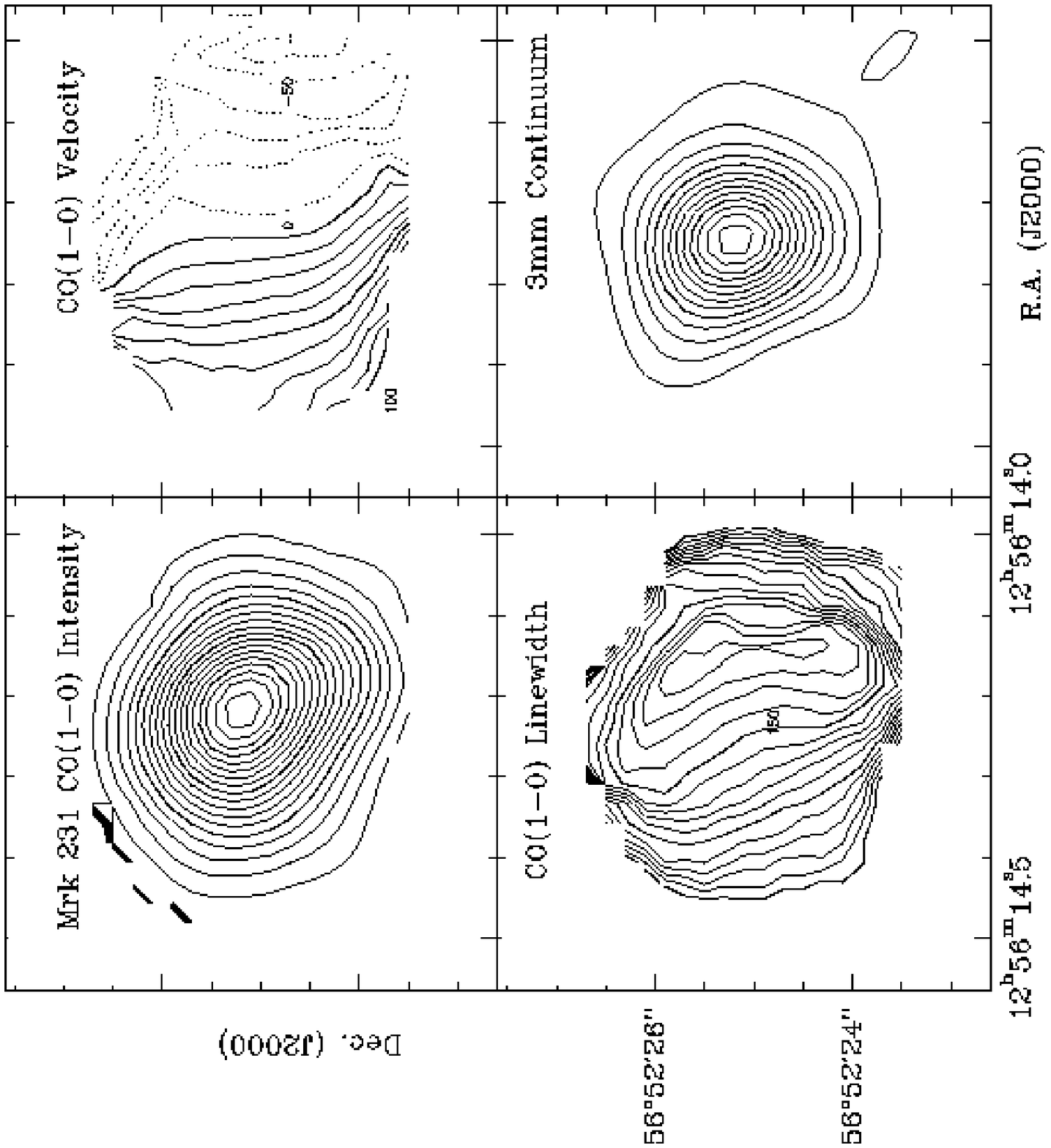}{\hsize}{-90}{56}{56}{-230}{500}
\plotfiddle{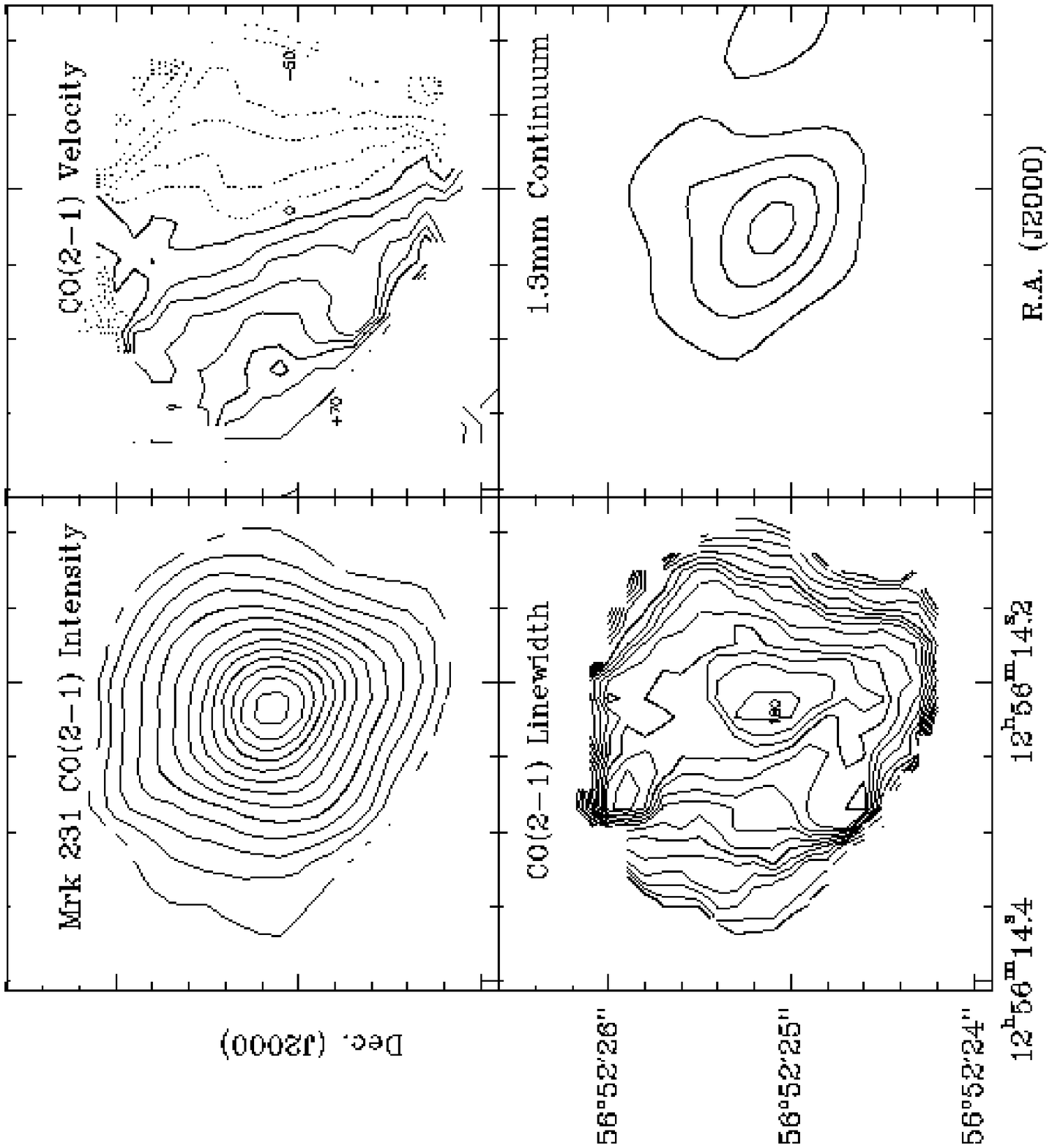}{\hsize}{-90}{56}{56}{-230}{656}
\end{figure}
\begin{figure}
\vspace{-5cm}
\caption[{\bf Mrk 231 :} CO(1--0) and (2--1) Moment maps]
{{\bf Mrk 231 :} 

{\it --- upper:} CO(1--0) integrated intensity, velocity,  
linewidth (FWHM), and the 3\,mm continuum.  The continuum
source has been subtracted from the CO data.
Integration limits:  ($-180, +200$\,\kms ).
Beam $= 1''.3\times 1''.1$. Contours: 

\noindent
{\it  integrated CO:} 
1 to 18 by 1, in units of 2\,Jy beam$^{-1}$ \kms , 
with $T_b/S = 70$\,K/Jy;

\noindent
{\it  CO velocity:}
$-60$ to $+100$\,\kms , in steps of 10\,\kms\ relative to 110.602\,GHz
($cz_{\rm lsr} =$ 12656\,\kms ).  Labels are in \kms ;

\noindent
{\it  CO linewidth:} 50 to 190\,\kms\ in steps of 10\,\kms .  
Labels are in \kms .

\noindent
{\it 3\,mm continuum:} 1 to 13 by 1, in units of 4\,mJy.
%%% }\end{figure}

%FIGURE 6b ---------------------------------------------------------
%%% \begin{figure}
%%% \plotfiddle{f07.ps}{\hsize}{-90}{65}{65}{-220}{520}
%%% \end{figure}
%%% \begin{figure}
%%% \vspace{-5cm}

%%% \caption[{\bf Mrk 231 :} CO(2--1) Moment maps]
%%% {{\bf Mrk 231 :}  

\noindent
{\it --- lower:} CO(2--1) integrated intensity, velocity, 
linewidth (FWHM), and the 1.3\,mm continuum.  The continuum
source has been subtracted from the CO data.
Integration limits:  ($\pm 180$\,\kms ).
Beam $= 0''.7\times 0''.5$. Contours: 

\noindent
{\it integrated CO:} 
1 to 13 by 1, in units of 5\,Jy beam$^{-1}$ \kms , 
with $T_b/S = 75$\,K/Jy;

\noindent
{\it CO velocity:} $-60$ to $+70$\,\kms , 
in steps of 10\,\kms\ relative to 221.204\,GHz 
($cz_{\rm lsr} =$ 12650\,\kms ).  Labels are in \kms ;

\noindent
{\it CO linewidth:} 50 to 180\,\kms\ in steps of 10\,\kms . Labels are in \kms .

\noindent
{\it  1.3\,mm continuum:} 1 to 4 by 1, in units of 8\,mJy.}\end{figure}

\clearpage
%FIGURE 7 ----------------------------------------------------------
\begin{figure}
\plotfiddle{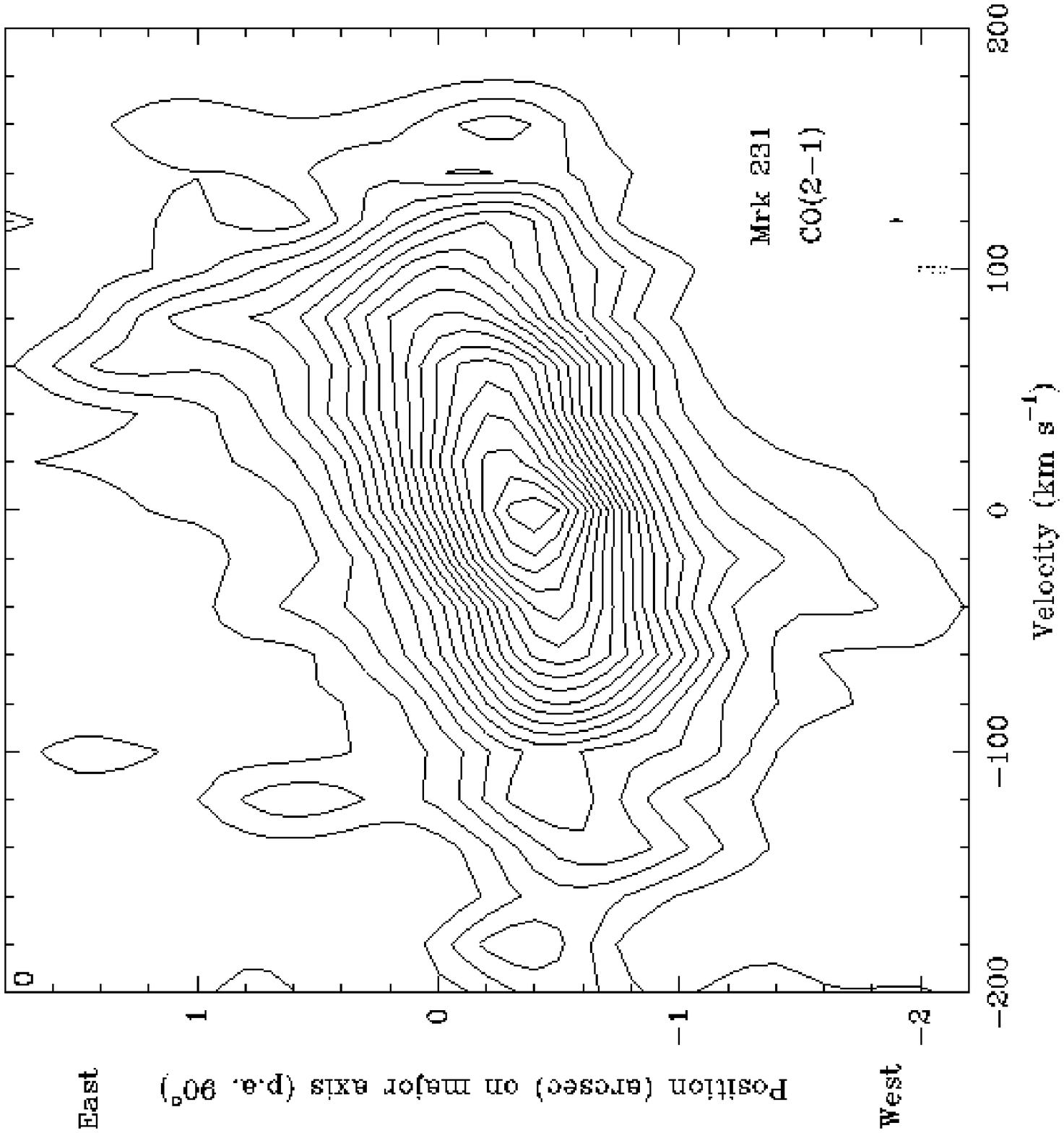}{\hsize}{-90}{65}{65}{-220}{520}
\end{figure}
\begin{figure}
\vspace{-5cm}
\caption[{\bf Mrk 231 :} CO(2--1) Position-velocity diagram] 
{{\bf Mrk 231 :} 
CO(2--1) position-velocity diagram along the line of nodes (p.a. 90$^\circ$).  
Contour levels are 1 to 19, in units of 20\,mJy beam$^{-1}$, 
with $T_b/S = 75$\,K/Jy. Beam $= 0''.7\times 0''.5$.  
The (0,0) position is 12$^{\rm h}$56$^{\rm m}$14.$^{\rm s}$260,
56$^\circ$52$'$25.$''13$ (J2000).
Velocity is relative to 221.204\,GHz
($cz_{\rm lsr} =$ 12650\,\kms ). 
The continuum source has been subtracted from the CO data.}\end{figure}

%FIGURE 8 ----------------------------------------------------------
\begin{figure}
\plotfiddle{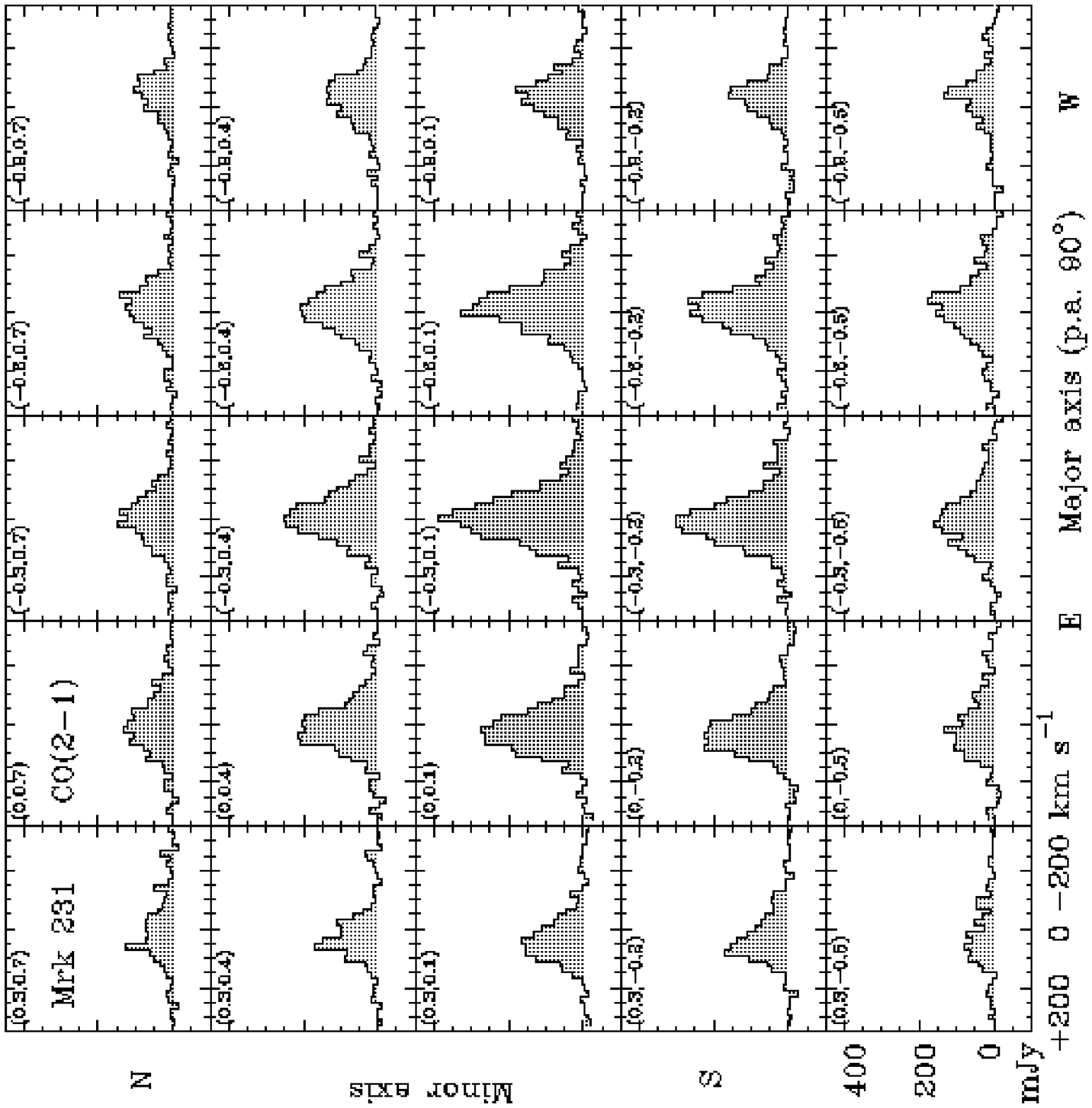}{\hsize}{-90}{65}{65}{-230}{500}
\end{figure}
\begin{figure}
\vspace{-5cm}
\caption[{\bf Mrk 231 :} CO(2--1) spectra.]
{{\bf Mrk 231 :} CO(2--1) spectra vs.\ 
radial velocity relative to 221.204\,GHz ($cz_{\rm lsr} =$ 12650\,\kms ;  
the velocity scale is reversed to emphasize the symmetry of the profiles).
In the upper left of each box are offsets (arcsec) 
on the kinematic major and minor axes. 
The (0,0) position is 12$^{\rm h}$56$^{\rm m}$14.$^{\rm s}$260,
56$^\circ$52$'$25.$''13$ (J2000).
Beam $=  0''.7\times 0''.5$, with $T_b/S =  75$\,K/Jy.
The continuum source has been subtracted from the CO data.
}\end{figure}

\clearpage

%FIGURE 9 ---------------------------------------------------------
\begin{figure} 
\plotfiddle{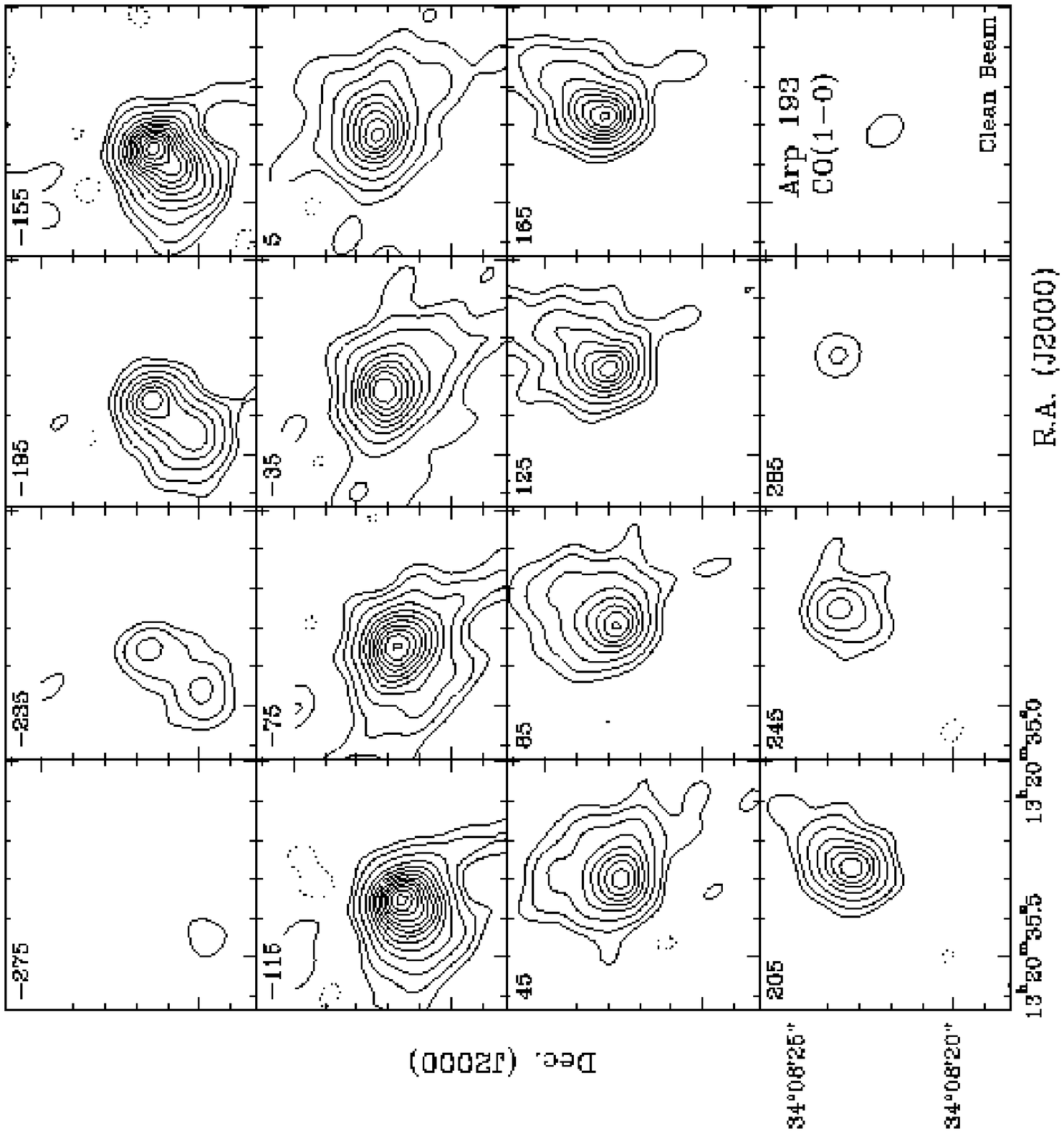}{\hsize}{-90}{56}{56}{-230}{510}
\plotfiddle{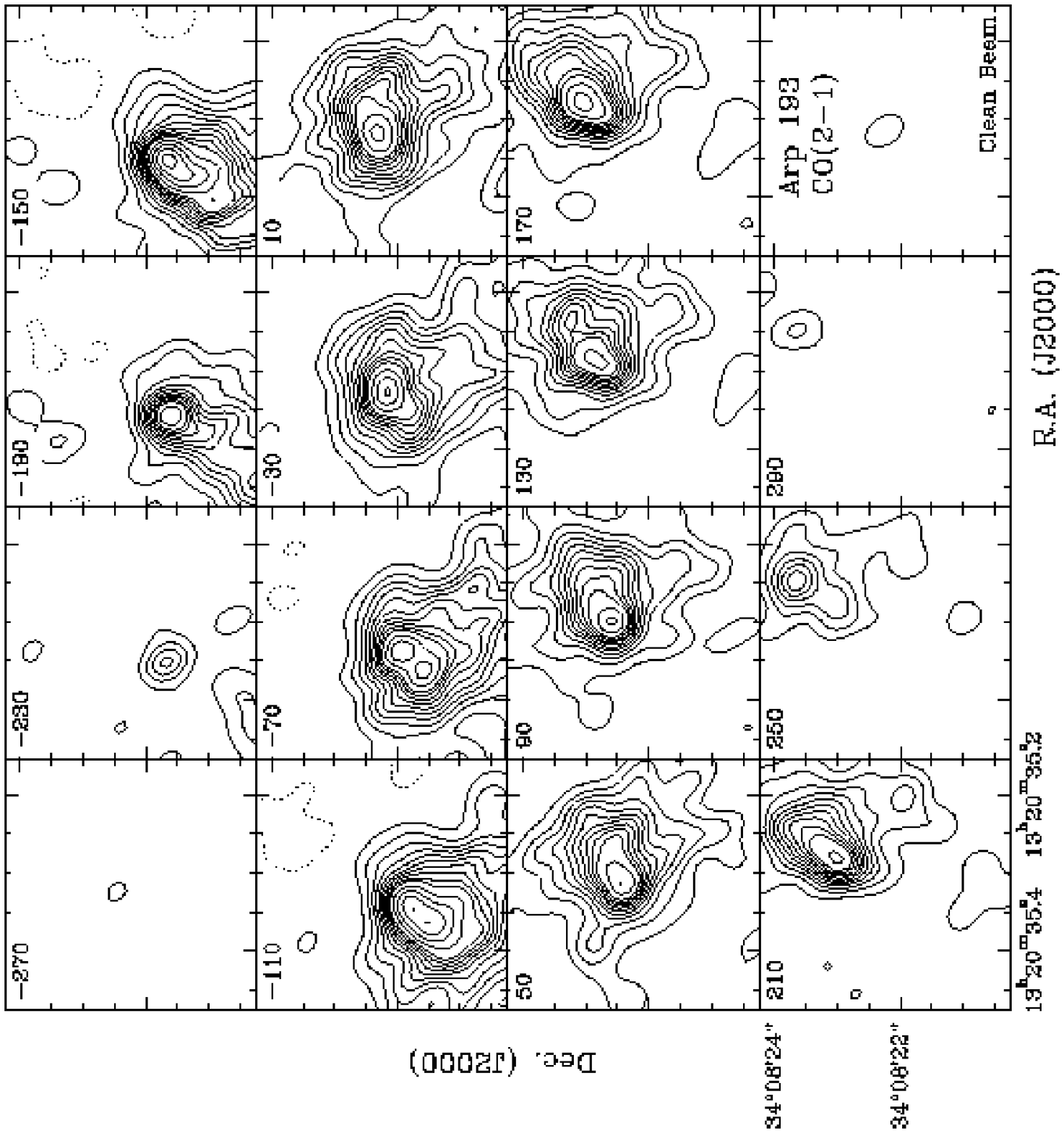}{\hsize}{-90}{56}{56}{-230}{666}
\end{figure}
\begin{figure}
%\vspace{-8cm}
\caption[{\bf Arp 193 :} 40\,\kms\ Channel maps (a) CO(1--0);
(b) CO(2--1)]
{ {\bf Arp 193 :}   

{\it --- upper:} CO(1--0) maps in 40\,\kms\ channels.
Radial velocities (\kms , upper left of each box) are relative to 
112.641\,GHz ($cz_{\rm lsr} =$ 7000\,\kms ).
Contour unit: $-$5, 5, 10, then 20 to 170 by 15\,mJy beam$^{-1}$.
Beam $= 1''.3\times 0''.9$ with $T_b/S = 83$\,K/Jy.  

{\it --- lower:} CO(2--1) maps in 40\,\kms\ channels.
Radial velocities (\kms , upper left of each box) are relative to 
225.282\,GHz ($cz_{\rm lsr} =$ 6994\,\kms ).
Contour interval $-$10, 10 to 100 in steps of 10, then 
10 to 180 in steps of 20\,mJy beam$^{-1}$.
Beam $= 0''.6\times 0''.4$ with $T_b/S = 84$\,K/Jy.  }\end{figure}

%FIGURE 10 ---------------------------------------------------------
\begin{figure}
\plotfiddle{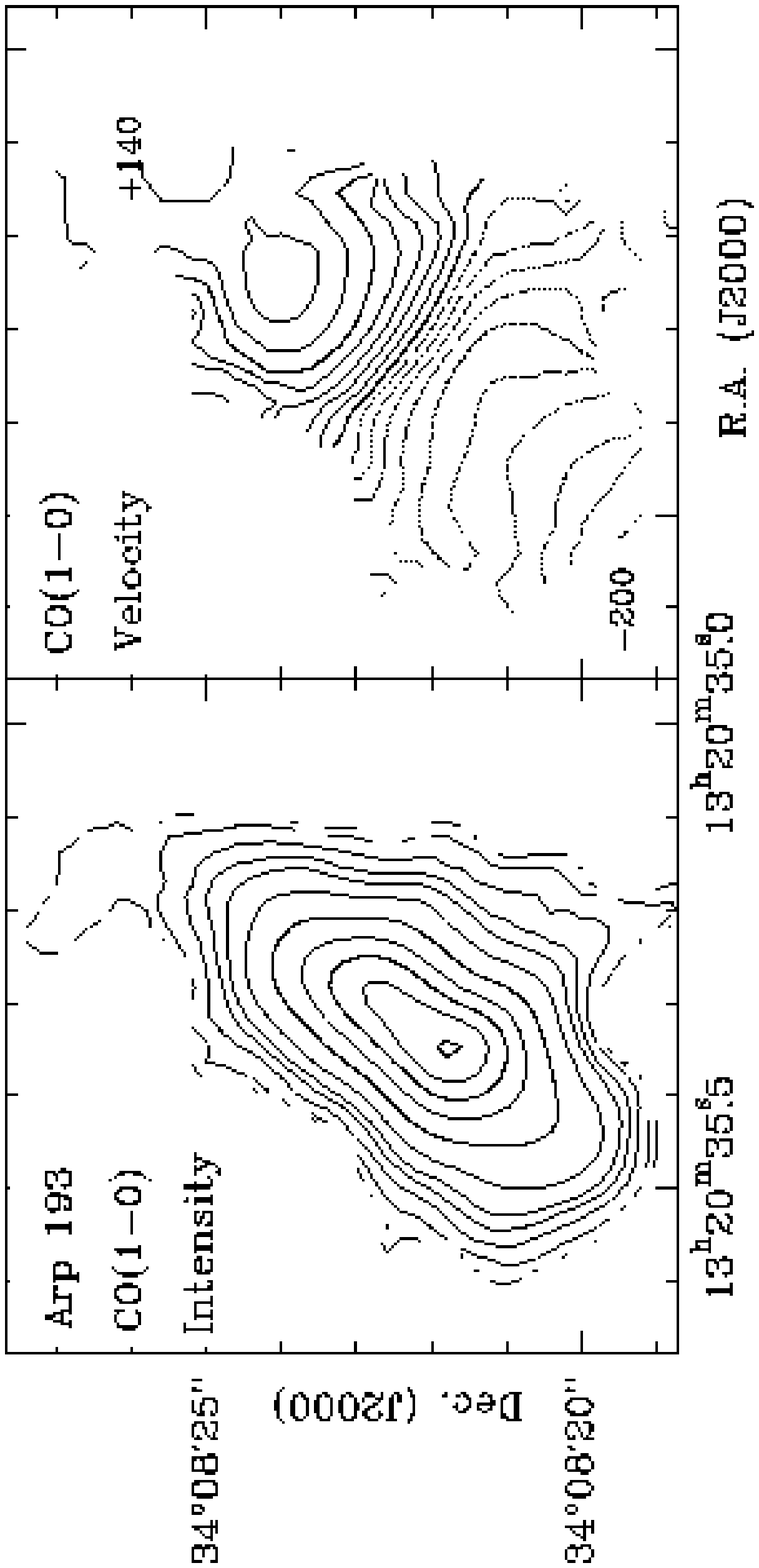}{\hsize}{-90}{60}{60}{-230}{520}
\plotfiddle{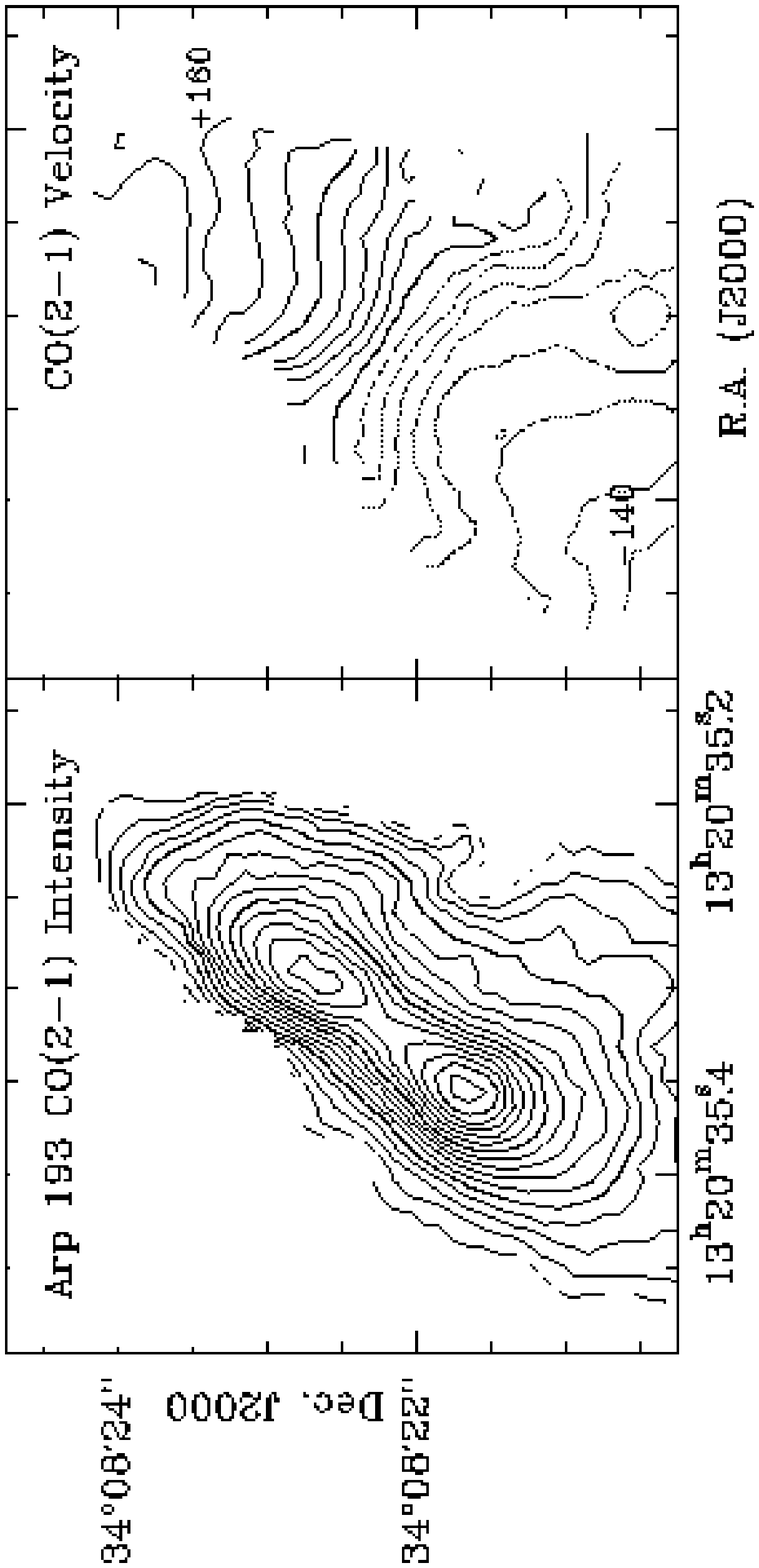}{\hsize}{-90}{60}{60}{-230}{810}
\end{figure}
\begin{figure}
\vspace{-8cm}
\caption[{\bf Arp 193:} CO(1--0) and (2--1) moment maps]
{
{\bf Arp 193 :}  
{\it  Upper left: }
CO(1--0) integrated from $-245$ to $+270$\,\kms .
Beam $= 1''.3\times 0''.9$. Contours: 
0.5,  1, 2, 3, 4, 6, then 10 to 45 by 
5\,Jy beam$^{-1}$ \kms , with $T_b/S = 83$\,K/Jy.
{\it Upper right:} Velocity contours 
%%  $-200$ to $+140$\,\kms , 
in steps of 20\,\kms\ relative to 112.641\,GHz 
($cz_{\rm lsr} =$ 7000\,\kms ).   Labels are in \kms .
{\it Lower left: } 
CO(2--1) integrated from  $-210$ to $+$270\,\kms .
Beam $ = 0''.6\times 0''.4$. Contour step 2\,Jy beam$^{-1}$ \kms , 
with $T_b/S = 84$\,K/Jy.
{\it Lower right:} Velocity contours  
in steps of 20\,\kms\ relative to 
225.282\,GHz ($cz_{\rm lsr} =$ 6994\,\kms ).  Labels are in \kms ;
}
\end{figure}

\clearpage
%FIGURE 11 ---------------------------------------------------------
\begin{figure} 
\plotfiddle{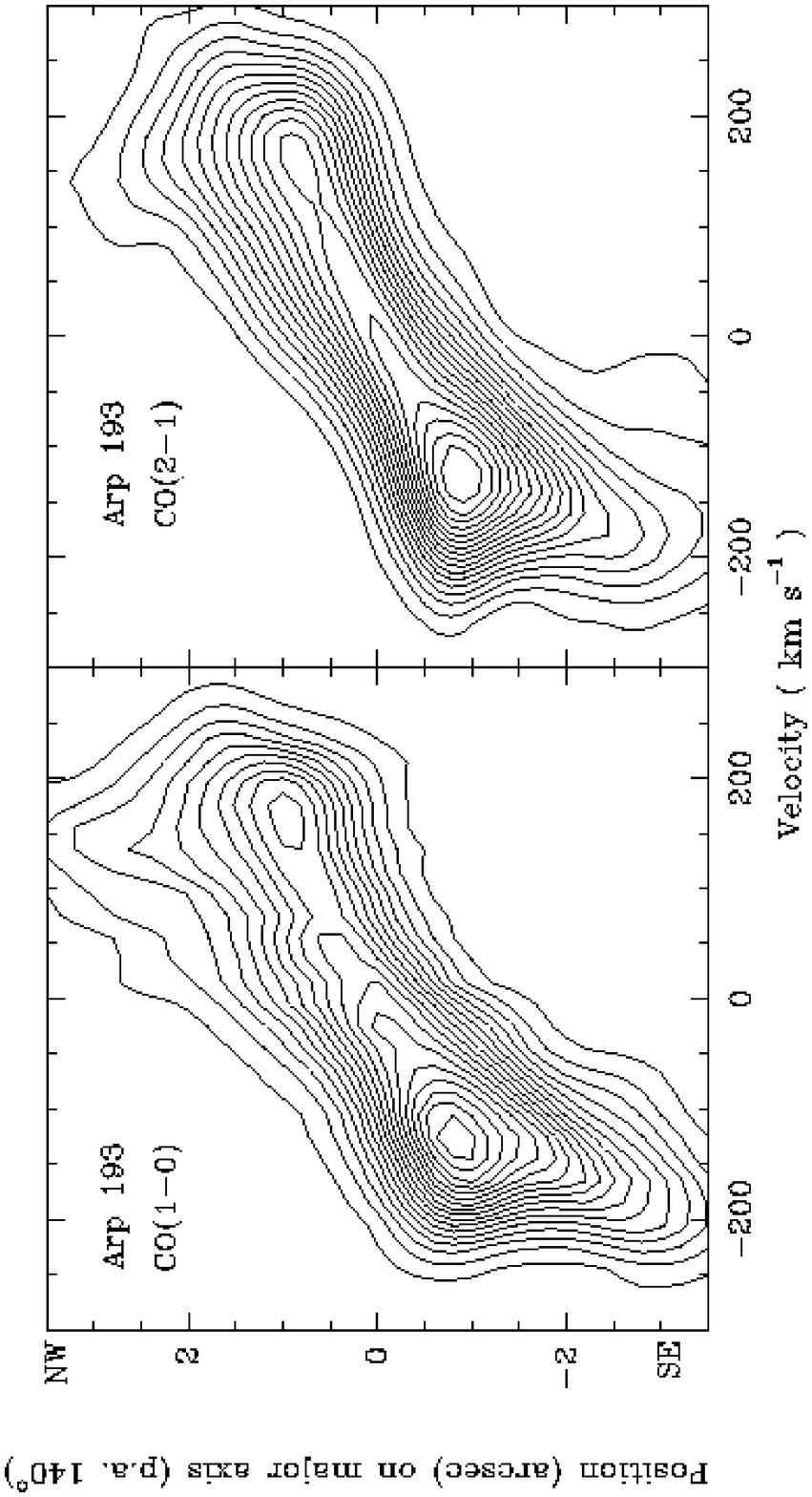}{\hsize}{-90}{65}{65}{-270}{580}
\end{figure}
\begin{figure}
\vspace{-8cm}
\caption[{\bf Arp 193:}  Position-velocity diagrams]
{
{\bf Arp 193 :}  
CO position-velocity diagrams along the line of nodes (p.a.\ 140$^\circ$),

{\it left:} CO(1--0), contour unit 10\,mJy beam$^{-1}$, 
with $T_b/S = 83$\,K/Jy. Beam $= 1''.3\times 0''.9$.  
Velocity is relative to relative to 112.641\,GHz 
($cz_{\rm lsr} =$ 7000\,\kms ). 

{\it right:} CO(2--1), contour unit 10\,mJy beam$^{-1}$, 
with $T_b/S = 84$\,K/Jy. Beam $= 0''.6\times 0''.4$.  
Velocity is relative to relative to 225.282\,GHz ($cz_{\rm lsr} =$ 6994\,\kms ).
The (0,0) position in both diagrams is 
13$^{\rm h}$20$^{\rm m}$35.$^{\rm s}$315,
34$^\circ$08$'$22.$''20$ (J2000).
}
\end{figure} 

\clearpage
%FIGURE 12 ---------------------------------------------------------
\begin{figure}
\plotfiddle{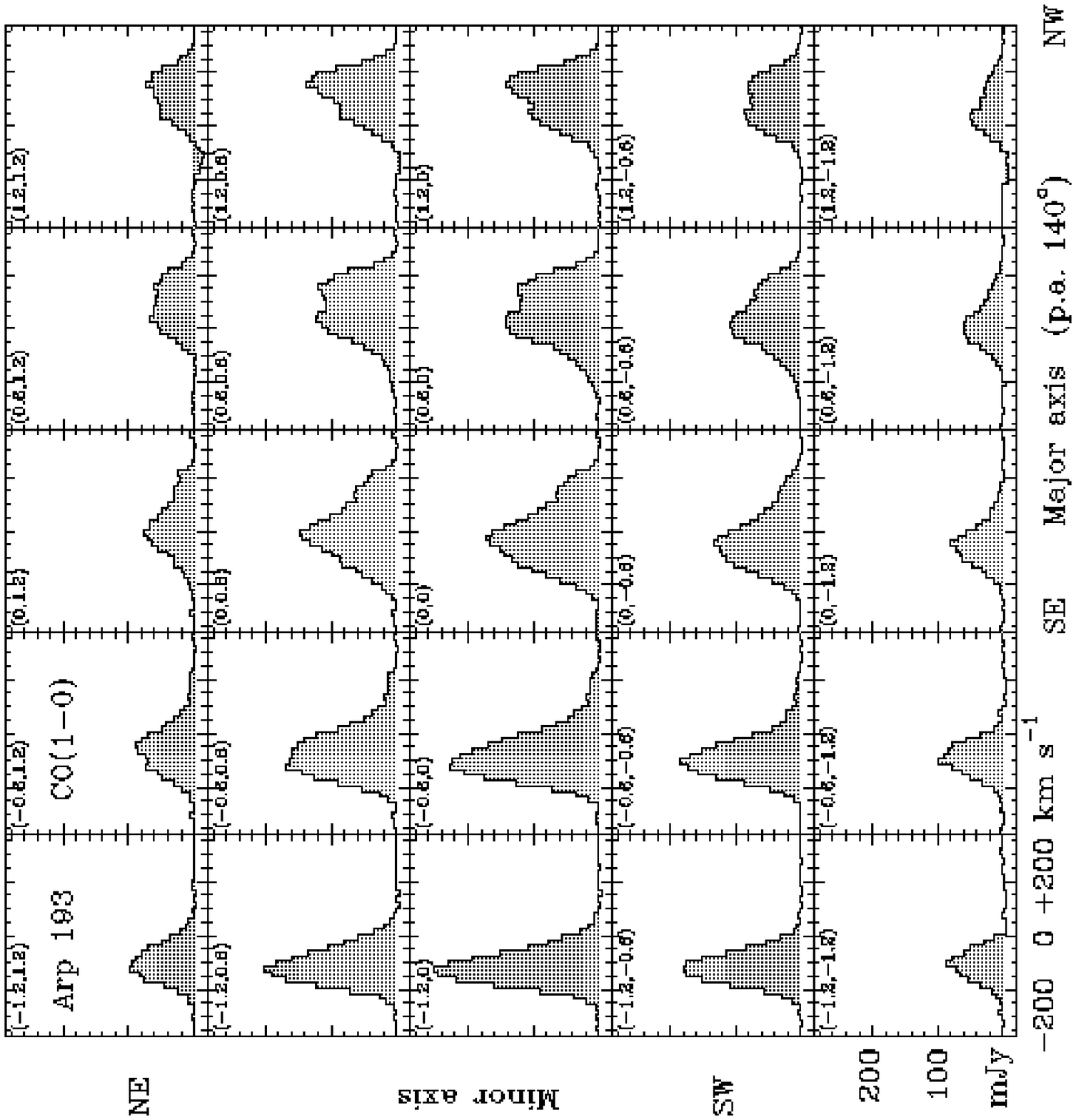}{\hsize}{-90}{56}{56}{-230}{510}
\plotfiddle{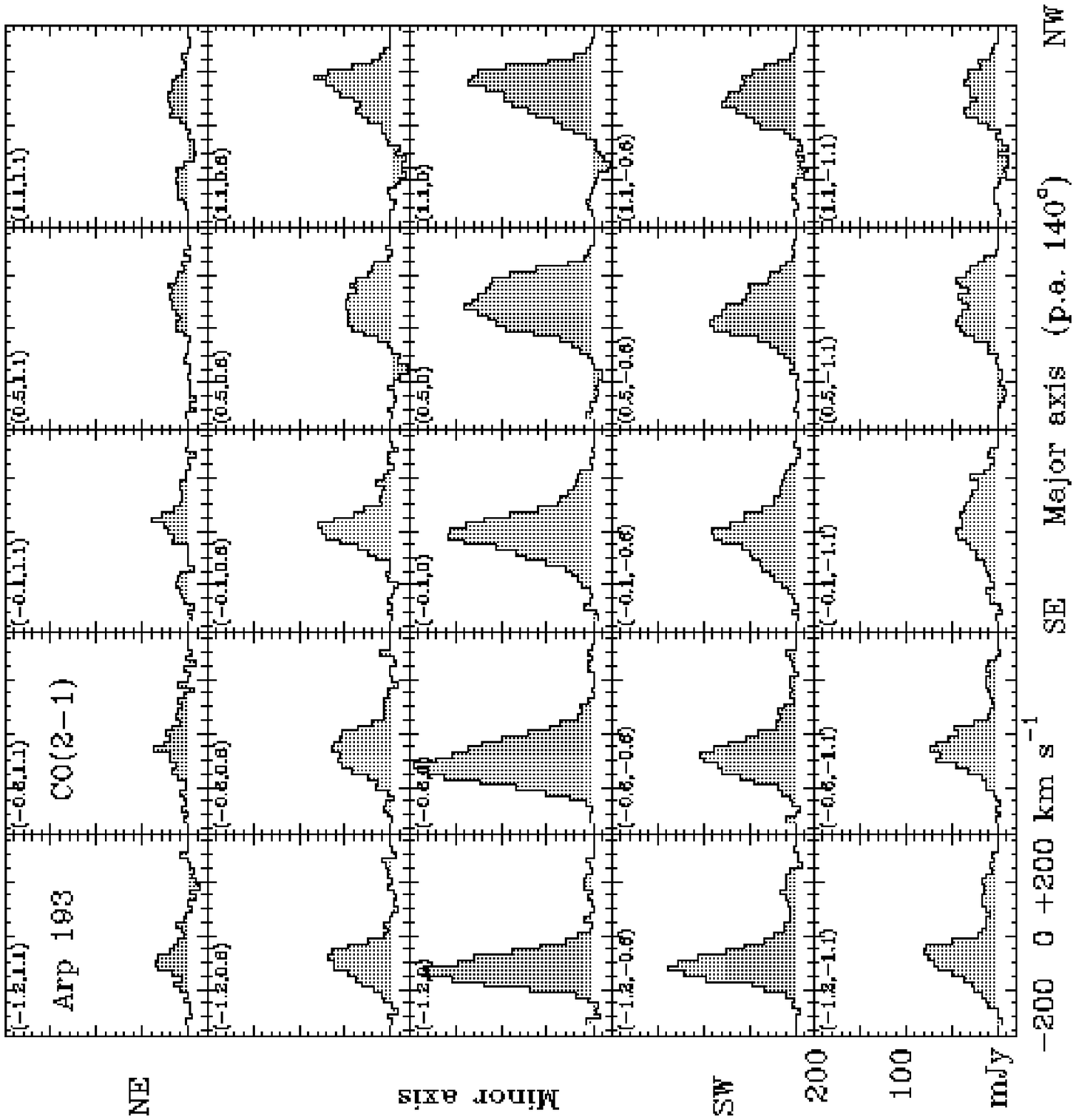}{\hsize}{-90}{56}{56}{-230}{666}
\end{figure}
\begin{figure}
%\vspace{-5cm}
\caption[{\bf Arp 193:}  CO(1--0) and (2--1) Spectra]
{
{\bf Arp 193 :}  
{\it upper:}
CO(1--0) spectra.
In each box vertical axis is CO intensity and 
horizontal axis is radial velocity relative to 
112.641\,GHz 
($cz_{\rm lsr} =$ 7000\,\kms ).
In the upper left of each box are offsets (arcsec) 
on the kinematic major and minor axes.
Beam smoothed to $1''.7\times 1''.6$.  

{\it lower:}
CO(2--1) spectra.  Beam $= 0''.6\times 0''.4$.  
Velocity is relative to relative to 225.282\,GHz ($cz_{\rm lsr} =$ 6994\,\kms ).
The (0,0) position in both diagrams is 
13$^{\rm h}$20$^{\rm m}$35.$^{\rm s}$315,
34$^\circ$08$'$22.$''20$ (J2000).
}
\end{figure}

%FIGURE 13 ---------------------------------------------------------
\begin{figure} 
\plotfiddle{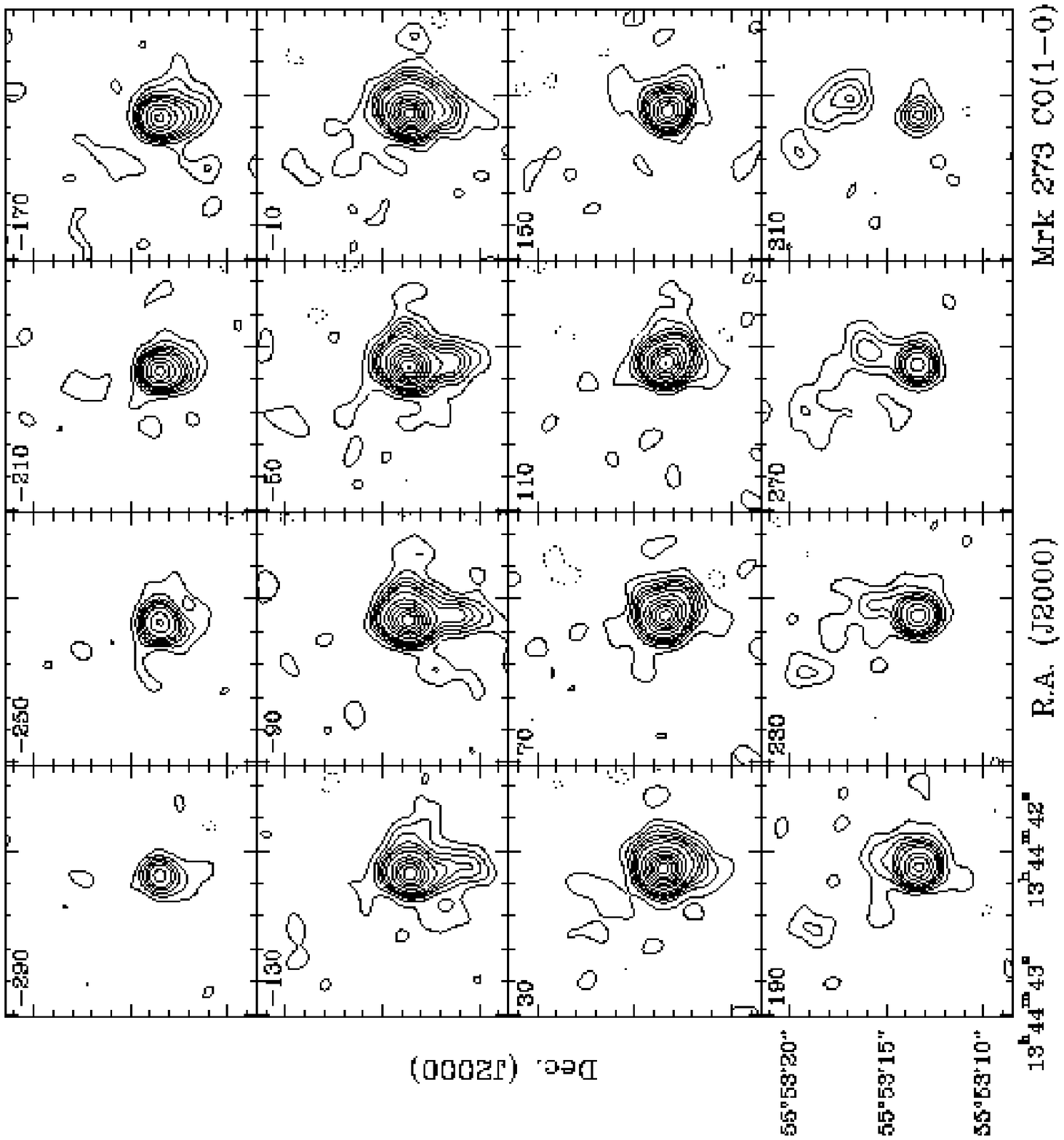}{\hsize}{-90}{56}{56}{-230}{510}
\plotfiddle{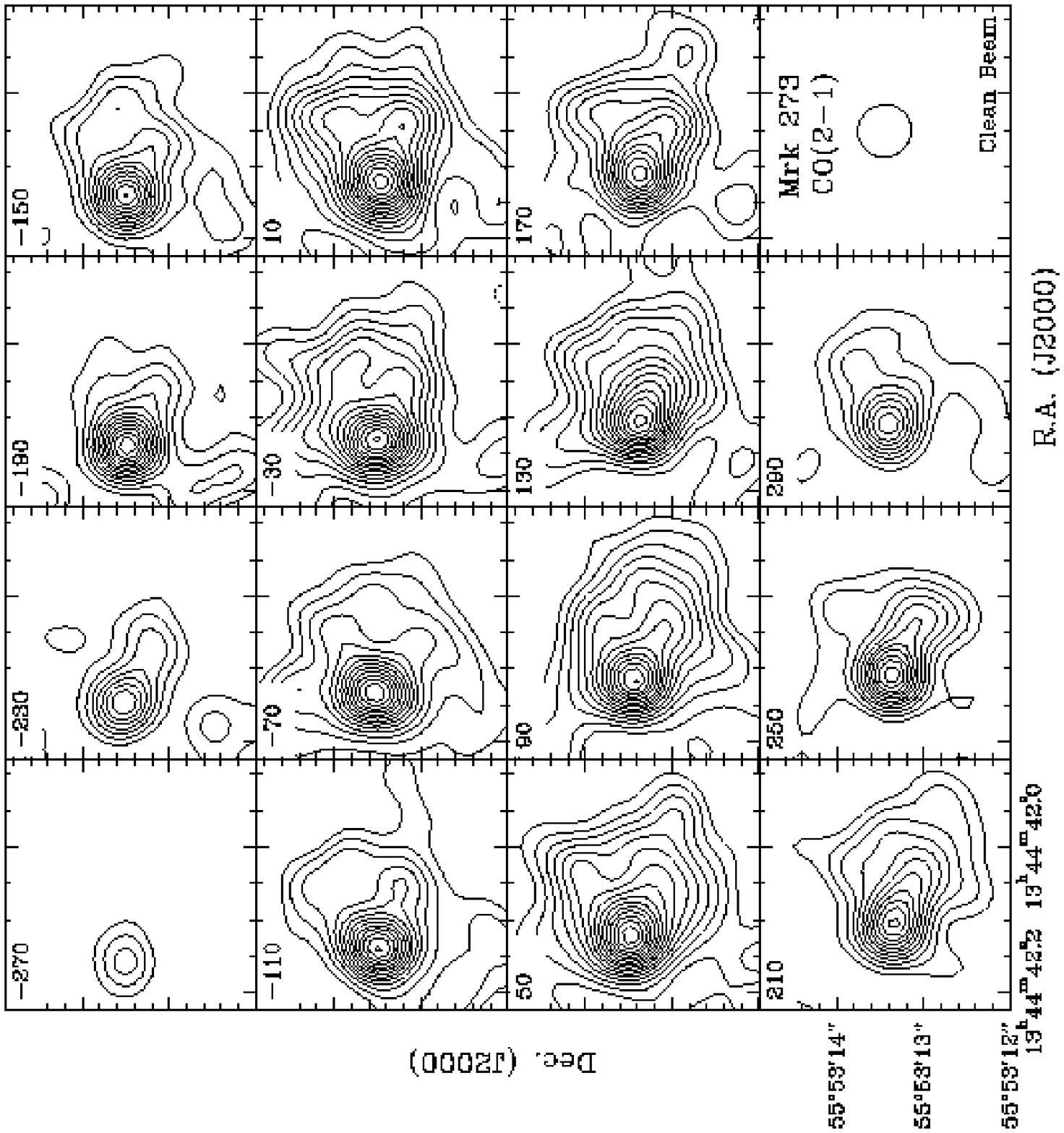}{\hsize}{-90}{56}{56}{-230}{666}
\end{figure}
\begin{figure}
%\vspace{-8cm}
\caption[{\bf Mrk 273 :} 40\,\kms\ Channel maps (a) CO(1--0);
(b) CO(2--1)]
{ {\bf Mrk 273 :}   

{\it a)} CO(1--0) maps in 40\,\kms\ channels.
Radial velocities (\kms , upper left of each box) are relative to 
111.076\,GHz ($cz_{\rm lsr} =$ 11323\,\kms ).
Contour unit: $-$3, $+$3 to 15 by 5, then 20 to 80 by 10\,mJy beam$^{-1}$, 
negative contours are dashed, zero contour omitted.
Beam $= 1''.4\times 1''.3$ with $T_b/S = 56$\,K/Jy.  

{\it b)} CO(2--1) maps in 40\,\kms\ channels.
Radial velocities (\kms , upper left of each box) are relative to 
222.176\,GHz ($cz_{\rm lsr} =$ 11283\,\kms ).
Contour interval 10 to 85, in steps of 5\,mJy beam$^{-1}$.
Beam $= 0''.6$ with $T_b/S = 66$\,K/Jy.  }\end{figure}

%FIGURE 14 --------------------------------------------------------- 
\begin{figure} 
\plotfiddle{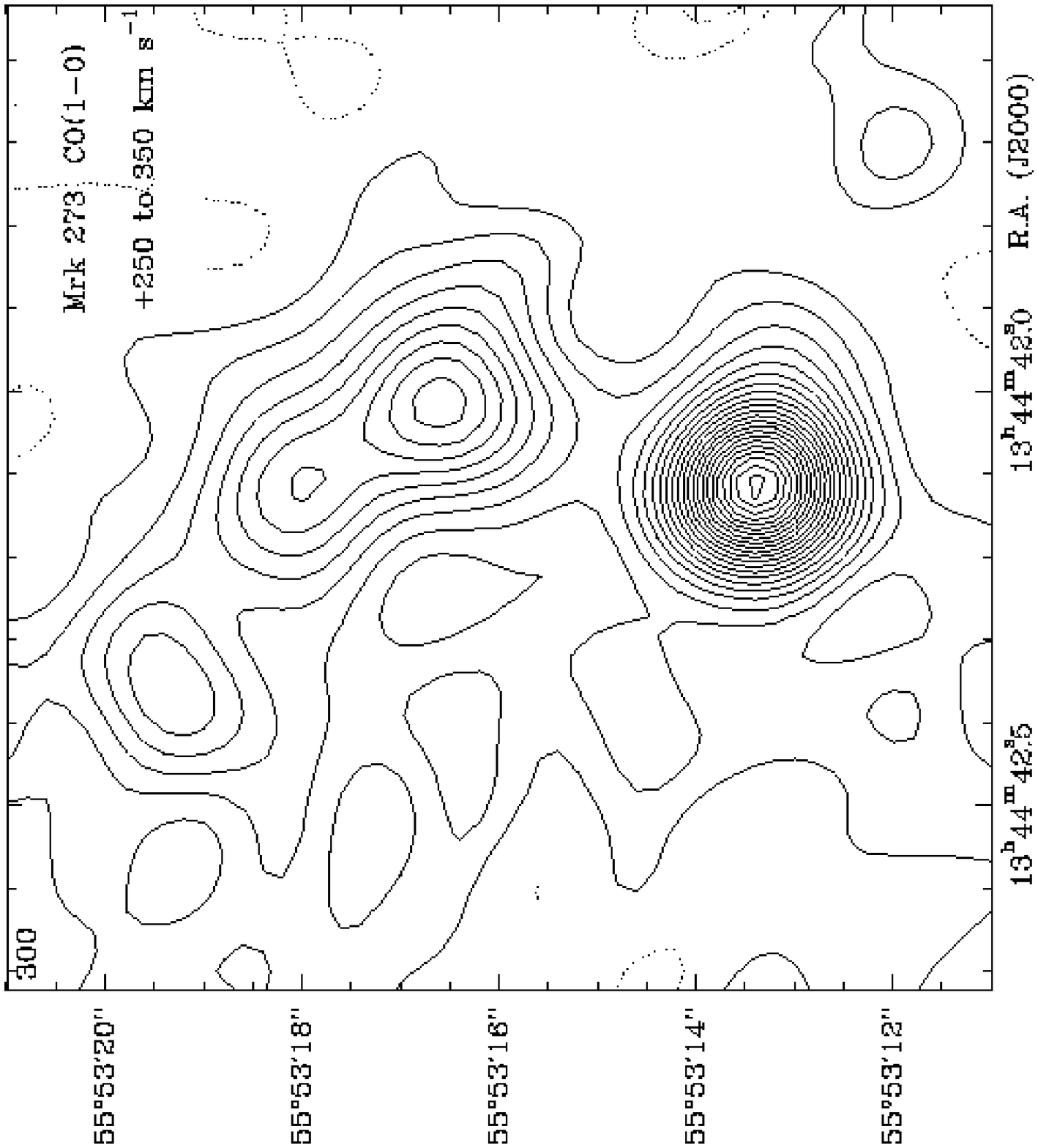}{\hsize}{-90}{65}{65}{-230}{520}
\end{figure}
\begin{figure}
\vspace{-8cm}
\caption[{\bf Mrk 273 :} CO(1--0) red vels w. N component]
{{\bf Mrk 273 :}  CO(1--0), integrated over 
250 to 350\,\kms , relative to 111.076\,GHz ($cz_{\rm lsr} =$ 11323\,\kms ).
Beam $= 1''.4\times 1''.3$ with $T_b/S = 56$\,K/Jy.  Contour step 1\,mJy
beam$^{-1}$,  negative contours are dashed, zero contour omitted. 
} \end{figure}

%FIGURE 15 --------------------------------------------------------- 
\begin{figure} 
\plotfiddle{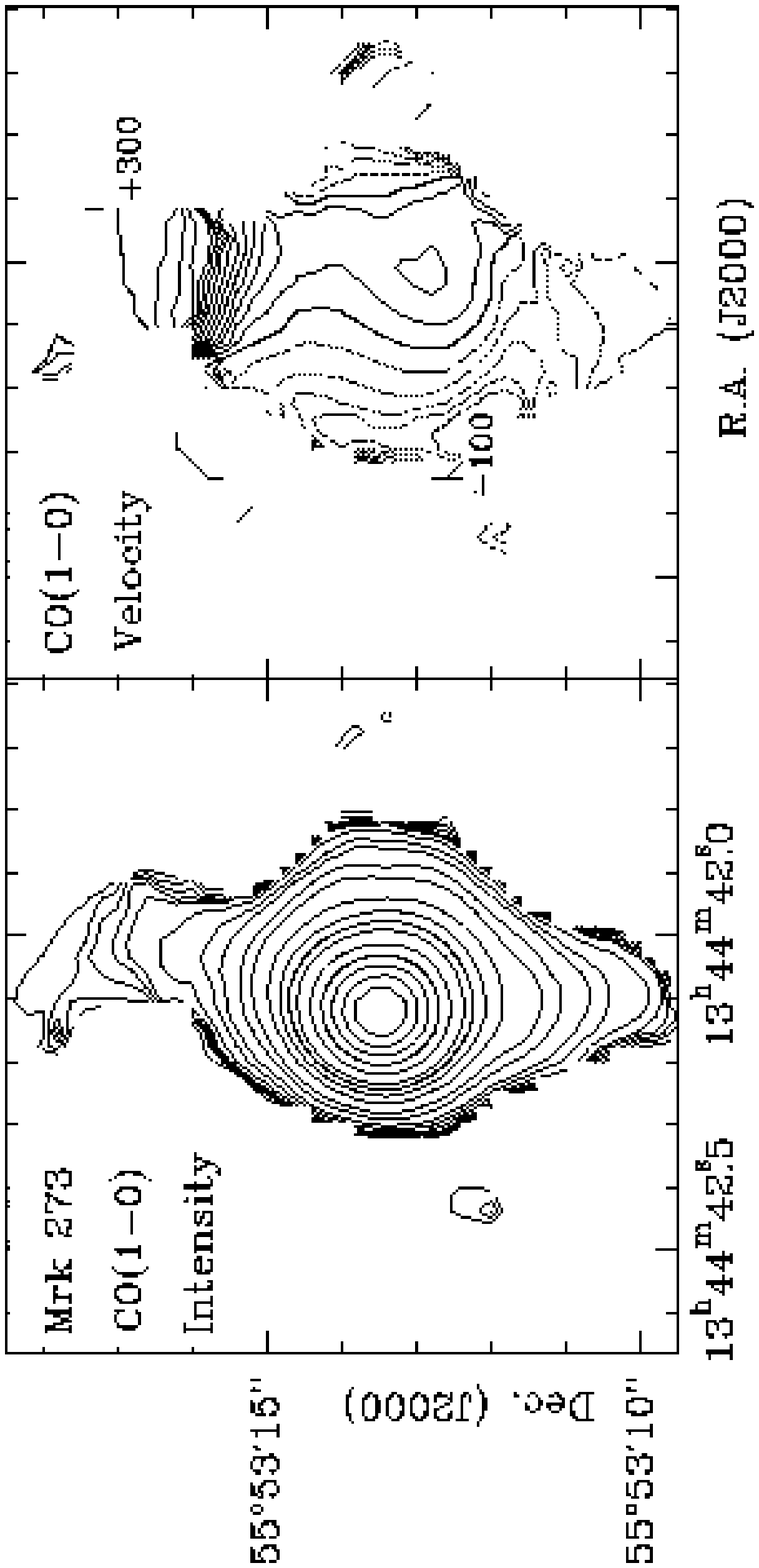}{\hsize}{-90}{60}{60}{-230}{520}
\plotfiddle{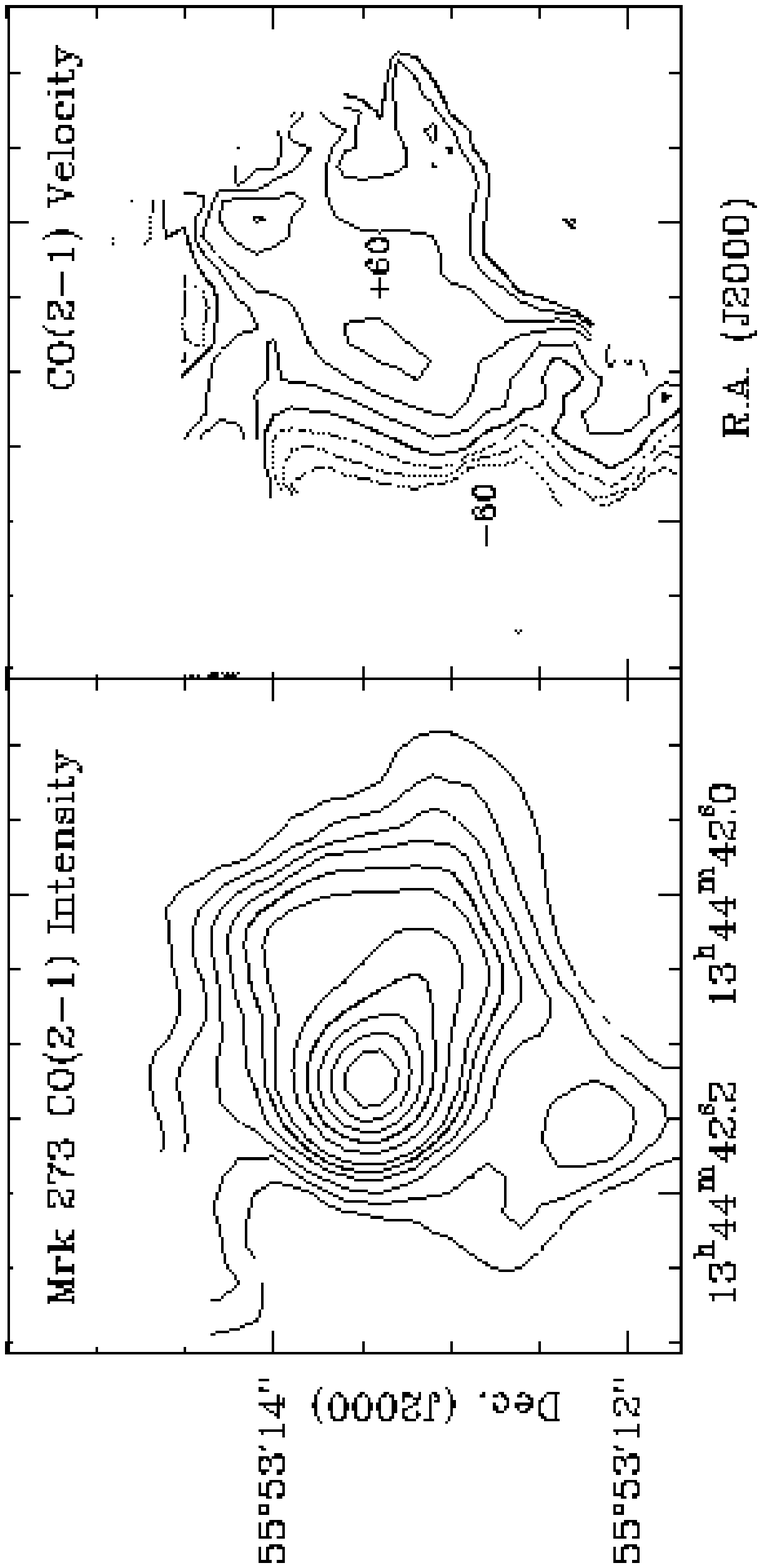}{\hsize}{-90}{60}{60}{-230}{810}
\end{figure}
\begin{figure}
\vspace{-8cm}
\caption[{\bf Mrk 273 :} CO(1--0) and (2--1) Moment maps]
{
{\bf Mrk 273 :}  
{\it Upper left: }
CO(1--0) integrated over  $\pm 340$\,\kms .
Beam $= 1''.4\times 1''.3$. Contours:   
0.3 to 0.7 by 0.1, then 1, 1.5, 2, 3, 4, 6, then 8 to 36 by 
4\,Jy beam$^{-1}$ \kms , with $T_b/S = 56$\,K/Jy.
{\it Upper right: } Velocity contours  
in steps of 20\,\kms\ relative to 111.076\,GHz 
($cz_{\rm lsr} =$ 11323\,\kms ).   Labels are in \kms ;
{\it Lower left: } 
CO(2--1) integrated over  $\pm 300$\,\kms .
Beam $ = 0''.6\times 0''.6$. Contours:  
2, 4, 6, then 8 to 32 in steps of 4\,Jy beam$^{-1}$ \kms , 
with $T_b/S = 66$\,K/Jy.
{\it  Lower right:} Velocity contours in steps of 20\,\kms\ relative to 
222.176\,GHz ($cz_{\rm lsr} =$ 11283\,\kms ).  Labels are in \kms .
}\end{figure}

\clearpage
%FIGURE 16-----------------------------------------------------
\begin{figure}
\plotfiddle{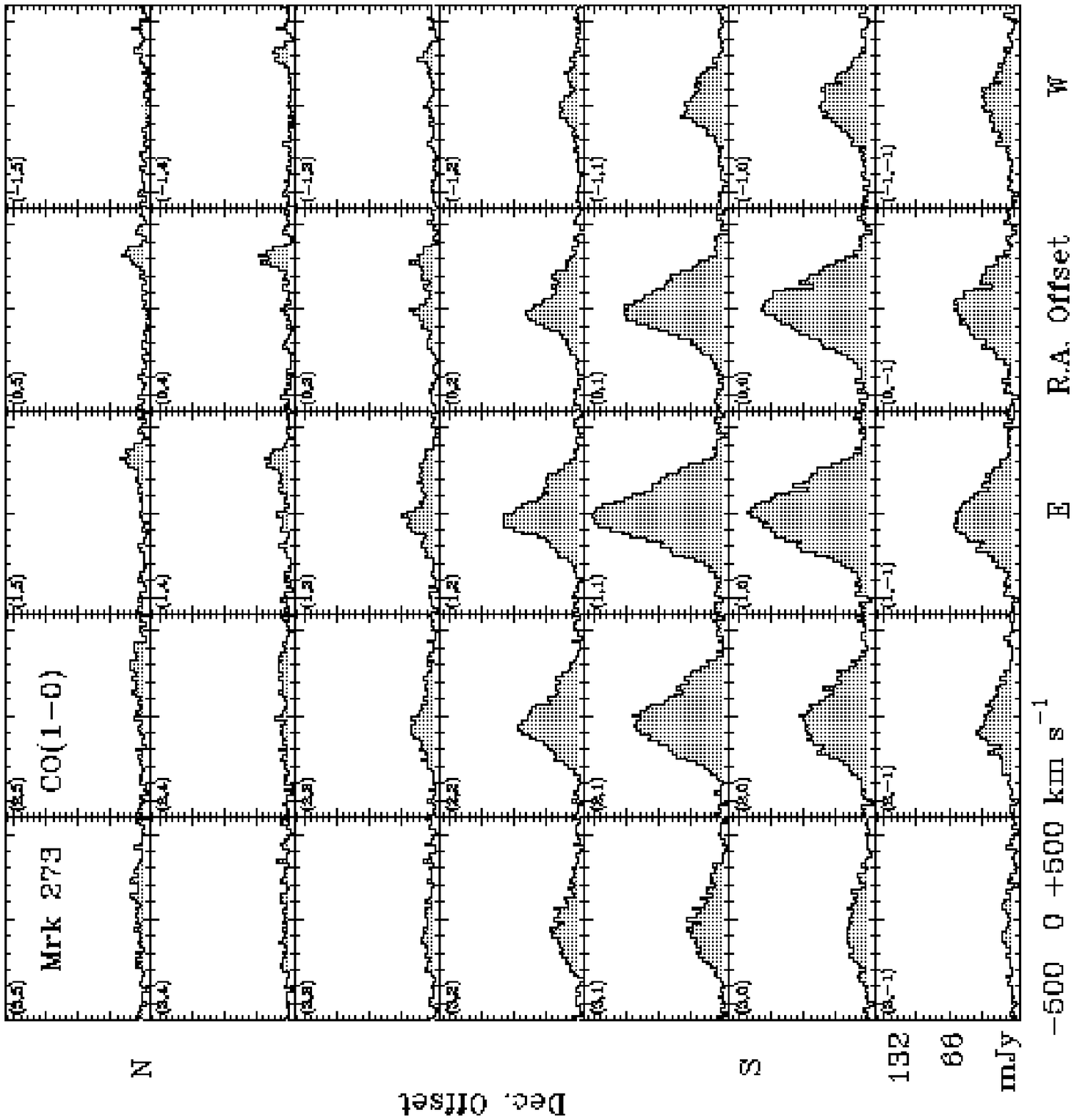}{\hsize}{-90}{65}{65}{-230}{520}
\end{figure}
\begin{figure}
\vspace{-8cm}
\caption[{\bf Mrk 273 :} CO(1--0) spectra.]
{{\bf Mrk 273 :} CO(1--0) spectra.  
In each box vertical axis is CO intensity, 
% with large ticks at 0, 33, 66, 99, and 132\,mJy  beam$^{-1}$; 
horizontal axis is radial velocity relative to 111.076\,GHz
($cz_{\rm lsr} =$ 11323\,\kms ). 
% with large ticks at $-$500, 0, and $+$500\,\kms\ (left to right).
In the upper left of each box are R.A.\ and Dec.\ offsets (arcsec); 
the (0,0) position is 13$^{\rm h}$44$^{\rm m}$42.$^{\rm s}$01,
55$^\circ$53$'$13.$''0$ (J2000).
Beam $=  2''.9\times 2''.1$, with $T_b/S = 16$\,K/Jy.
}\end{figure}

\clearpage
%FIGURE 17--------------------------------------------------------------
\begin{figure}
\plotfiddle{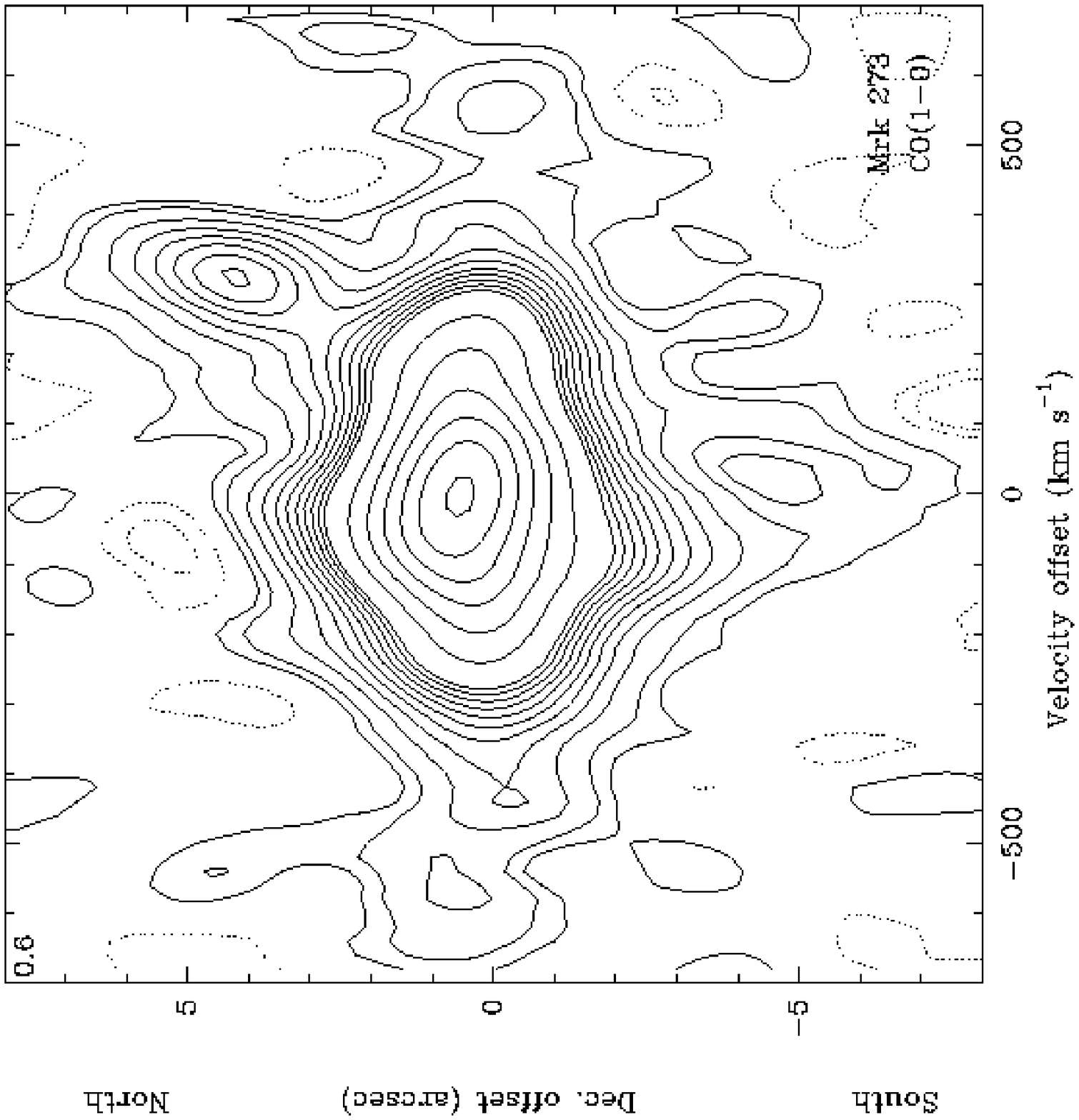}{\hsize}{-90}{56}{56}{-230}{510}
\plotfiddle{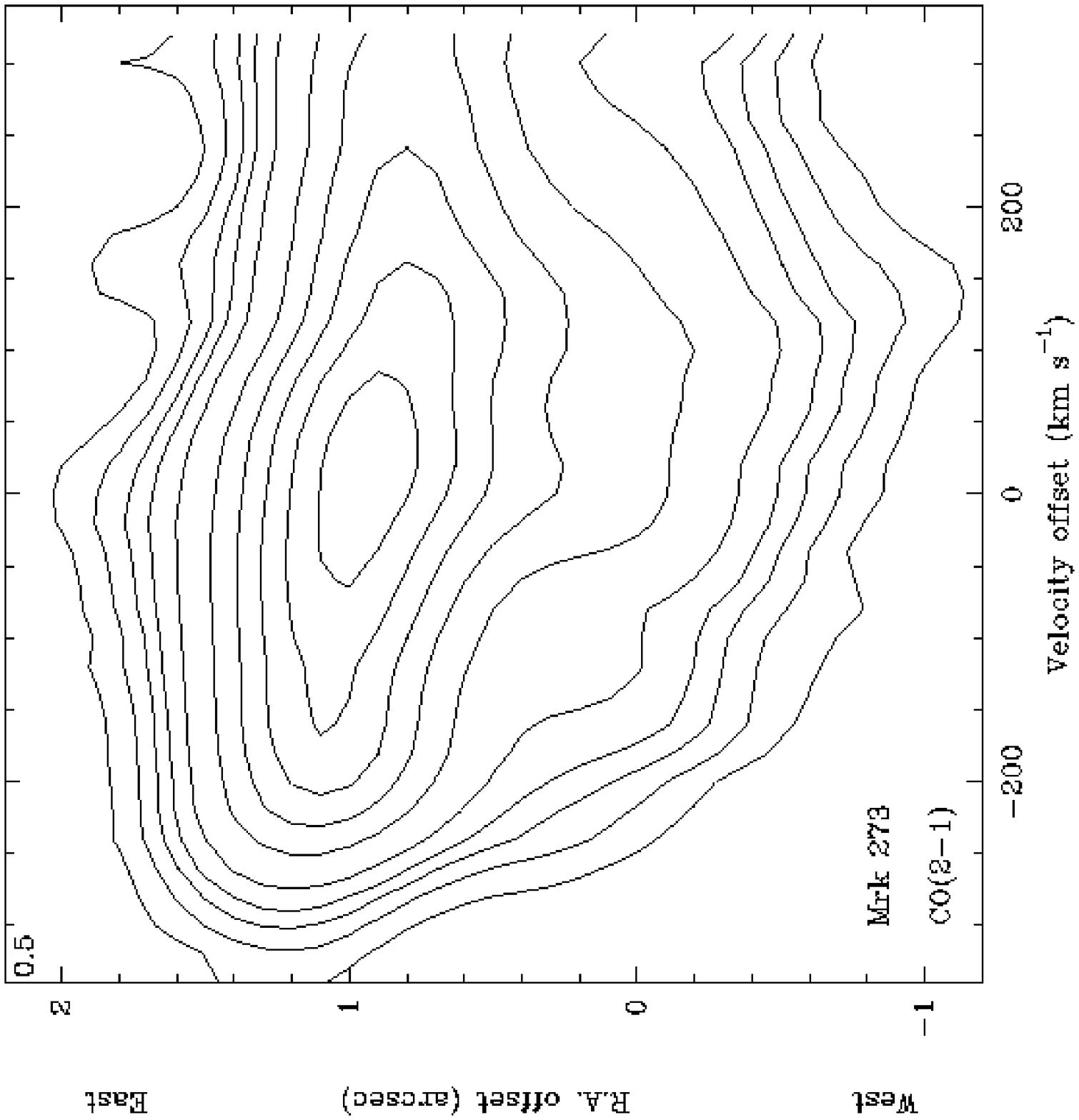}{\hsize}{-90}{56}{56}{-230}{666}
\end{figure}
\begin{figure}
\vspace{-5cm}
\caption[{\bf Mrk 273 :} Position-velocity cuts (a) CO(1--0) vel--Dec.;
(b) CO(2--1) vel--R.A.]
{{\bf Mrk 273 :} 

{\it (a)} CO(1--0) position-velocity diagram in declination, 
through the center of Mrk~273.  Note  at $+300$\,\kms\ the second CO 
complex $4''$ north of the main CO peak.
Contour levels:
$-2$, $-1$, $+1$, then 2 to 20 by 2, and 20 to 160 by 10, in units of
1.7\,mJy beam$^{-1}$, with $T_b/S = 16$\,K/Jy.
Beam $= 2''.9\times 2''.1$.  Velocity is relative to 111.076\,GHz
($cz_{\rm lsr} =$ 11323\,\kms ).   
The (0,0) position is 13$^{\rm h}$44$^{\rm m}$42.$^{\rm s}$01,
55$^\circ$53$'$13.$''0$ (J2000).

{\it (b)}
CO(2--1) position-velocity diagram along the line of nodes of the 
nuclear disk (p.a. 90$^\circ$, i.e., the R.A. axis).  
Contour levels: 3 to 15 by 3, then 20 to 70 by 10\,mJy beam$^{-1}$, 
with $T_b/S = 66$\,K/Jy.
Beam {\bf $= 0''.6\times 0''.6$}.  Velocity is relative to 222.176\,GHz
($cz_{\rm lsr} =$ 11283\,\kms ).
The (0,0) position is 13$^{\rm h}$44$^{\rm m}$42.$^{\rm s}$01,
55$^\circ$53$'$13.$''5$ (J2000).
}\end{figure}

\clearpage
%FIGURE 18 --------------------------------------------------------- 
\clearpage						% for 1 on a page
\begin{figure}
\vspace{-8cm}
\caption[{\bf Arp 220 : } CO(2--1) and (1--0) Moment maps]
{
{\bf Arp 220 :}  
{\it a):} CO(2--1) integrated intensity, velocity, 
linewidth (FWHM), and the 1.3\,mm continuum.  
CO integration limits:  ($-320, +300$\,\kms ). 
Beam $= 0''.7\times 0''.5$ .  Contours: 

\noindent
{\it integrated CO:} 1 to 10 by 1, in units of 11.4\,Jy beam$^{-1}$ \kms , 
with $T_b/S = 69$\,K/Jy.

\noindent
{\it  CO velocity:}
$-225$ to $+225$\,\kms , in steps of 25\,\kms\ relative to 226.422\,GHz
($cz_{\rm lsr} =$ 5450\,\kms ).  
Labels are in \kms ;

\noindent
{\it  CO linewidth:}
25 to 350\,\kms\ in steps of 25\,\kms .  Labels are in \kms .

\noindent
{\it 1.3\,mm continuum:} 
1 to 12 by 1, in units of 5.7\,mJy beam$^{-1}$.		

\noindent
{\it --- b):} CO(1--0) integrated intensity, velocity, 
linewidth (FWHM), and east streamer. 
For all maps, the beam is $1''.6\times 1''.1$,  with $T_b/S = 55$\,K/Jy,
and velocities are relative to 113.228\,GHz ($cz_{\rm lsr} =$ 5410\,\kms ). 
Contours: 

\noindent
{\it CO integrated from $-$320 to $+$300\,\kms  :} 0.1, 0.25, 0.5, then 
1 to 12 by 1, in units of 11.4\,Jy beam$^{-1}$ \kms ;

\noindent
{\it  CO velocity:}
$-$200 to $+$260\,\kms , in steps of 20\,\kms\   Labels are in \kms ;

\noindent
{\it  CO linewidth:}
25 to 350\,\kms\ in steps of 25\,\kms .  Labels are in \kms .

\noindent
{\it  CO east streamer, integrated from $+$85 to $+$165\,\kms :}
 0.025 to 0.2 by 0.025, then 0.25, then 0.4 to 2.4 by 0.4, in
units of 11.4\,Jy beam$^{-1}$ \kms .
}
\end{figure}

\clearpage
%FIGURE 19 ---------------------------------------------------------
\begin{figure}
\plotfiddle{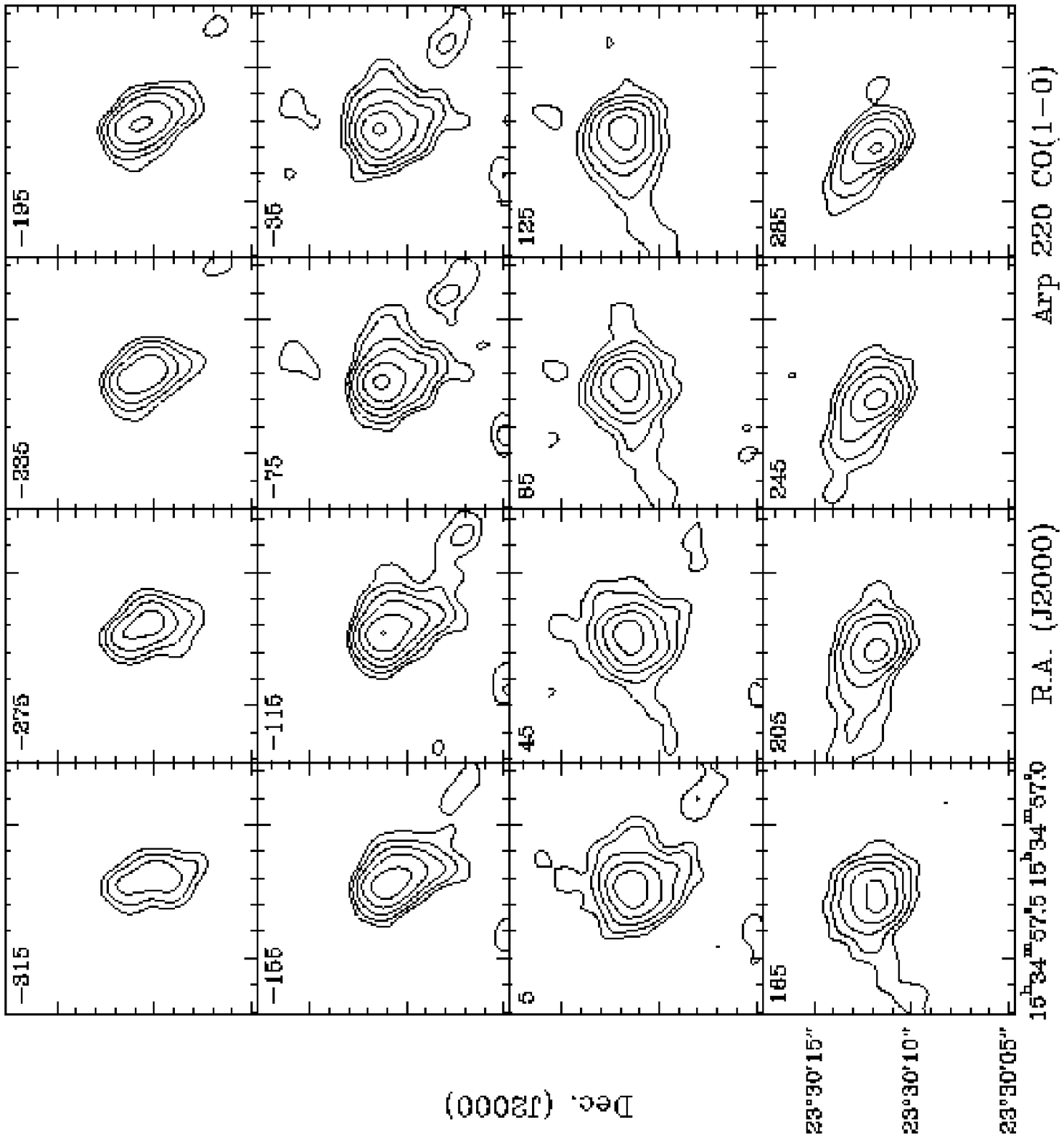}{\hsize}{-90}{56}{56}{-230}{510}
\plotfiddle{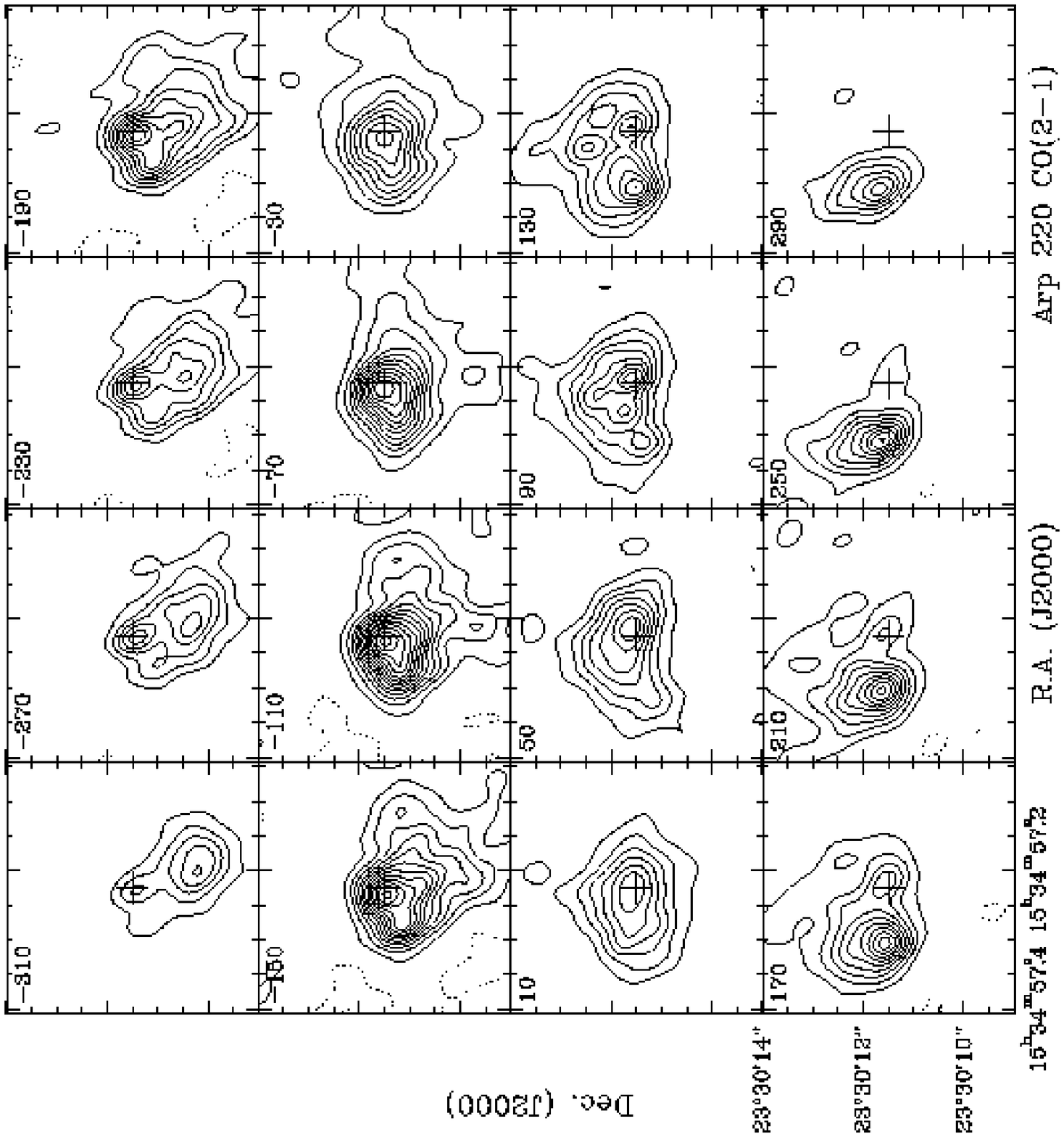}{\hsize}{-90}{56}{56}{-230}{666}
\end{figure}
\begin{figure}
%\vspace{-8cm}
\caption[{\bf Arp 220 :} Channel maps]
{ {\bf Arp 220 :}  {\it --- (a)} CO(1--0) maps in 40\,\kms\ channels.
Radial velocities (\kms , upper left of each box) are relative to 
113.228\,GHz ($cz_{\rm lsr} =$ 5410\,\kms ).
Contours: 1, 2, 4, 8, 16, 32, in units of 11.4\,mJy beam$^{-1}$, 
Beam $= 1''6\times 1''.1$ with $T_b/S = 55$\,K/Jy.

{\it --- (b)} CO(2--1) maps in 40\,\kms\ channels.
Radial velocities (\kms , upper left of each box) are relative to 
226.422\,GHz ($cz_{\rm lsr} =$ 5450\,\kms ).
The cross marks the position of the integrated CO(2--1) west peak
(Table~2). 
Beam $= 0''7\times 0''.5$ with $T_b/S = 69$\,K/Jy. 
 The continuum has been subtracted from the CO maps.
Contour unit: 50\,mJy beam$^{-1}$.
On this contour scale, the continuum from the west peak 
would be two contours. 	
}\end{figure}
\clearpage

%FIGURE 20----------------------------------------------------
\begin{figure}
\plotfiddle{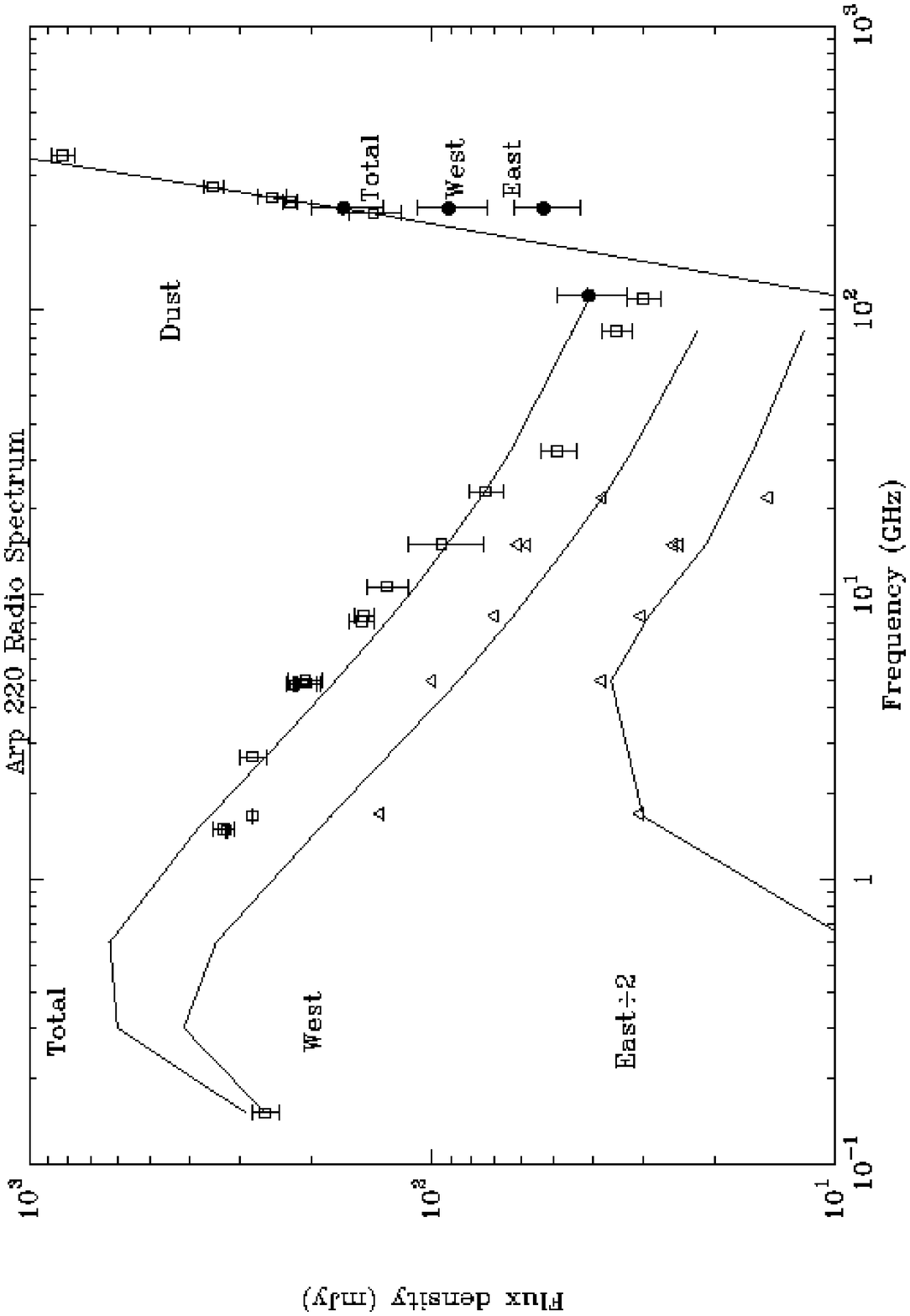}{\hsize}{-90}{65}{65}{-230}{520}
\end{figure}
\begin{figure}
\vspace{-8cm}
\caption[{\bf Arp 220 :} Radio continuum spectrum]
{{\bf Arp 220 :}  Radio continuum spectrum.  The emission below 30\,GHz
is nonthermal (Norris et al. 1985; 
Becklin \& Wynn-Williams 1987; Baan \& Haschick 1987; Baan et al. 1987;
Norris 1988; Condon et al. 1991; Sopp \& Alexander 1991
and other references in their Table~3). Millimeter and sub-mm data show 
the increasing flux of dust emission (this paper (solid circles); 
Radford et al. 1991b; Scoville et al. 1991, 1997; Woody et al. 1989; 
Carico et al. 1992;  Eales et al. 1989; and Rigopoulou et al. 1996.
The curves are spectra of the form $\nu^{2.1+\alpha}(1+{\rm e}^{-\tau})$,
normalized to the observed fluxes. The free-free opacity,
$\tau$, has the form $(\nu/\nu_0)^{-2.1}$, where $\nu_0$ is the 
turnover frequency.  The curve for the dust spectrum is for an emissivity
index $n=1.5$.
}\end{figure}

\clearpage
%FIGURE 21----------------------------------------------------
\begin{figure}
\plotfiddle{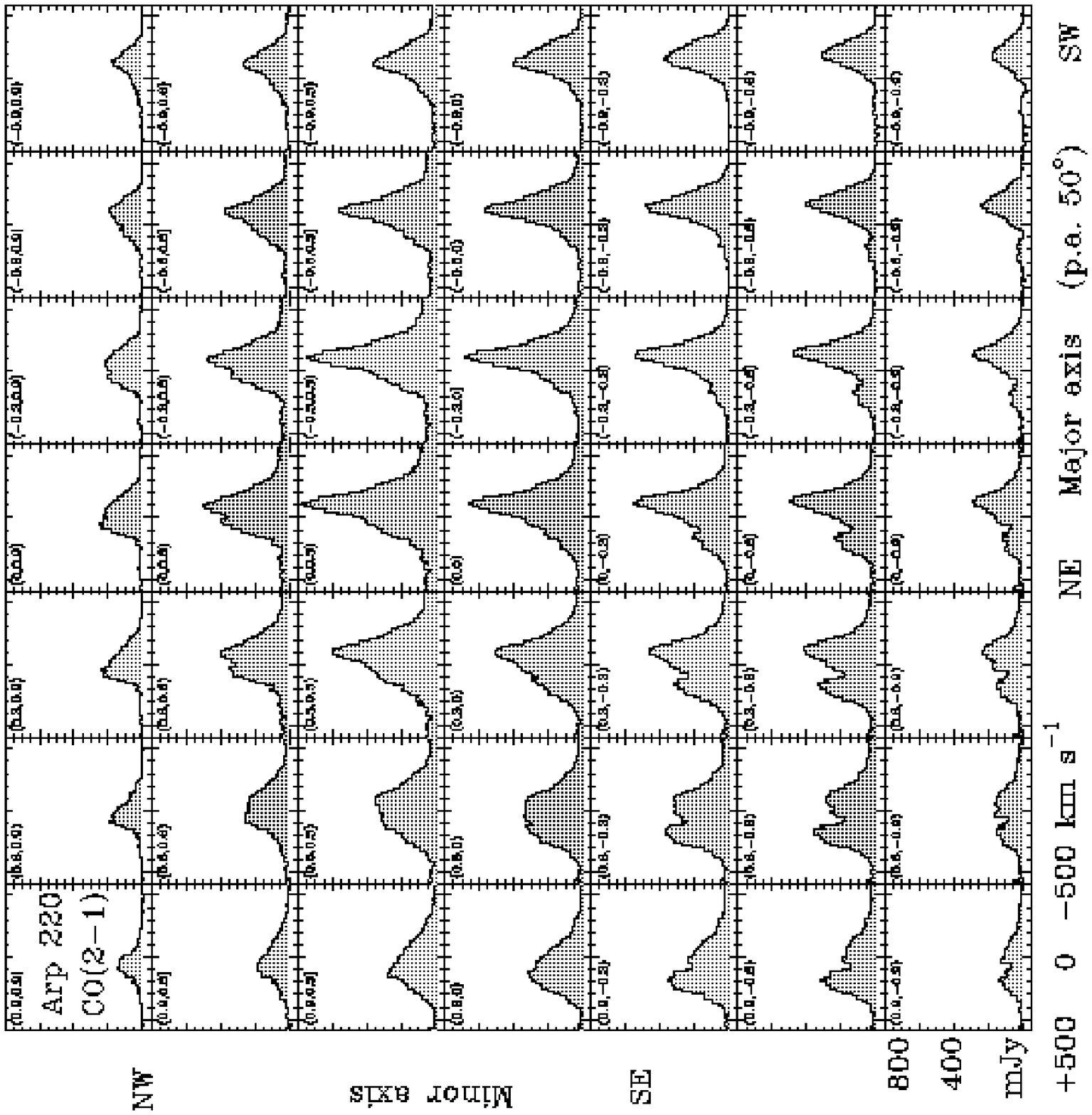}{\hsize}{-90}{65}{65}{-230}{520}
\end{figure}
\begin{figure}
\vspace{-8cm}
\caption[{\bf Arp 220 :} Map of CO(2--1) Spectra.]
{{\bf Arp 220 :}  CO(2--1) spectra.  
In each box vertical axis is CO intensity,  
horizontal axis is radial velocity relative to  
226.422\,GHz ($cz_{\rm lsr} =$ 5450\,\kms ).
In the upper left of each box are offsets (in steps of 0.$''$3) 
on the kinematic major and minor axes, relative to
15$^{\rm h}$34$^{\rm m}$57.$^{\rm s}$24,
23$^\circ$30$'$11.$''2$ (J2000).   
Beam smoothed to $= 1''.1\times 0''.6$ at p.a. 34$^\circ$, 
with $T_b/S = 34$\,K/Jy.
}\end{figure}

%FIGURE 22-----------------------------------------------------OLD F33
\begin{figure}
\plotfiddle{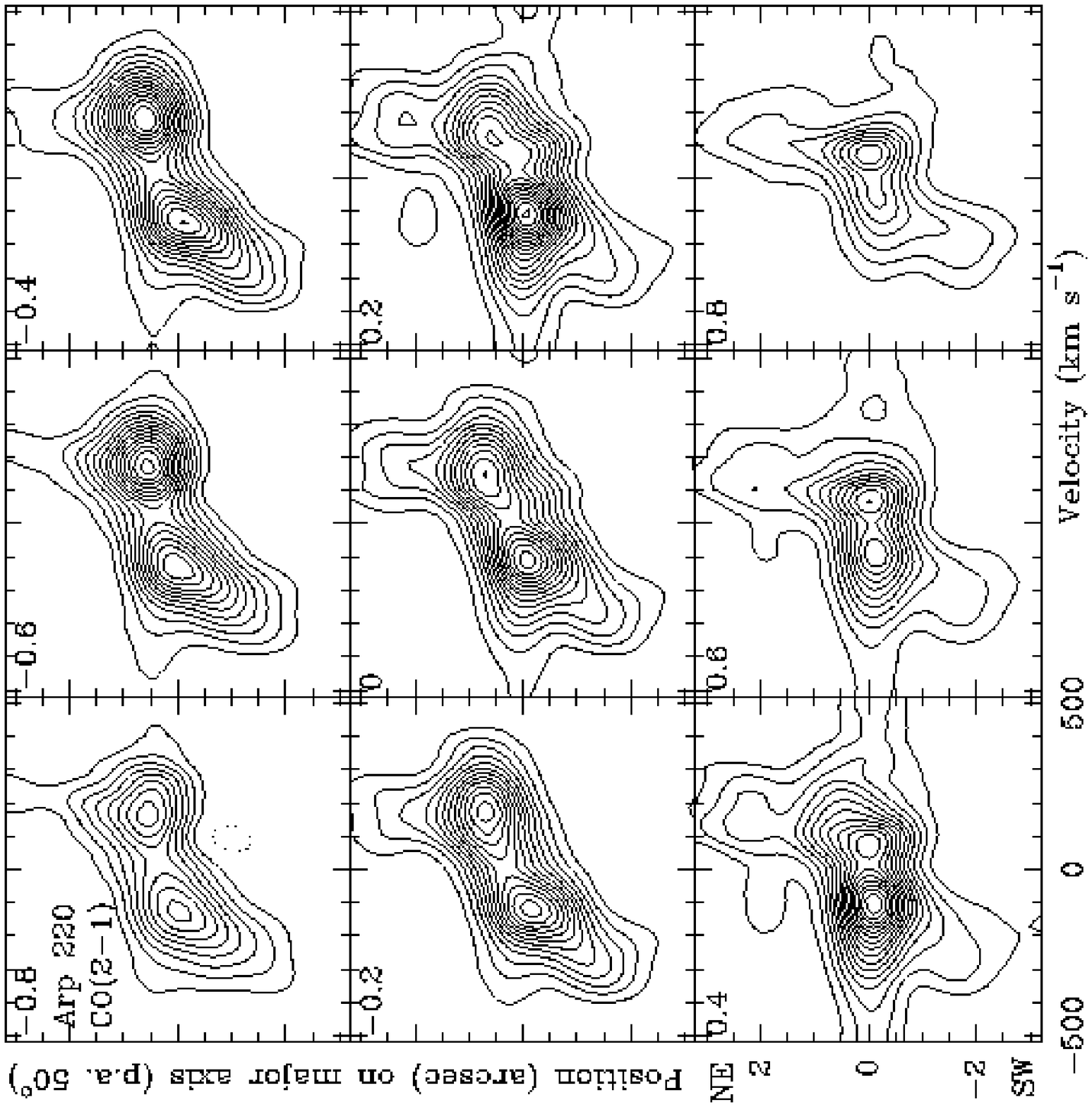}{\hsize}{-90}{65}{65}{-230}{520}
\end{figure}
\clearpage						% for 1 on a page
\begin{figure}
\plotfiddle{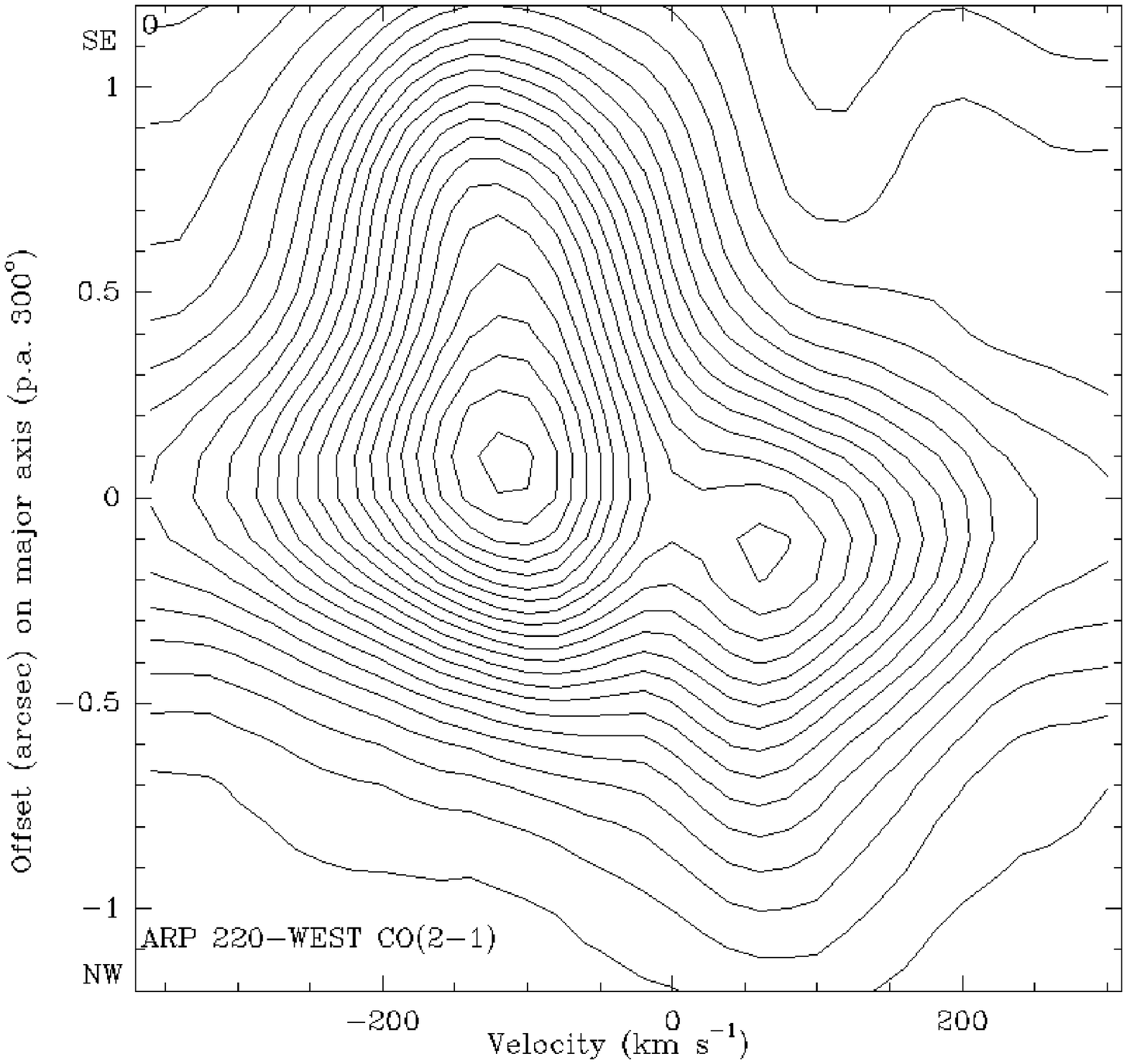}{\hsize}{-90}{85}{85}{-300}{520}	% 
\end{figure}
\clearpage
\begin{figure}
\plotfiddle{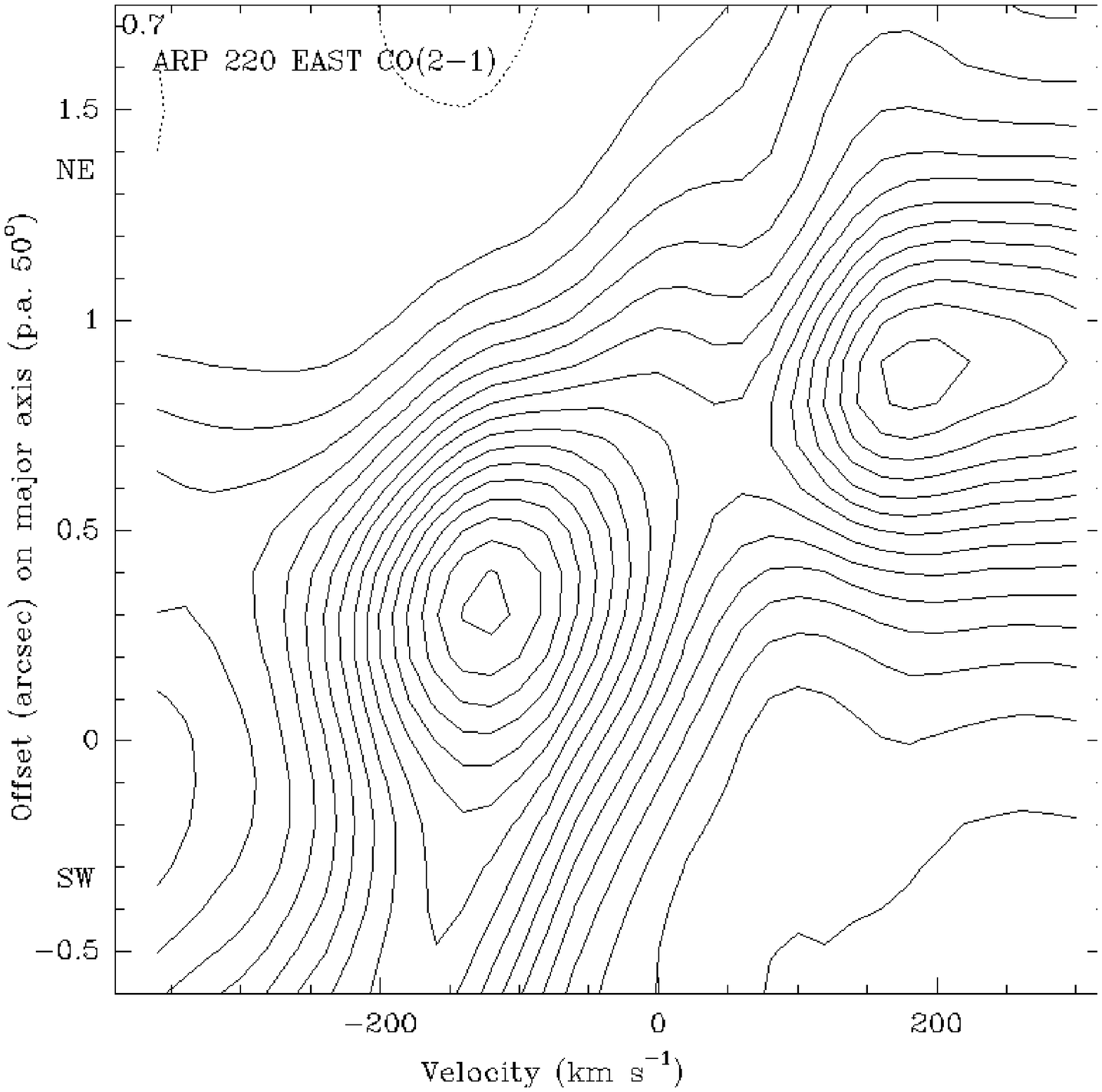}{\hsize}{-90}{85}{85}{-300}{520}	% 
\end{figure}
\clearpage						
\begin{figure}
%\vspace{-8cm}
\caption[{\bf Arp 220 :} CO(2--1) position-velocity diagrams]
{ {\it a)} {\bf Arp 220-disk :} CO(2--1) position-velocity cuts
along the line of nodes (p.a. 50$^\circ$).  Labels in the upper left of
each box are offsets (arcsec) along the minor axis (northwest is positive,
southeast is negative). The (0,0) position is 
15$^{\rm h}$34$^{\rm m}$57.$^{\rm s}$24,
23$^\circ$30$'$11.$''2$ (J2000).   
Contour levels are 1 to 17, in units of 
30\,mJy beam$^{-1}$, with $T_b/S = 44$\,K/Jy.
Beam smoothed to $= 1''.0\times 0''.55$ at p.a. 45$^\circ$.  

{\it b)} {\bf Arp~220-west}: CO(2--1) position-velocity cut
through the Arp~220-west, along a possible
major axis at p.a. 300$^\circ$ (see text).  Contour interval 
17\,mJy beam$^{-1}$, with $T_b/S = 80$\,K/Jy.  The source peak is 
280\,mJy beam$^{-1}$.

{\it c)} {\bf Arp~220-east}: CO(2-1) position-velocity cut
through Arp~220-east at p.a. 50$^\circ$.  
Contour interval 
17\,mJy beam$^{-1}$, with $T_b/S = 80$\,K/Jy.  The source peak is 
380\,mJy beam$^{-1}$.
For diagrams {\it b)}
and {\it c}, the restoring beam was $0''.6\times 0''.5$ at p.a. 120$^\circ$, 
and the (0,0) position is 
15$^{\rm h}$34$^{\rm m}$57.$^{\rm s}$225,
23$^\circ$30$'$11.$''5$ (J2000) (the western continuum peak).
In all these position-velocity diagrams,
velocity is relative to  
226.422\,GHz ($cz_{\rm lsr} =$ 5450\,\kms ).

}\end{figure}

\clearpage
%FIGURE 23----------------------------------------------------
\clearpage						% for 1 on a page
\begin{figure}
\vspace{-5cm}
\caption[(Plate) {\bf Arp 220 :} CO superposed on optical image. ]
{(Plate) {\bf Arp 220 :} CO contours superposed on 
a false-color presentation of the HST $V$ band image of
Arp~220 (Shaya et al. 1994).  

\noindent
--- {\it a)} CO(1--0) emission mapped with a beam of $1''.6\times 1''.1$,  
with $T_b/S = 55$\,K/Jy,  integrated in two different velocity ranges,
relative to $cz_{\rm lsr}$ = 5410\,\kms :
{\it --- left:} redshifted CO(1--0) emission integrated from
$+$205 to $+$325\,\kms ; contours  
 0.75 to 1.75 by 0.25, 2, 2.5, 3, 4.5, 6, 9, 12, 15, 20, 25, in
units of 1.14\,Jy beam$^{-1}$ \kms ; 
{\it --- right:} blueshifted CO(1--0) emission integrated from
$-$235 to $+$5\,\kms ,  with contours  
 0.75, 2.5, 3.5,  6, 12, 24, 36, 50, 60, 70, in
units of 1.14\,Jy beam$^{-1}$ \kms .

\noindent
{\it b):} CO(1--0) emission in the east streamer and CO(2--1) emission: 
{\it --- left:}  CO(1--0) emission in the east streamer integrated from
$+$85 to $+$165\,\kms ; contours and beam as in Fig.~23{\it a), left}.
{\it --- right:} CO(2--1) emission integrated from $-$320 to $+$300\,\kms , 
relative to $cz_{\rm lsr}$ = 5450\,\kms , 
with contours  20, 30 to 180 by 30, then 240, 
in units of 0.57\,Jy beam$^{-1}$ \kms ;
beam $= 0''.7\times 0''.5$, with $T_b/S = 69$\,K/Jy.
}\end{figure}

\clearpage
%FIGURE 24-----------------------------------------------------
\begin{figure}
\epsscale{0.9}
\plotone{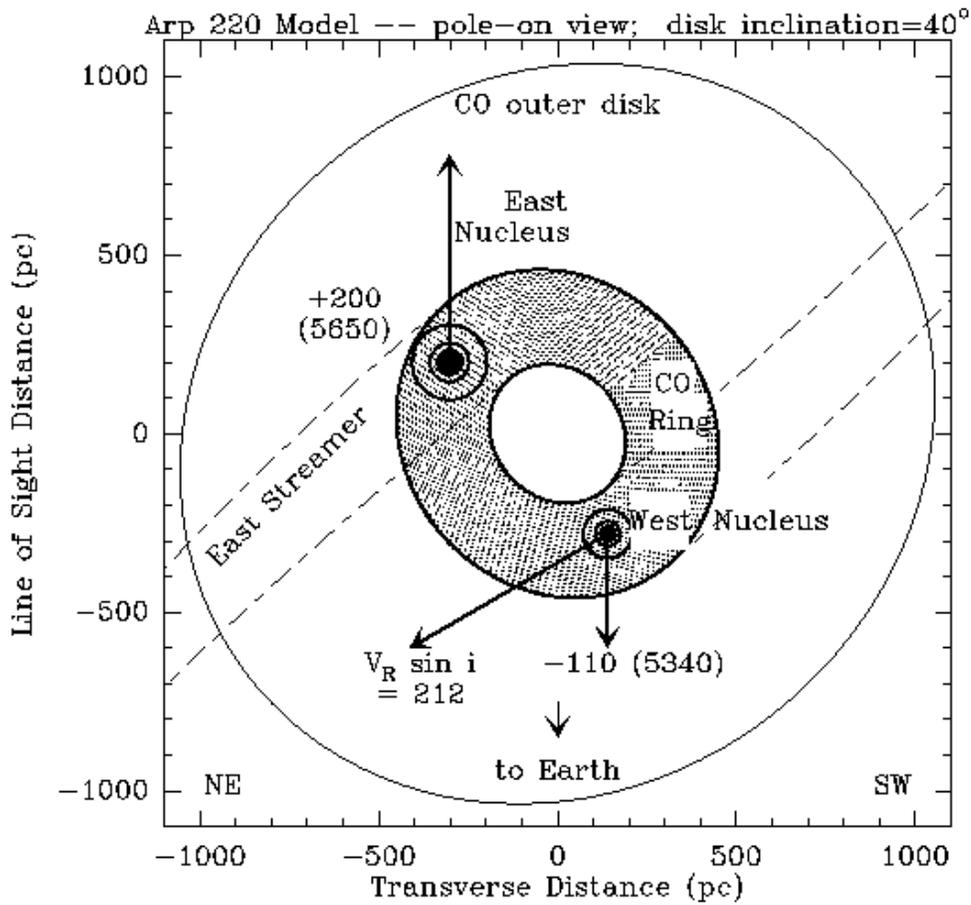}
\caption[{\bf Arp 220 :} Sketch]
{{\bf Arp 220 :} Model of the eastern and western
nuclei in the molecular disk.  The rotation velocity is 330\,\kms, and 
the view is pole-on.  The arrows indicate  
the observed radial components along the line of sight, 
at a disk inclination of 40$^\circ$ from face-on.  Velocities
are relative to a systemic velocity of $cz_{\rm lsr} =$ 5450\,\kms .
Labels are \kms , values in parentheses are  $cz_{\rm lsr}$.
}\end{figure}

\clearpage
%FIGURE 25 ------------------------------------------------------- 
\begin{figure}
\plotfiddle{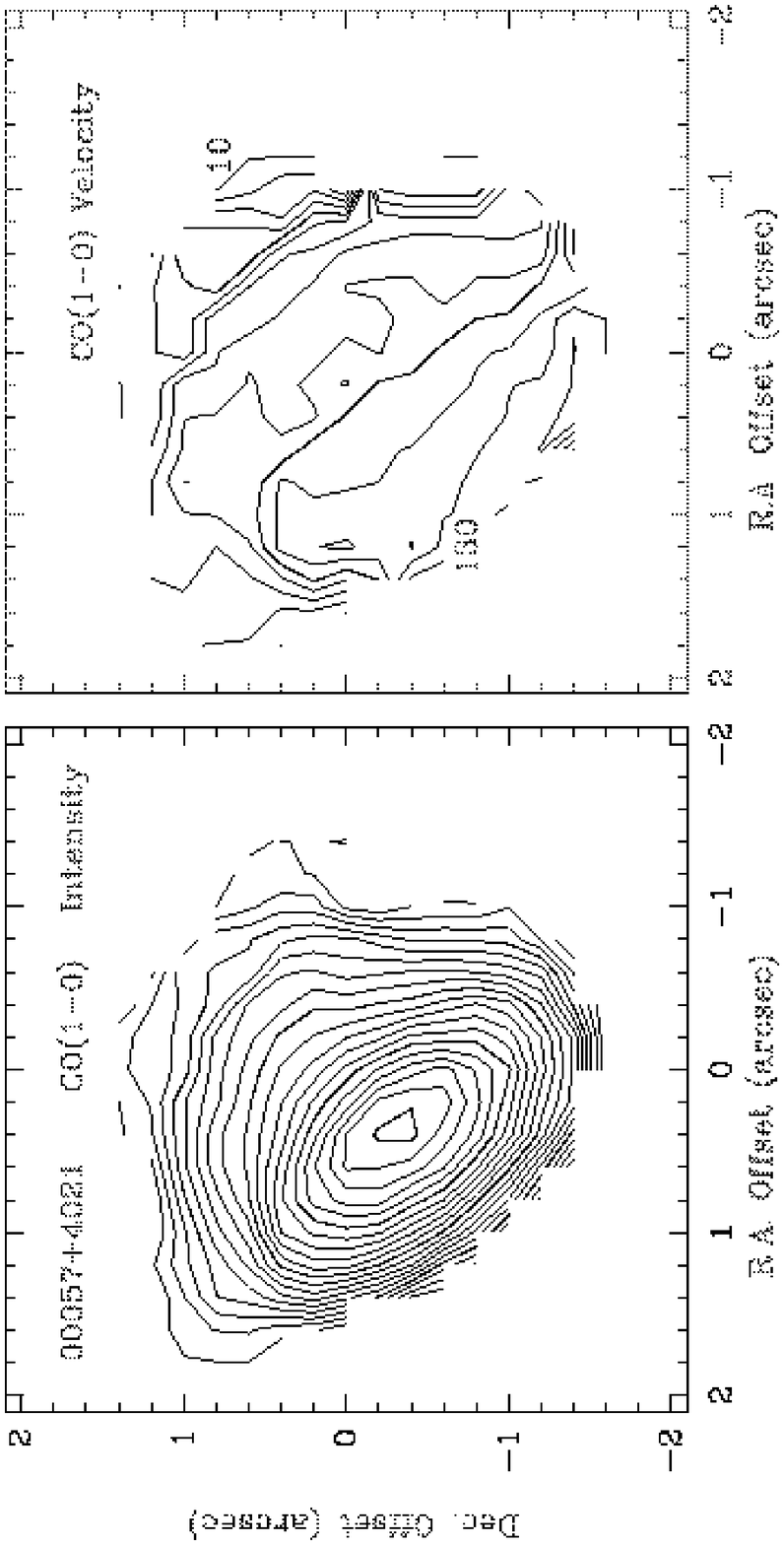}{\hsize}{-90}{65}{65}{-270}{620}
\plotfiddle{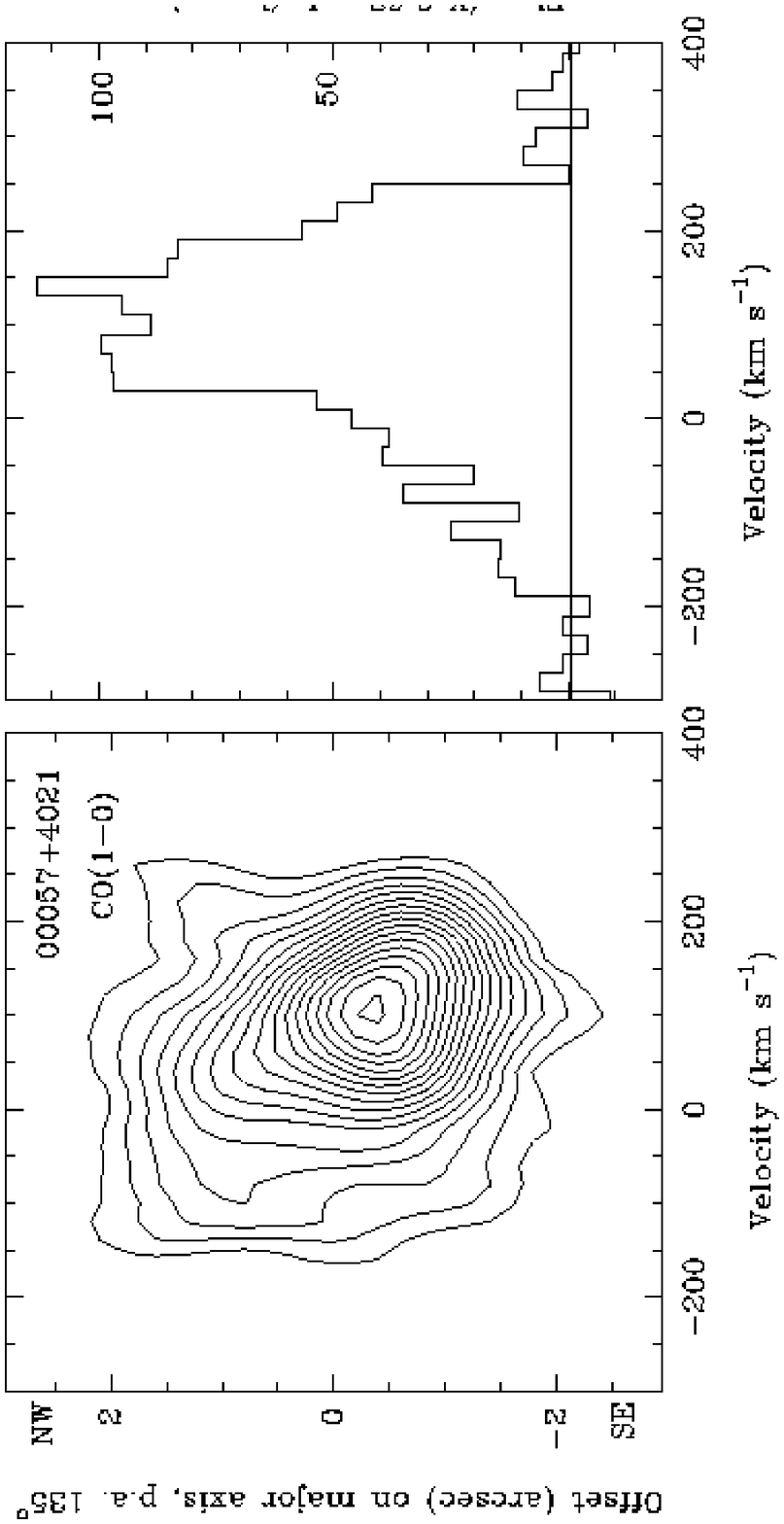}{\hsize}{-90}{65}{65}{-270}{860}
\end{figure}
\begin{figure}
\vspace{-5cm}
\caption[{\bf 00057$+$4021 :} CO(1--0) moments, posn-vel, spectrum]
{ {\bf 00057$+$4021:}
{\it  upper left:}
CO(1--0) integrated over ($-$160, $+$240\,\kms ).
Beam $ = 2''.2\times 1''.1$. 
Contour step = 0.63\,Jy beam$^{-1}$ \kms , 
with $T_b/S = 40$\,K/Jy.
{\it upper right:}
 CO velocity contours in steps of 10\,\kms .  
Labels are in \kms .
{\it lower left:}
CO(1--0) position-velocity diagram at p.a.\
135$^\circ$. Contours: 3 to 19 by 1, in units of 3.1\,mJy beam$^{-1}$,
with $T_b/S = 40$\,K/Jy.  
{\it lower right:} CO spectrum in the 
$2''.2\times 1''.1$ beam at the source peak.
Position offsets are relative to 00$^{\rm h}$08$^{\rm m}$20.$^{\rm s}$58,
$+40^\circ$37$'$55.$''5$ (J2000), and velocity scales are relative to 
110.360\,GHz ($cz_{\rm lsr} =$ 13341\,\kms ). 
}\end{figure}

\clearpage
%FIGURE 26 ---------------------------------------------------  
\begin{figure}
\plotfiddle{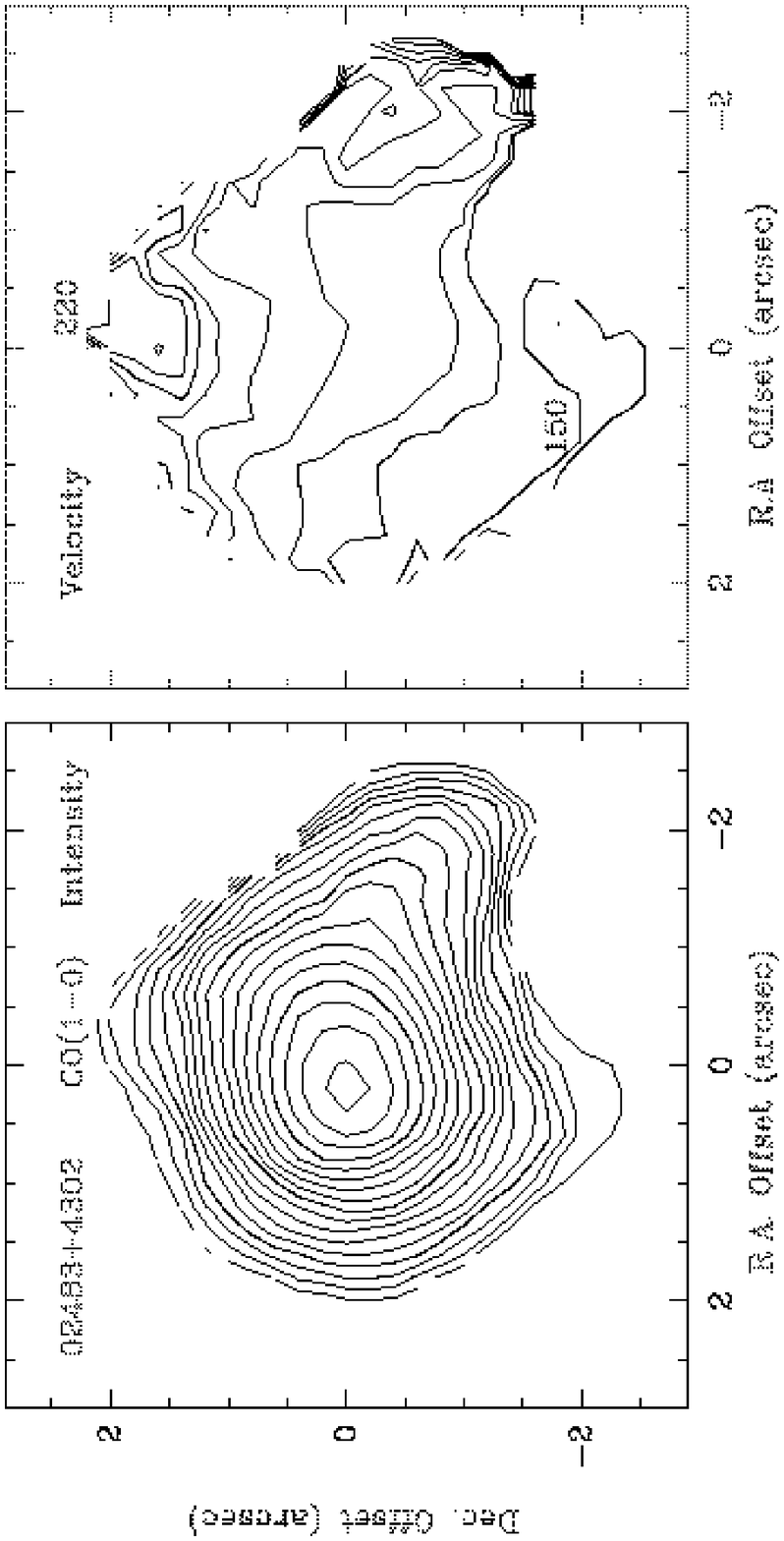}{\hsize}{-90}{65}{65}{-270}{620}
\plotfiddle{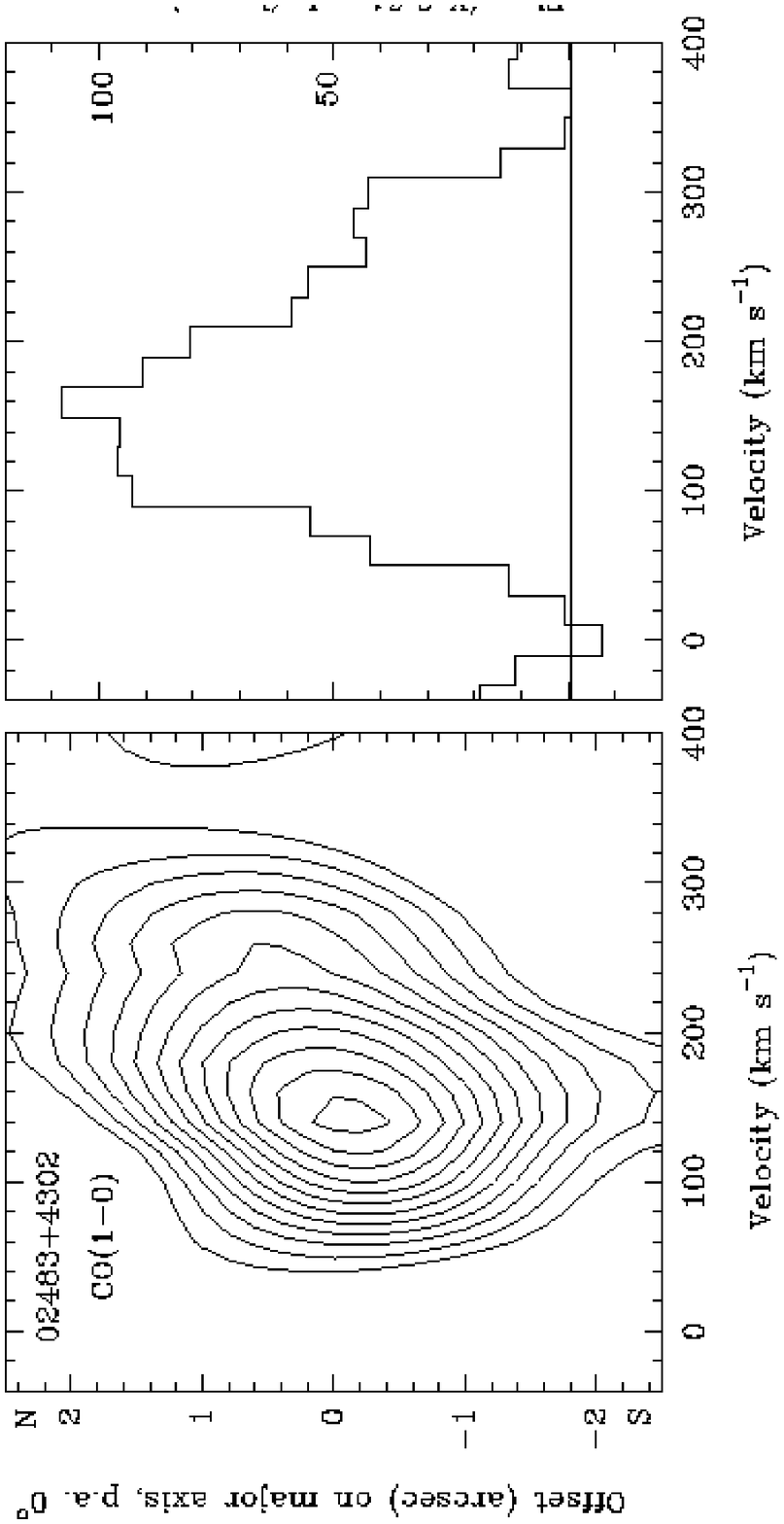}{\hsize}{-90}{65}{65}{-270}{860}
\end{figure}
\begin{figure}
\vspace{-6cm}
\caption[{\bf 02483$+$4302 :} CO(1--0) Ico, V, L-V, spectrum]
{ {\bf 02483$+$4302 :} 
{\it upper left:}
CO(1--0)  integrated over ($+40, +340$\,\kms ).
Beam $ = 2''.0\times 2''.0$.  Contour step =
% 1 to 9 by 1, in units of 1.22\,Jy beam$^{-1}$ \kms , with $T_b/S = 25$\,K/Jy.
% 1 to 18 by 1,
 0.61\,Jy beam$^{-1}$ \kms , with $T_b/S = 25$\,K/Jy.
{\it upper right:}
 CO isovelocity contours from $+$140 to $+$220\,\kms , in steps of 10\,\kms .  
Labels are in \kms .
{\it lower left:}
CO(1--0) position-velocity diagram at p.a.\
0$^\circ$. Contours: 3 to 14 by 1,  in units of 
3\,mJy beam$^{-1}$, with $T_b/S = 32.5$\,K/Jy. Beam $= 2''.5\times 1''.3$.    
{\it lower right:} CO spectrum at the source peak, in the
$2''.0\times 2''.0$ beam, with $T_b/S = 25$\,K/Jy.
Position offsets are relative to
02$^{\rm h}$51$^{\rm m}$36.$^{\rm s}$01,
$+43^\circ$15$'$10.$''8$ (J2000), and 
velocity scales are relative to  
109.700\,GHz ($cz_{\rm lsr} =$ 15225\,\kms ).
}\end{figure}

\clearpage
%FIGURE 27 ---------------------------------------------------------OLD F15
\begin{figure}
\plotfiddle{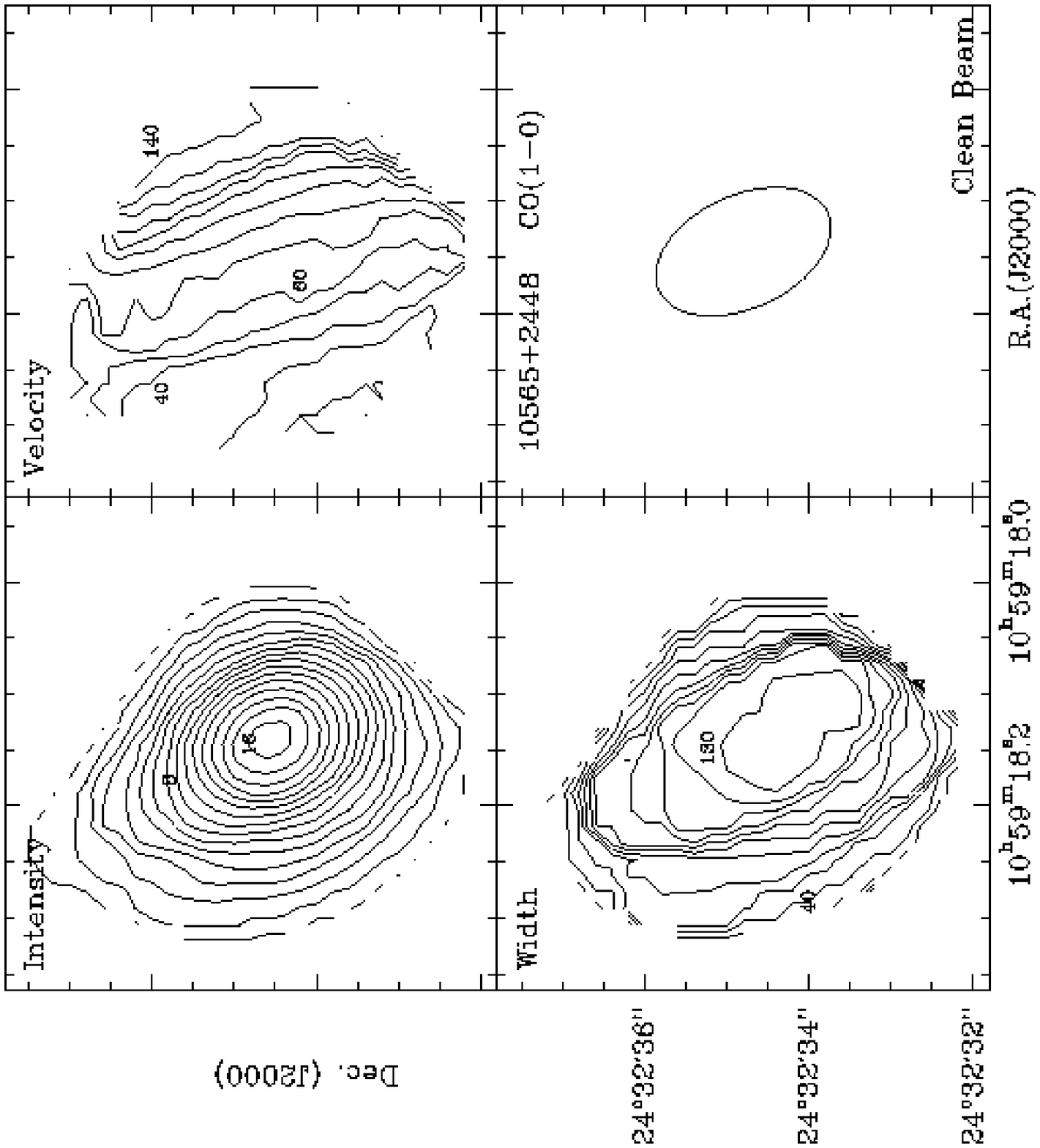}{\hsize}{-90}{65}{65}{-230}{520}
\end{figure}
\begin{figure}
\vspace{-8cm}
\caption[{\bf 10565$+$2448 :}  Moment maps.]
{{\bf 10565$+$2448 :} 
CO(1--0) integrated intensity, velocity, and linewidth (FWHM).  
Integration limits:  ($-100, +220$\,\kms ).
Beam $= 2''.3\times 1''.4$ (lower right). Contours: 

\noindent
{\it integrated CO:} 
% 1 to 16 by 1  
1 to 16 by 1, in units of 1.47\,Jy beam$^{-1}$ \kms , with 
$T_b/S = 32$\,K/Jy;

\noindent
{\it CO velocity:}
$+20$ to $+140$\,\kms , in steps of 10\,\kms\ relative to 110.535\,GHz
($cz_{\rm lsr} =$ 12846\,\kms ).  
Labels are in \kms ;

\noindent
{\it CO linewidth:} 30 to 130\,\kms\ in steps of 10\,\kms . 
Labels are in \kms .
}\end{figure}

\clearpage
%FIGURE 28 --------------------------------------------------------- 
\begin{figure}
\plotfiddle{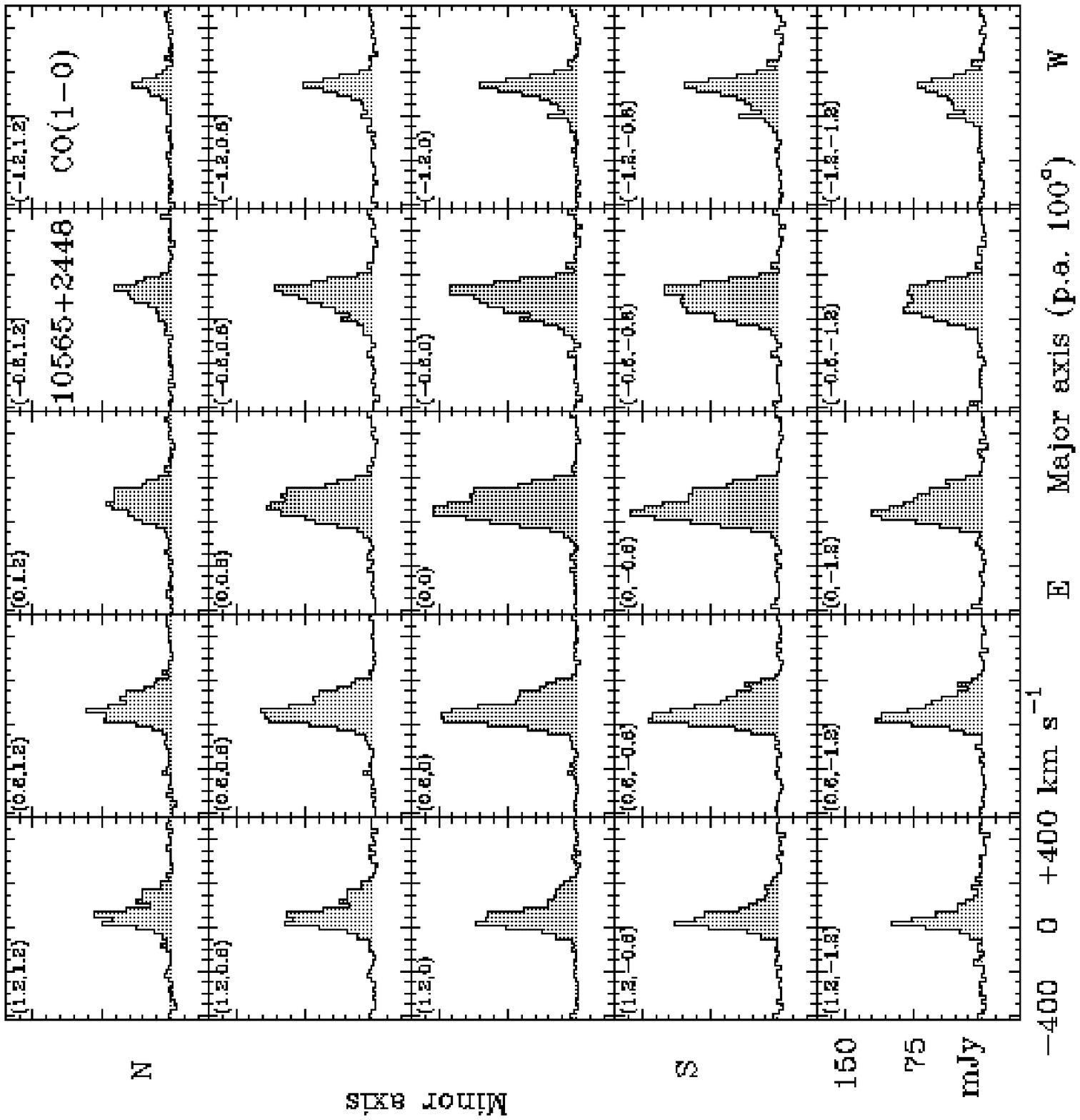}{\hsize}{-90}{65}{65}{-230}{520}
\end{figure}
\begin{figure}
\vspace{-8cm}
\caption[{\bf 10565$+$2448 :} Map of spectra]
{{\bf 10565$+$2448 :} CO(1--0) spectra.  
In each box vertical axis is CO intensity and  
% with large ticks at 0, 7.5, and 15\,mJy  beam$^{-1}$; 
horizontal axis is radial velocity relative to 110.535\,GHz
($cz_{\rm lsr} =$ 12846\,\kms ). 
% with large ticks at $-$400, $-$200, 0, $+200$, and $+$400\,\kms\ 
% (left to right).
In the upper left of each box are offsets (arcsec) 
on the kinematic major and minor axes, relative to
the CO centroid listed in Table~2.  
Beam $=  2''.3\times 1''.4$, with $T_b/S = 32$\,K/Jy.
}\end{figure}

\clearpage
%FIGURE 29 ---------------------------------------------------------
\begin{figure}
\plotfiddle{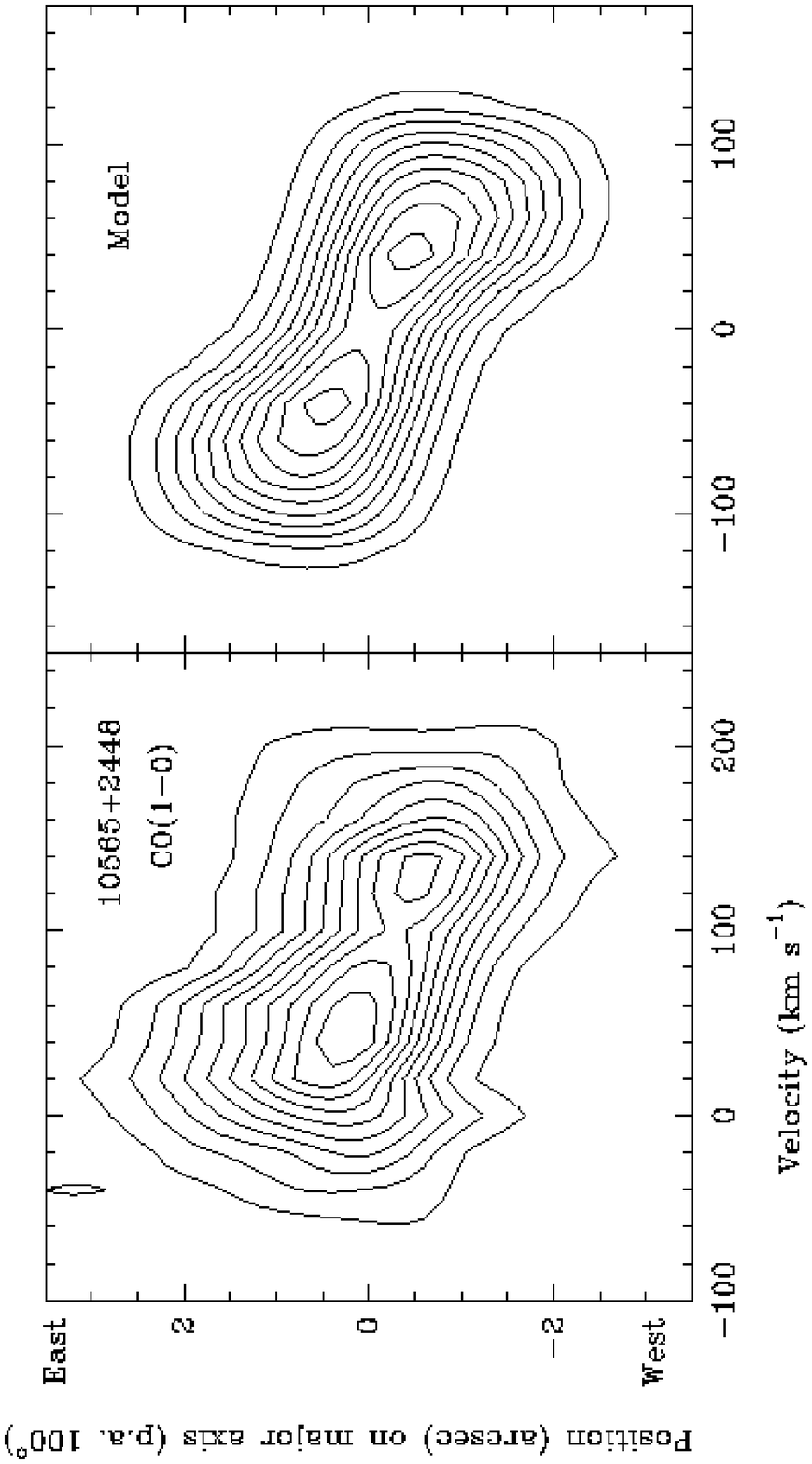}{\hsize}{-90}{65}{65}{-270}{580}
\end{figure}
\begin{figure}
\vspace{-8cm}
\caption[{\bf 10565$+$2448 :} Position-velocity diagram: data \& model]
{{\bf 10565$+$2448 :} 
CO(1--0) position-velocity diagrams 
along the line of nodes (p.a. 100$^\circ$).  
{\it left:} --- Observed data: Contour levels are 1 to 10, in units of 
15\,mJy beam$^{-1}$, with $T_b/S = 32$\,K/Jy.
Beam  $= 2''.3\times 1''.4$.  The (0,0) position is listed
in Table~2. Velocity is relative to 110.535\,GHz
($cz_{\rm lsr} =$ 12846\,\kms ).
{\it right:} --- Ring/disk model with the parameters listed in the Tables.
Contours and resolution are the same as for the observed data at the left.
}\end{figure}

\clearpage
%FIGURE 30 -------------------------------------------------- 
\begin{figure}
\plotfiddle{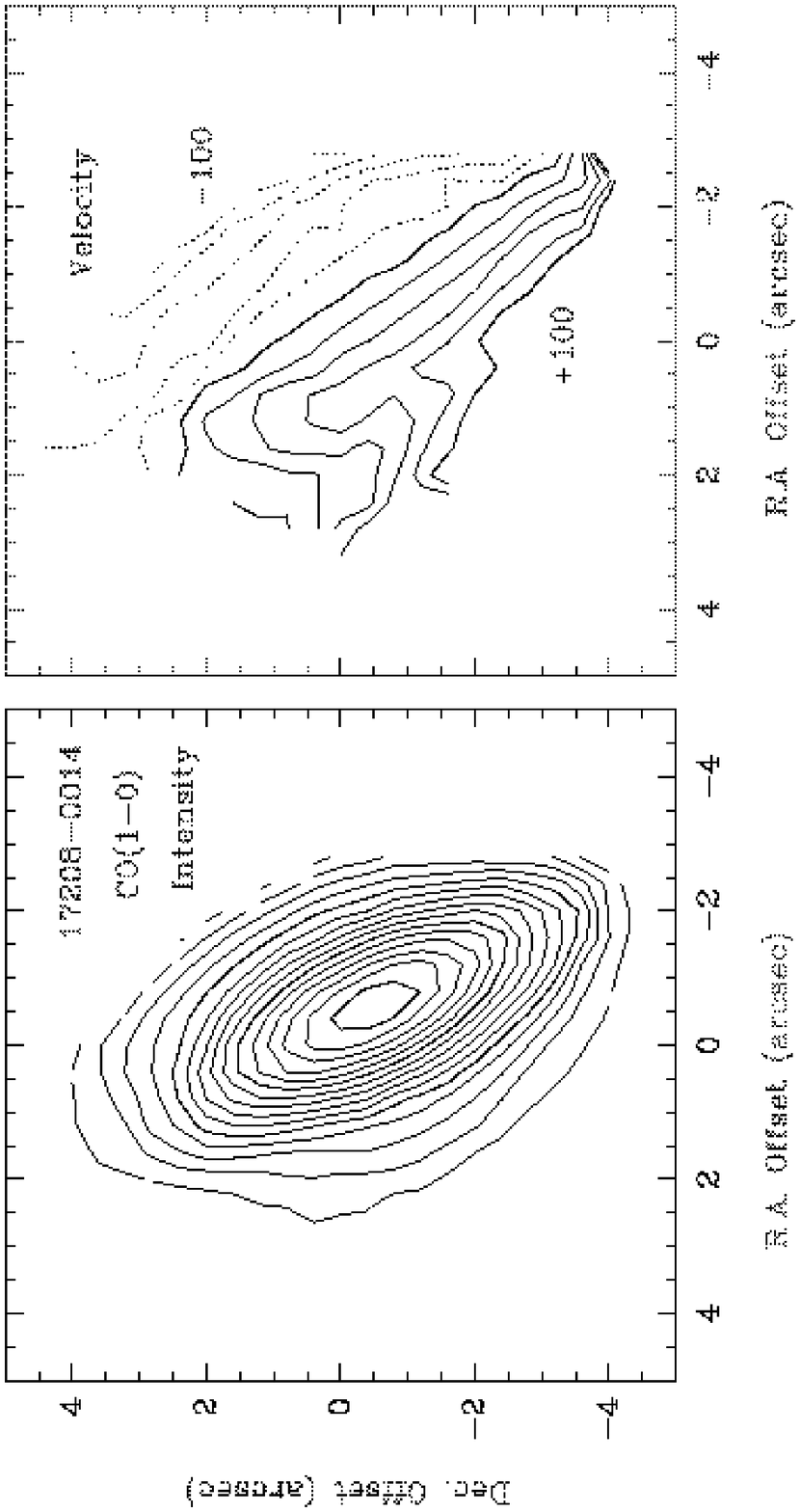}{\hsize}{-90}{65}{65}{-270}{620}
\plotfiddle{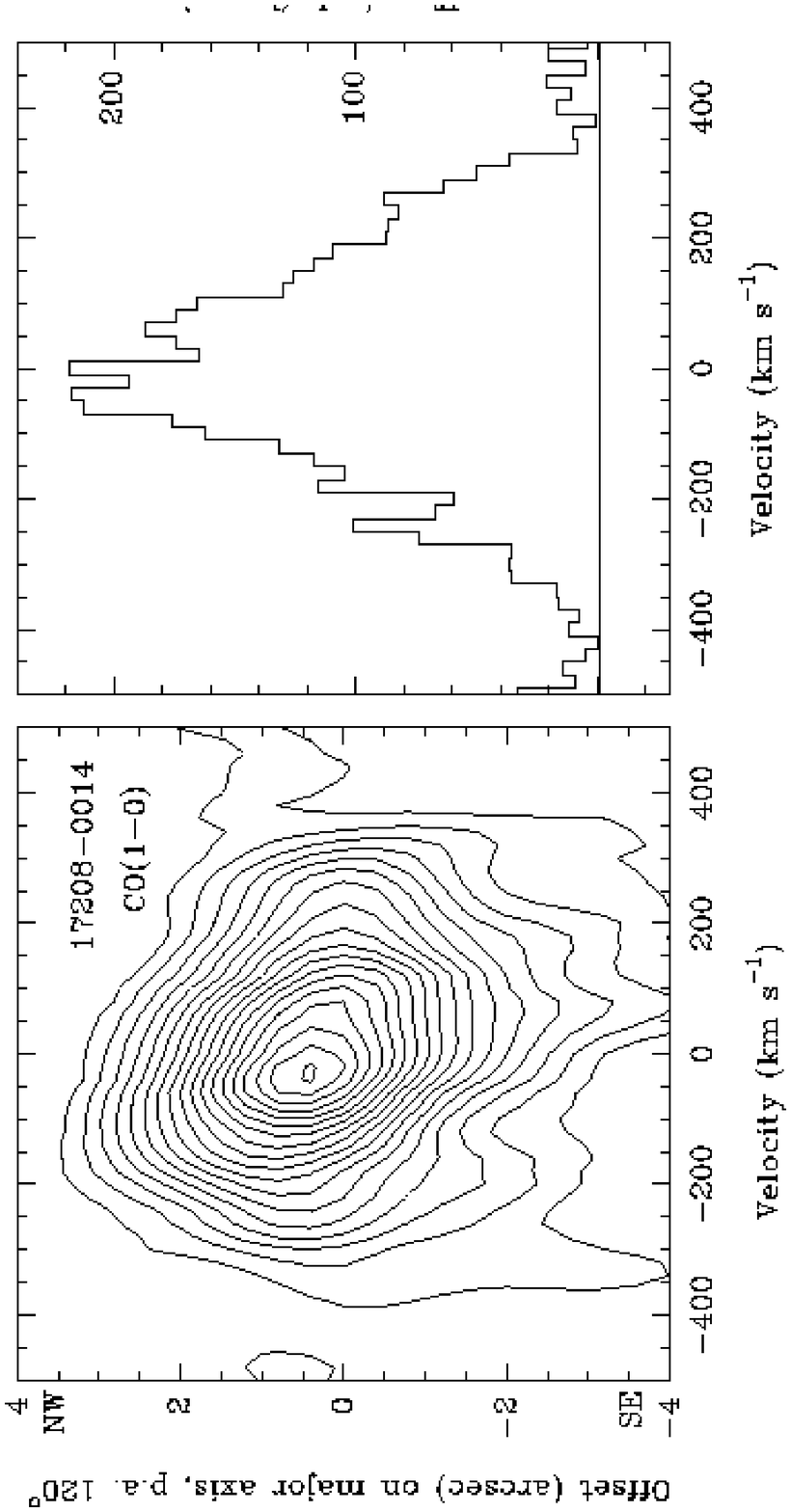}{\hsize}{-90}{65}{65}{-270}{860}
\end{figure}
\begin{figure}
\vspace{-5cm}
\caption[{\bf 17208$-$0014 :} CO(1--0) moments, posn-vel, spectrum]
{{\bf 17208$-$0014 :}  
{\it upper left:}
CO(1--0) integrated over ($-$360, $+$360\,\kms ).
Beam $ = 5''.1\times 1''.6$. 
Contour step = 5\,Jy beam$^{-1}$ \kms , 
with $T_b/S = 12$\,K/Jy.
{\it  upper right:} CO velocity contours in steps of 20\,\kms .
Labels are in \kms . 
{\it lower left:}
CO(1--0) position-velocity diagram at p.a.\
120$^\circ$. Contours: 10 to 180, in steps of 10\,mJy beam$^{-1}$,
with $T_b/S = 12$\,K/Jy.  
{\it lower right:} CO spectrum in the 
$5''.1\times 1''.6$ beam at the source peak (Table~2).  
Position offsets are relative to
17$^{\rm h}$23$^{\rm m}$21.$^{\rm s}$99,
$-00^\circ$17$'$00.$''2$ (J2000), and 
velocity scales are relative to 110.535\,GHz
($cz_{\rm lsr} =$ 12846\,\kms ). 
}\end{figure}

\clearpage
%FIGURE 31 ---------------------------------------------------------
\begin{figure}
\plotfiddle{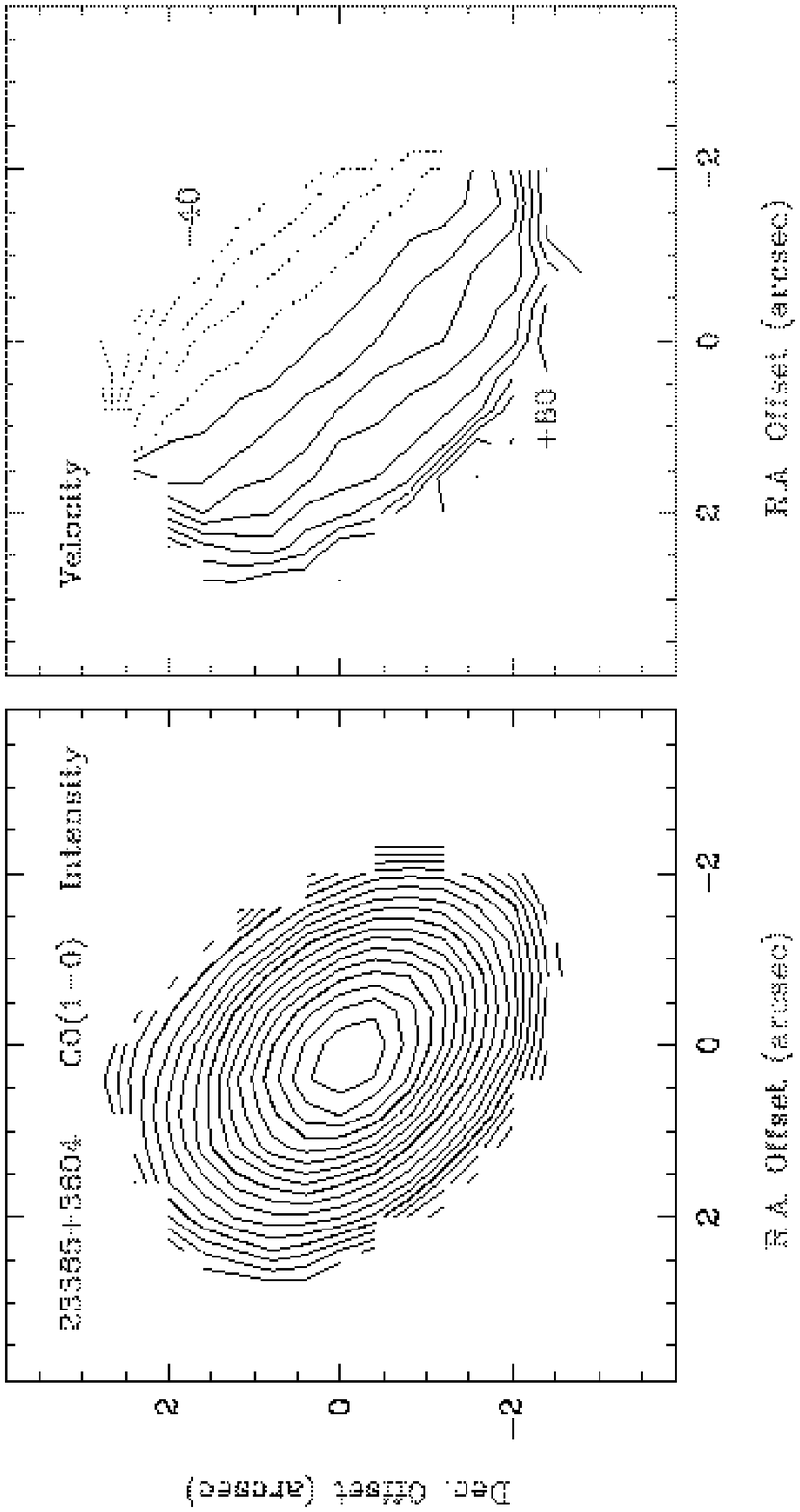}{\hsize}{-90}{65}{65}{-270}{625}
\plotfiddle{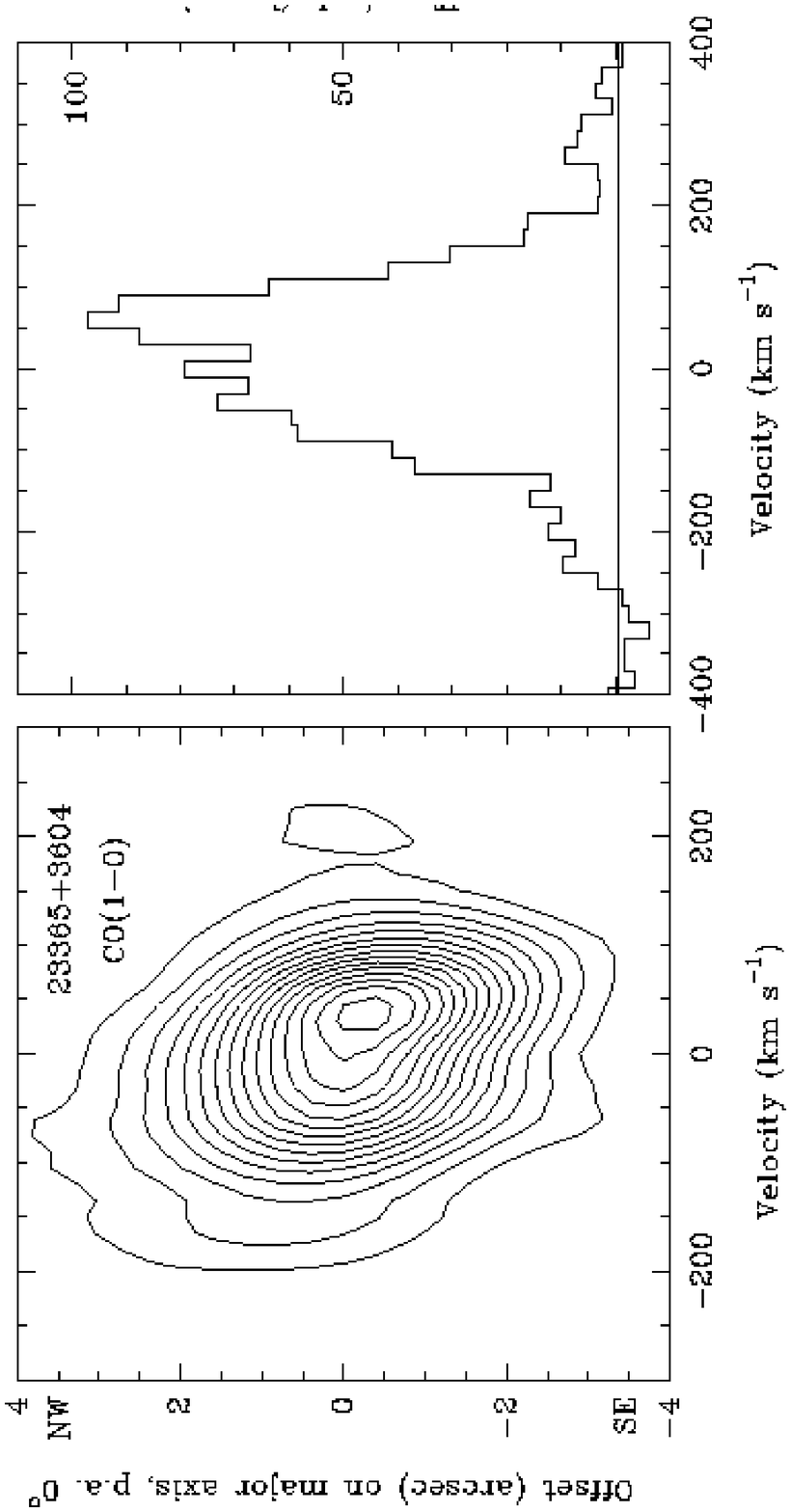}{\hsize}{-90}{65}{65}{-270}{865}
\end{figure}
\begin{figure}
\vspace{-5.25cm}
\caption[{\bf 23365$+$3604 :} CO(1--0) Ico, V, L-V, and profile]
{{\bf 23365$+$3604 :}  
{\it upper left:}
CO(1--0) integrated over ($-$240, $+$180\,\kms ).
Beam $ = 3''.9\times 2''.5$. 
Contour step - 2\,Jy beam$^{-1}$ \kms , 
with $T_b/S = 11$\,K/Jy.
{\it upper right:} CO velocity contours in steps of 10\,\kms . 
Labels are in \kms .
{\it lower left:}
CO(1--0) position-velocity diagram at p.a.\
135$^\circ$. Contour interval: 3\,mJy beam$^{-1}$.
{\it lower right:} CO spectrum at the source peak, in the 
$3''.9\times 2''.5$ beam.  
Position offsets are relative to
23$^{\rm h}$39$^{\rm m}$01.$^{\rm s}$25,
$36^\circ$21$'$08.$''4$ (J2000), and velocity scales are
relative to 108.292\,GHz ($cz_{\rm lsr} =$ 19321\,\kms ). 
}\end{figure}

\clearpage

%------------------------------------------------------
% TABLES: before each table we use \clearpage for page break.  Also use
% a \clearpage after the last table so that it gets its own page, too.

\clearpage
%TABLE-1-----------------------------------------------------------TABLE 1 ---
%\tighten
%\tt apjpt4
%\tt aaspp4
\begin{deluxetable}{llllccc}
\tablecaption{OBSERVING PARAMETERS}
\tablehead{
\colhead{Source}&\colhead{Date} &\colhead{Freq.}   	&\colhead{Phase}  
	&\colhead{Flux}	&\colhead{Clean beam}  &\colhead{Flux scale}	\nl
\colhead{name}	&\colhead{} 	&\colhead{(GHz)}	&\colhead{calibrator}	
	&\colhead{(Jy)}	&\colhead{(arcsec, p.a.)} &\colhead{(K Jy$^{-1}$)} }
\startdata
00057$+$4021 	&06/94--11/94 &110.360 
			&0003$+$380 &0.5 &$2.2\times 1.1$, \ 47$^\circ$ &40\nl
02483$+$4302 	&06/94--11/94 &109.700 
			&0234$+$285 &1.4 &$2.2\times 1.7$, \ \ 0$^\circ$ &27\nl
% VII\,Zw\,31  	&06/94--11/94 &109.380 
%			&0355$+$508 &1.9 &$2.2\times 1.8$, \ 40$^\circ$ &26\nl
VII\,Zw\,31	&03/96--05/96 &109.380 
			&0355$+$508 &2.6 &$1.4\times 0.8$, \ 58$^\circ$ &88			
\nl	     	&03/96--05/96 &218.650 
			&0355$+$508 &1.0 &$0.7\times 0.4$, \ 46$^\circ$ &78
%			&0355$+$508 &1.0 &$1.1\times 0.8$, \ 57$^\circ$ &31
\nl
%10565$+$2448 	&11/91--01/92 &110.508 
%			&0923$+$392  &3.2 &$2.1\times 1.4$, \ 36$^\circ$ &36
%\nl
10565$+$2448   	&06/94--11/94 &110.535 
			&0923$+$392 &5.5 &$2.3\times 1.4$, \ 25$^\circ$ &32
%\nl		&&	&1012$+$232 &0.5 
\nl	
%Mrk\,231     	&06/94--11/94 &110.535 
%			&1044$+$719 &0.7 &$2.8\times 1.1$, 104$^\circ$ &31\nl
Mrk\,231	&11/95--02/96 &110.602 
			&0923$+$392 &4.4 &$1.3\times 1.1$, \ \ 0$^\circ$  &70 
\nl          		&11/95--02/96 &221.204 
			&0923$+$392 &2.7 &$0.7\times 0.5$, \ 52$^\circ$ &75           
%		&	&	 &1308$+$326 &0.7
\nl
Arp\,193	&06/96--02/98 	&112.641
			&1308$+$326 &1.5 &$1.3\times 0.9$, \ \ 37$^\circ$ &83
\nl
		&05/97--02/98	&225.282  
			&1308$+$326 &0.8 &$0.6\times 0.4$, \ \ 33$^\circ$ &84
\nl
% Mrk\,273     	&04/95--06/95 	&111.076 
%			&1418$+$546 &0.5 &$2.9\times 2.1$, \ 57$^\circ$ &16\nl
Mrk\,273	&03/96--05/96 	&111.076 
			&1308$+$326 &1.4 &$1.4\times 1.3$,  116$^\circ$ &56	
\nl	     	&03/96--05/96 	&222.176 
			&1308$+$326 &0.7 &$0.6\times 0.6$, \ \ 0$^\circ$ &66	
\nl		
% Arp\,220     	&03/94--11/94 &113.228 
%			&3C345	&4.5      &$2.5\times 1.7$, \ 51$^\circ$ &22\nl
Arp\,220        &11/95--03/96 &113.228 
			&3C345 &4.8   &$1.6\times 1.1$, \ 25$^\circ$ &55
\nl 
             		&11/95--03/96 &226.422 
			&3C345 &2.4   &$0.7\times 0.5$, \ 28$^\circ$ &69
\nl		
17208$-$0014 	&06/94--10/94 &110.535 
			&1741$-$038 &2.6 &$5.1\times 1.6$, \ 47$^\circ$ &12
\nl		
23365$+$3604 	&05/96--08/96 &108.292
			&3C454.3 &5.0 &$3.9\times 2.5$, \ 45$^\circ$ &11
\nl
\tablenotetext{}{
Note: The frequency is the zero of our velocity scales, not 
necessarily the CO line center.
}
\enddata
\end{deluxetable}
\clearpage

%TABLE-2-----------------------------------------------------------TABLE 2 ---
\tighten
\begin{deluxetable}{llllccc}
\tablecaption{CO POSITIONS AND FLUXES}
\tablehead{
\colhead{Source}&\colhead{CO} 
		&\colhead{R.A.} &\colhead{Dec.}  
		&\colhead{CO flux}	&\colhead{CO flux}  
		&\colhead{Continuum}
\nl
\colhead{name}	&\colhead{line}
		&\colhead{J2000}	&\colhead{J2000} 
		&\colhead{interferom.} &\colhead{30\,m tel.} 
		&\colhead{freq.:  \ flux}  
\nl
\colhead{}	&\colhead{}
		&\colhead{(\ \,h\ \ m\ \ s)}
		&\colhead{(\ \ \ $^\circ$\ \ \ $'$\ \ \ $''$)}
		&\colhead{(Jy\,km\ts s$^{-1}$)}
	        &\colhead{(Jy\,km\ts s$^{-1}$)} 
		&\colhead{(GHz): \ (mJy)}
}
%\tableline
\startdata
00057$+$4021 &(1--0) &00 08 20.58 &$+$40 37 55.5 &47  &45  &110: \ $<10$ \nl
02483$+$4302 &(1--0) &02 51 36.01 &$+$43 15 10.8 &28  &25  &109: \ $<10$ \nl 
VII\,Zw\,31 &(1--0) &05 16 46.49 &$+$79 40 12.6	 &87  &92  &109: \ $<\ 2$\nl	
	    &(2--1) &05 16 46.53 &$+$79 40 12.4	 &135 &280 &219: \ $<10$ \nl
10565$+$2448&(1--0) &10 59 18.15 &$+$24 32 34.4	 &68  &71  &111: \ $< 2$ \nl		
Mrk\,231    &(1--0) &12 56 14.21 &$+$56 52 25.1	 &68  &97  &111: \ 63    \nl	
	    &(2--1) &12 56 14.22 &$+$56 52 25.1	 &276 &280 &221: \ 36    \nl 		
Arp\,193    &(1--0) &13 20 35.32 &$+$34 08 22.2  &161 &162 &110: \ $< 5$  \nl
\phantom{Arp} disk	    
	    &(2--1) &13 20 35.32 &$+$34 08 22.2  &360 &--- &228: \ \ 7	 \nl 		
\phantom{A} SE core
	    &(2--1) &13 20 35.35 &$+$34 08 21.7  &90  &--- &228: \ \ 3	 \nl
\phantom{Arp} sum	    
	    &(2--1) &--- 	 &---		&450  &415 &228: \ 10   \nl
Mrk\,273    &(1--0) &13 44 42.12 &$+$55 53 13.5	 &78  &86  &111: \ 11   \nl
\phantom{Mrk}disk
	&(2--1)     &13 44 42.10 &$+$55 53 13.4	&197 &--- &222: \ $<$10 \nl
\phantom{Mrk}core
	&(2--1)     &13 44 42.12 &$+$55 53 13.4	&34 &--- &222: \ \ 8 \nl
\phantom{Mrk}sum
	&(2--1)     &--- 	 &---		&231  &--- &222: \ \ 8 
\nl		
Arp\,220 
	&(1--0) &15 34 57.24	&+23 30 11.2 &410 &490 &113: \ 41 \nl
\phantom{Arp} west 
	&(2--1) &15 34 57.221	&+23 30 11.5 &130 &--- &229: \ 90\nl
% \phantom{Arp} east-I 
%	&(2--1) &15 34 57.279	&+23 30 11.4 &70 &--- &229: \ 30\nl
\phantom{Arp} east
	&(2--1)	&15 34 57.29    &+23 30 11.3 &220  &--- &229: \ 30\nl
\phantom{Arp} disk
 	 &(2--1) &15 34 57.24	&+23 30 11.2 &750 &--- &229: \ 55\nl
\phantom{Arp} sum
	&(2--1) &---		&--- 	     &1100 &1100 &229: 175\nl
17208$-$0014 &(1--0) &17 23 21.95 &$-$00 17 00.7 &132 &162 &111: \ $<5$ \nl		
23365$+$3604 &(1--0) &23 39 01.25 &$+$36 21 08.4 & 30 & 47 &108: \ $<5$ \nl
\tablenotetext{}{
Errors:  Positions at CO(1--0): $\pm 0''.2$, at CO(2--1): $\pm 0''.1$ ;
 Fluxes:  $\pm 20\%$;
}  
\enddata
\end{deluxetable}
\clearpage

%TABLE-3-------------------------------------------------------TABLE 3 ---
\tighten
\begin{deluxetable}{lccc cc cc}
\tablecaption{MEASURED DIAMETERS OF THE {\it INTEGRATED} CO EMISSION}
\tablehead{
\colhead{Source} &\colhead{Angular}  	&\colhead{P.A.} 
	    	&\colhead{Ang.size} 	&\colhead{Redshift}
		&\colhead{Linear} &\colhead{Semi-}  &\colhead{Semi-}
\nl
\colhead{name}	&\colhead{diameter}	  		&\colhead{E of N}
                &\colhead{distance} 
		&\colhead{$cz_{\rm lsr}$}	
		&\colhead{scale} &\colhead{maj.axis}  &\colhead{min.axis}
\nl
\colhead{}	&\colhead{(arcsec)}  &\colhead{(deg.)}  &\colhead{(Mpc)}
&\colhead{(\kms )}  &\colhead{( pc/$''$ ) }&\colhead{(pc)}  &\colhead{(pc)} }
\startdata
00057$+$4021	&$1.1\times <0.6$ &$-22$  &165  &13390  &802  &440  &$<$240\nl
02483$+$4302	&$1.8\times 1.7$  &$-4$	  &188  &15419  &913  &820  &780   \nl 
VII\,Zw\,31	&$2.4\times 2.1$  &$-$5	  &196  &16260  &949  &1140 &1000  \nl		
10565$+$2448	&$1.6\times 1.4$  &$-26$  &160  &12923  &776  &620  &540   \nl		
Mrk\,231    	&$0.9\times 0.8$  &77     &157  &12650  &761  &340  &300   \nl
Arp\,193--disk   &$2.8\times 0.8$  &$-40$  &\phantom{1}90  &\phantom{1}7000  
		&436  &610  &170   \nl		
Arp\,193-SE core  &$1.1\times 0.5$  &$-40$  &\phantom{1}90  &\phantom{1}7000  
		&436  &240  &100   \nl		
Mrk\,273--tail	&$3.1\times 2.8$  &5      &142  &11324  &686  &1060 &960   \nl
Mrk\,273--disk	&$0.9\times 0.6$  &18     &142	&11324	&686  &310  &210   \nl
Mrk\,273--core	&$0.35\times <$0.2  &76    &142	&11324	&686  &120  &$<$69 \nl
Arp\,220-disk	&$2.0\times 1.6$  &50  &\phantom{1}70  &\phantom{1}5450
			&341 &340 &270  \nl
Arp\,220-west	&$0.31\times 0.28$  &$-14$  &\phantom{1}70  &\phantom{1}5340  
			&341  &\phantom{1}53 &\phantom{1}48  \nl
% Arp\,220-east-I	&$0.59\times 0.50$  &$-36$  &\phantom{1}70  &\phantom{1}5330 
%			&341	&101 &\phantom{1}85 	    \nl 
Arp\,220-east &$0.9\times 0.9$   &---    &\phantom{1}70  &\phantom{1}5650
			&341	&150    &150
\nl
17208$-$0014	&$1.8\times 1.6$  &7     &159  &12837  &771   &690  &620  \nl		
23365$+$3604   	&$1.0\times 0.9$ &$-15$  &231  &19330  &1120  &560  &500  \nl
\tablenotetext{}{Diameters are FWHM from elliptical Gaussian fits to $u,v$ data.}
\tablenotetext{}{Diameters: $\pm 0''.2$ ; position angles:  $\pm 10^\circ$. }  
\enddata\end{deluxetable}\clearpage

% TABLE-4-------------------------------------------------TABLE 4 ---
\tighten
\begin{deluxetable}{lccc cc}
\tablecaption{DISK SIZES, FROM POSITION--VELOCITY DATA}
\tablehead{
\colhead{Source}&\colhead{Linear}  	&\colhead{Rotation curve} 
	&\colhead{half-power} 		&\colhead{Outer disk}	&\colhead{Disk} 
\nl
\colhead{name}	&\colhead{scale}	&\colhead{turnover radius}
   	&\colhead{radius} 		&\colhead{boundary}	
					&\colhead{thickness} 
\nl
\colhead{}	&\colhead{(pc/$''$)}	&\colhead{$R_0$ (pc)}
	&\colhead{$R_1$ (pc)}	&\colhead{$R_{\rm max}$ (pc)}
	&\colhead{$H$(pc)}
}
\startdata
00057$+$4021 &\phantom{1}802  &240  		&\phantom{1}480  &1000  &50 \nl
02483$+$4302 &\phantom{1}913  &270  		&\phantom{1}730  &1400  &54 \nl
VII\,Zw\,31  &\phantom{1}949  &290  		&1100 		 &3300  &58 \nl		
10565$+$2448 &\phantom{1}776  &230  		&\phantom{1}700  &1600  &47 \nl		
Mrk\,231     &\phantom{1}761  &\phantom{1}75    &\phantom{1}460  &1100  &23 \nl
Arp\,193-disk&\phantom{1}436  &220              &\phantom{1}740  &1300  &65 \nl 		
Mrk\,273-disk&\phantom{1}686  &\phantom{1}70   	&\phantom{1}400  &1900  &42 \nl
Arp\,220-disk&\phantom{1}341  &200  		&\phantom{1}480  &1400  &80 \nl		
Arp\,220-west&\phantom{1}341  &\phantom{1}35   	&\phantom{12}68  &---   &68 \nl
Arp\,220-east&\phantom{1}341  &\phantom{1}50   	&\phantom{1}110  &--- 	&80\nl
17208$-$0014 &\phantom{1}771  &310  		&\phantom{1}540  &1500 &62  \nl		
23365$+$3604 &1120  	      &340  		&\phantom{1}670  &2200 &67  \nl
\tablenotetext{}{Disk thickness $H$ is FWHM perpendicular to the equator.}  
\enddata\end{deluxetable}\clearpage

%TABLE-5-------------------------------------------------------TABLE 5 ---
\tighten
\begin{deluxetable}{lccccc}
\tablewidth{0pc}
\tablecaption{ROTATION CURVES, FROM POSITION--VELOCITY DATA}
\tablehead{
\colhead{}	&\colhead{Data:} 	
		&\colhead{Data:} 	
		&\colhead{Model:}  
		&\colhead{Model:}  	
		&\colhead{Model:}
\nl
\colhead{}	&\colhead{line of nodes} 	
		&\colhead{apparent} 	
		&\colhead{inclination}  
		&\colhead{rotation}  	
		&\colhead{turbulent}
\nl
\colhead{Source}&\colhead{E of N}
		&\colhead{rotation}   
		&\colhead{0$^\circ$=face-on}  
		&\colhead{velocity}   	  
		&\colhead{velocity}
\nl
\colhead{name}	&\colhead{p.a.}
		&\colhead{$V_{\rm rot}\sin i$}				
		&\colhead{$i$}
                &\colhead{$V_{\rm rot}$}
		&\colhead{$\Delta V$}	
\nl
\colhead{}	&\colhead{(deg)}
		&\colhead{(\kms )}	
		&\colhead{(deg)}
		&\colhead{(\kms )}
		&\colhead{(\kms )}	
}
\startdata
00057$+$4021 	&135 
		&\phantom{1}75
		&20
		&$250\,r^\beta$ 
 		&$100\,r^{-0.5}$ 
\nl
02483$+$4302 	&\phantom{9}0
		&\phantom{1}70
		&15  
		&$270\,r^\beta$ 	
		&$40\,r^{0.0}$ 
\nl
VII\,Zw\,31  	&\phantom{9}0
		&100
		&20 
		&$290\,r^\beta$
 		&$30\,r^{0.0}$

\nl		
10565$+$2448 	&100
		&\phantom{1}75
		&20 
		&$220\,r^\beta$ 
		&$40\,r^{0.0}$
\nl		
Mrk\,231     	&90
		&\phantom{1}60
		&10
		&$345\,r^\beta$ 	 			
		&$60\,r^{-0.3}$
\nl
Arp\,193-disk	&140
		&175
		&50
		&$230\,r^\beta$
		&$40\,r^{0.0}$
\nl 		
Mrk\,273--disk  &90
		&200
		&45 
		&$280\,r^\beta$ 	  		
		&$140\,r^{0.0}$ 
\nl
Arp\,220--disk  &40
		&212
		&40 
		&$330\,r^\beta$		
		&$140\,r^{-0.2}$
\nl
Arp\,220--west  &300
		&103
		&20		
		&$300\,r^\beta$	 
		&$100\,r^{-0.3}$
\nl		
Arp\,220--east  &50
		&300
		&60 
		&$350\,r^\beta$				
		&$100\,r^{-0.3}$
\nl		
17208$-$0014 	&120 
		&130
		&30
		&$260\,r^\beta$  	     		
		&$150\,r^{-0.2}$ 
\nl		
23365$+$3604	&135
		&130
		&30 
		&$260\,r^\beta$  	 	 		
		&$100\,r^{-0.2}$
\nl 
\tablenotetext{}{Model rotation curve: $\beta = 1$ for $r<1$ and
$\beta = 0$ for $r\geq 1$, and $r=R/R_0$, where $R_0$ is
the rotation curve turnover radius (Table~4).} 
\tablenotetext{}{Turbulent velocity $\Delta V$  = halfwidth to $(1/e)$ level. 
$\Delta V$ = 0.6 FWHM = 1.4 $\sigma$, where $\sigma$ = 1-D 
velocity dispersion.}
\tablenotetext{}{Turbulent velocities  
are valid for $R_0 \leq R \leq R_{\rm max}$ (Table~4).}
\enddata
\end{deluxetable}
\clearpage

%TABLE 6------------------------------------------------------TABLE 6 ---
\begin{deluxetable}{lccc}
\tablewidth{0pc}
\tablecaption{MODEL TEMPERATURES AND DENSITIES}
\tablehead{
\colhead{Source}&\colhead{Gas kinetic}  	&\colhead{$<n$(H$_2)>$} 
		&\colhead{$<n$(H$_2)>$}		\nl
\colhead{name}	&\colhead{temperature}		&\colhead{at $R_0$}
                &\colhead{outer disk} 		\nl
\colhead{}	&\colhead{(K)}			&\colhead{(cm$^{-3}$)}
		&\colhead{(cm$^{-3}$)}		}
\startdata
00057$+$4021	&$72\ r^{-0.5}$ 		&400 	&40 	\nl
02483$+$4302	&$50\ r^{-0.5}$		&250 	&20 	\nl
VII\,Zw\,31	&$50\ r^{-0.5}$		&450 	&10	\nl		
10565$+$2448	&$60\ r^{-0.5}$		&1200 	&30	\nl		
Mrk\,231	&$100\ r^{-0.5}$	&3600 	&60	\nl
Arp\,193--disk	&$40\ r^{-0.5}$		&550	&10	\nl 		
Mrk\,273--disk	&$70\ r^{-0.5}$ 	&1800   &60     \nl
Arp\,220--disk	&$64\ r^{-0.5}$		&900 	&30	\nl
Arp\,220--west	&$150\ r^{-0.5}$	&22000 	&900	\nl
Arp\,220--east	&$100\ r^{-0.5}$	&20000 	&900	\nl
17208$-$0014	&$70\ r^{-0.5}$ 		&600 	&60 	\nl
23365$+$3604   	&$67\ r^{-0.5}$ 		&200 	&20     \nl
\tablenotetext{}{In the scaling laws, $r=R/R_0$ where $R_0$ is
the turnover radius of the rotation curve (Table~4).  At large radii, 
we adopted minimum gas kinetic temperatures of 10\,K.}
\tablenotetext{}{Gas densities $<n({\rm H}_2)>$ are averaged over 
the inner and outer disk volumes.}
\enddata
\end{deluxetable}
\clearpage

%TABLE 7-------------------------------------------------------TABLE 7 ---
\begin{deluxetable}{l ccccc}
\tablewidth{0pc}
\tablecaption{CO(1--0) LINE PARAMETERS}
\tablehead{
\colhead{}	&\colhead{Model:} 	&\colhead{Model:} 	& 
		&\colhead{Model:}  	&\colhead{Data:}
\nl
\colhead{Source}&\colhead{CO(1--0)}	&\colhead{True $T_b$}	&\colhead{Beam} 
		&\colhead{Predicted}  	&\colhead{Observed}
\nl
\colhead{name}	&\colhead{opacity}	&\colhead{at $R_0$}	&\colhead{FWHM}
		&\colhead{$T_b$ in beam}&\colhead{$T_b$ in beam}
\nl
\colhead{}	&\colhead{at $R_0$} 	&\colhead{(K)} 	&\colhead{(arcsec)}
		&\colhead{(K)}  	&\colhead{(K)}	}
\startdata
00057$+$4021	&4.8    &16 	&$2.2\times 1.1$	&2.7 	&2.4	\nl
02483$+$4302	&9.7   	&11 	&$2.2\times 1.7$	&1.9 	&1.4   	\nl
VII\,Zw\,31	&7.8	&18 	&$1.4\times 0.8$	&7.2	&7.4  	\nl		
10565$+$2448	&9.8 	&36 	&$2.3\times 1.4$	&4.6	&4.5  	\nl		
Mrk\,231	&3.4 	&56 	&$1.3\times 1.1$	&13.5	&14.0  	\nl
Arp\,193--disk	&9.9	&25	&$1.7\times 1.6$	&5.0    &6.9	\nl 		
Mrk\,273--disk	&3.7    &36     &$1.4\times 1.3$	&4.8    &4.5    \nl
Arp\,220--disk	&5.9 	&29 	&$1.6\times 1.1$	&10.2	&10.2  	\nl
Arp\,220--west	&--- 	&50 	&$1.6\times 1.1$	&10	&---  	\nl
Arp\,220--east	&--- 	&50 	&$1.6\times 1.1$	&10	&---  	\nl
17208$-$0014	&3.9   	&21 	&$5.1\times 1.6$	&2.5 	&2.2    \nl
23365$+$3604   	&6.0   	&10 	&$3.9\times 2.5$	&1.0 	&0.8    \nl

\tablenotetext{}{Model CO abundance: $X_{\rm CO}/(dv/dr) = 8\times 10^{-5}$.}  
\tablenotetext{}{All brightness temperatures are Rayleigh-Jeans.}
\enddata
\end{deluxetable}
\clearpage

%TABLE 8-------------------------------------------------------TABLE 8 ---
\begin{deluxetable}{l cccccc }
\tablewidth{0pc}
\tablecaption{CO(2--1) LINE PARAMETERS}
\tablehead{
\colhead{}	&\colhead{Model:} 	&\colhead{Model:} 	&
		&\colhead{Model:}  	&\colhead{Data:}
		&\colhead{Data:}
\nl
\colhead{Source}&\colhead{CO(2--1)}	&\colhead{True $T_b$} 	&\colhead{Beam}
		&\colhead{Predicted}  	&\colhead{Observed}
		&\colhead{CO(2--1)}
\nl
\colhead{name}	&\colhead{opacity}	&\colhead{at $R_0$}	&\colhead{FWHM}
		&\colhead{$T_b$ in beam}&\colhead{$T_b$ in beam}
		&\colhead{to (1--0)}
\nl
\colhead{}	&\colhead{at $R_0$} 	&\colhead{(K) } &\colhead{(arcsec)}
		&\colhead{(K)}  	&\colhead{(K)}
		&\colhead{ratio in 3\,mm beam}		}
\startdata
VII\,Zw\,31	&25	&12 	&$1.1\times 0.8$
			&\phantom{3}5.4 &\phantom{2}5.4  &$0.7\pm 0.2$  \nl		
Mrk\,231	&12 	&51 		&$0.7\times 0.5$	
			&30.3	&28.5 	&$1.0\pm 0.2$ \nl 
Arp\,193--disk	&54	&13		&$0.7\times 0.6$
			&8.2	&8.1    &$0.5\pm 0.2$
\nl
Mrk\,273--disk	&11     &34  		&$0.6\times 0.6$   
			&\phantom{3}14    &\phantom{3}6.6  &--- 
\nl
Arp\,220--disk	&15 	&23		&$0.7\times 0.5$ 	
			&16.6	&10.0 & $0.6\pm 0.2$ \nl
Arp\,220--west	&17 	&53		&$0.7\times 0.5$ 	
			&18.8	&17.0 &--- \nl
Arp\,220--east	&\phantom{1}9 	&70		&$0.7\times 0.5$ 	
			&9.5	&9.1 &--- \nl
\tablenotetext{}{Model CO abundance, $X_{\rm CO}/(dv/dr) = 8\times 10^{-5}$.}  
\tablenotetext{}{All brightness temperatures are Rayleigh-Jeans.}
\enddata
\end{deluxetable}
\clearpage

%TABLE 9------------------------------------------------------TABLE 9 ---
\begin{deluxetable}{lccccccc}
\tablewidth{0pc}
\tablecaption{CO(1--0) LUMINOSITY, GAS MASS, AND DYNAMICAL MASS}
\tablehead{
\colhead{Source} 
	&\colhead{radius}
	&\colhead{$L^\prime_{\rm CO}$}		
	&\colhead{$M_{\rm gas}$}     
	&\colhead{$M_{\rm gas}/L^\prime_{\rm CO}$}
	&\colhead{$M_{\rm dyn}$} 	
	&\colhead{Mass ratio}
	&\colhead{$\mu/\mu_{\rm tot}$} 
\nl
\colhead{name}
	&\colhead{pc}
  	&\colhead{$10^9\,L_l$}
	&\colhead{$10^9$\,\Msun }
	&\colhead{\Msun /L$_l$}  
	&\colhead{$10^9$\,\Msun }  
  	&\colhead{$ M_{\rm gas}/M_{\rm dyn} $}
	&\colhead{max.}
}
\startdata
00057$+$40 to $R_1$ 	&480	&2.3	&0.6 	&0.3	&7.0 	&0.09  	&0.31
\nl 
\phantom{00057$+$40} to $R_{\rm max}$ 
			&1000  &4.0	&1.4	&0.4	&14.5    &0.10 	&--- 
\nl
\nl
02483$+$43 to $R_1$   	&730	&2.0	&1.2 	&0.6	&12.4 	&0.10 	&---	
\nl
\phantom{02483$+$43} to $R_{\rm max}$ 	
			&1400	&3.2	&1.9  	&0.6	&23.6   &0.08 	&---	
\nl
\nl
VII\,Zw\,31 to $R_1$  	&1100	&3.2	&3.0  	&0.9	&21.5 	&0.14 	&0.22	
\nl
\phantom{VII\,Zw\,31} to $R_2$  		
			&1900   &5.5	&5.2    &1.0	&37.3  	&0.14 	&---	
\nl
\phantom{VII\,Zw\,31} to $R_{\rm max}$	&3670   &8.5	&8.3    &1.0	&71.5  	&0.12 	&---	
\nl
\nl		
10565$+$24 to $R_1$	&700	 &2.3	&1.8  	&0.8	&7.8  	&0.23 	&0.28	
\nl 
\phantom{10565$+$24} to $R_2$  		
			&1150  &3.4	&3.0  	&0.9	&13.1  	&0.23	&---
\nl		
\phantom{10565$+$24} to $R_{\rm max}$	
			&1600  &5.5	&4.0  	&0.7	&17.4  	&0.23	&---
\nl
\nl		
Mrk\,231 to $R_1$ 	&460	&2.8	&1.8 	&0.7	&12.7  	&0.14  	&0.20
\nl
\phantom{Mrk\,231} to $R_2$  
			&850	&4.2	&3.1  	&0.7	&23.1  	&0.13	&---
\nl
\phantom{Mrk\,231} to $R_{\rm max}$ 	
			&1700	&5.1	&4.0  	&0.8	&31.5  	&0.13	&---
\nl
\nl 		
Arp\,193  to $R_1$ 	&740	&1.4	&1.7	&0.8	&9.1	&0.19	&---
\nl
\phantom{Arp\,193} to $R_2$	
			&1100	&2.1	&2.6	&0.8	&15.0	&0.16   &---
\nl
\phantom{Arp\,193} to $R_{\rm max}$
			&1300	&2.6	&3.4	&0.9	&16.1	&0.21	&---
\nl\nl
Mrk\,273  to $R_1$	&400	&4.1	&2.1 	&0.5	&8.7  	&0.24	&0.28
\nl
\phantom{Mrk\,273}  to $R_2$  
			&900	&4.6	&2.7  	&0.6	&16.2  	&0.17	&---
\nl
\phantom{Mrk\,273}  to $R_{\rm max}$	
			&1900	&5.6	&5.6  	&1.0	&34.6  	&0.16 	&---
\nl
\nl
Arp\,220 west 	  	&68	&0.6	&0.6 	&1.0 	&1.4  	&0.43  	&---
\nl
Arp\,220 east &150	&0.8	&1.1 	&1.3	&3.2 	&0.35    &---
\nl
Arp\,220 disk to $R_1$	&480	&2.7	&2.0  	&0.8	&12.1 	&0.17  	&0.30
\nl
Arp\,220 disk to $R_2$   		
			&760	&3.7	&3.0  	&0.8	&19.2  	&0.16	&---
\nl
%--- to $R_{\rm max}$   &1200	&2.8	&3.5  	&1.1	&---  	&---	&---
Arp\,220 total	  	&1360	&5.9	&5.2 	&0.9	&34.5  	&0.15	&---
\nl
\nl
\nl
17208$-$00 to $R_1$ &540	&3.3	&2.4 	&0.7	&8.5  	&0.28	&---\nl 
\phantom{17208$-$00}  to $R_{\rm max}$   
			&1500	&5.8	&6.1  	&1.0	&23.5   &0.26	&---\nl
\nl
23365$+$36 to $R_1$ &670  	&3.0	&1.5 	&0.5	&10.5  	&0.14    &---	\nl
\phantom{23365$+$36} to $R_{\rm max}$    
			&2200	&5.4	&3.8  	&0.7	&34.5 	&0.11    &\ \ ---
\tablenotetext{}{L$_l \equiv$ K \kms\,pc$^2$; \ \ 
$\mu =$gas (H$_2 +$ He) surface density in \Msun \,pc$^{-2}$;
$\mu_{\rm tot} =$ gas + stars; $\mu $/$\mu_{\rm tot}$ = peak ratio in the disk.}
\tablenotetext{}{$R_1=R_{\rm min}+ \Delta R$ is the inner disk's half-power
radius;  $R_2=R_{\rm min}+2\Delta R$ is its zero-power radius; 
$R_{\rm max}$ is the boundary of the outer disk (eq.~1; Table~4).} 
\enddata
\end{deluxetable}
\clearpage

%TABLE 10-------------------------------------------------------TABLE 10 ---
\tighten
\begin{deluxetable}{lccccc}
\tablecaption{ CONTINUUM EMISSION }
\tablehead{
\colhead{}	&\colhead{2.6\,mm }	&\colhead{2.6\,mm }
		&\colhead{1.3\,mm}	&\colhead{1.3\,mm} 
		&\colhead{Interpretation}	
\nl
\colhead{Source}&\colhead{Predicted}    &\colhead{Observed} 
		&\colhead{Predicted}    &\colhead{Observed}
		&\colhead{of observed} 
\nl
\colhead{name}	&\colhead{dust flux}	&\colhead{total flux}
                &\colhead{dust flux}	&\colhead{total flux}	
		&\colhead{flux at}
\nl
\colhead{}	&\colhead{(mJy)}	&\colhead{(mJy)}
		&\colhead{(mJy)}	&\colhead{(mJy)}
		&\colhead{1.3\,mm}
}
\startdata
00057$+$4021 	&0.9   &$<10$   &10.1   &---   &---	\nl  % for n=1.5
02483$+$4302 	&0.6   &$<10$   &6.1    &---   &---	\nl  % for n=1.5
VII\,Zw\,31  	&0.9   &$<2$    &17.0   &$<20$ &---  	\nl  % for n=1.5		
10565$+$2448 	&2.2   &$<2$    &24.9   &---   &---	\nl  % for n=1.5		
Mrk\,231     	&2.4   &63  	&7.0    &36    &80\% nt, 20\% d	\nl  % for n=2.0 at 1mm 		
Arp\,193	&3.8   &$<5$	&39	&10    &$>$90\% d	\nl
Mrk\,273--core  &1.5   &11 	&16	&8     &50\% nt, 50\% d   \nl
\nl
Arp\,220--west  &6.3 	&--- 	&71	&90 	&10\% nt, 10\% ff, 80\% d
\nl
Arp\,220--east  &2.3 	&--- 	&25	&30 	&10\% nt, 10\% ff, 80\% d	% for n=1.5
\nl
Arp\,220--disk  &4.8 	&--- 	&54	&55	&10\% nt, 90\% d% for n=1.5
\nl
Arp\,220--total &13.4 	&41  	&150	&175 	&15\%nt, 5\%ff, 80\% d 
\nl
\nl		
17208$-$0014 	&3.5   &5       &40     &---	&---\nl  % for n=1.5		
23365$+$3604	&1.0   &$<5$    &11 	&---	& ---\nl % for n=1.5	
\tablenotetext{}{Predicted fluxes are for the temperatures in Table~6 and dust
emissivity index 1.5 .} 
\tablenotetext{}{Errors in observed fluxes: $\pm 20$\%; 
nt = nonthermal, ff = free-free, d = dust.}
\enddata
\end{deluxetable}
\clearpage

%TABLE 11----------------------------------------------TABLE 11 ---
\begin{deluxetable}{l ccc }
\tablewidth{0pc}
\tablecaption{STARBURST CONSISTENCY CHECK: MASS IN GAS, NEW 
STARS, AND OLD BULGE STARS IN THE CENTRAL MOLECULAR DISKS}
\tablehead{
\colhead{}	&\colhead{Mrk 231} &\colhead{Mrk 273}  &\colhead{Arp 220}    \nl
\colhead{}	&\colhead{disk   } &\colhead{disk}     &\colhead{disk}   
}
\startdata
{\it Reference radius:}\nl
$R$(pc)						&460 	&400  	&480	\nl
\nl
{\it Luminosity:}\nl
$L_{\rm FIR}(<R)$ (10$^{12}$\,\Lsun )      	&1.9    &1.0    &0.7	\nl
$L_{\rm IR}/M_{\rm new\star}$ (\Lsun / \Msun)   &500	&500  	&300	\nl
\nl
{\it Deduced mass in new stars:}\nl
$M_{\rm new\star}(<R)$ (10$^9$\,\Msun )	    	&3.8    &2.0   	&2.3	\nl
\nl
{\it Mass from CO model:}\nl
$M_{\rm dyn}(<R)$ (10$^9$\,\Msun )	 	&12.7 	&8.7 	&12.1 	\nl 
$M_{\rm gas}(<R)$ (10$^9$\,\Msun )	    	&1.8    &2.1   	&1.9	\nl
\nl
{\it Deduced mass in old stars:}\nl
$M_{\rm old\star}(<R)$ (10$^9$\,\Msun )    	&7.1    &4.6  	&7.9	\nl
\nl
\tablenotetext{}{$L_{\rm IR}/M_{\rm new\star}=$ assumed
luminosity-to-mass ratio in a rapid starburst.
} 
\tablenotetext{}{$M_{\rm new\star} = L_{\rm FIR}(<R)
/(L_{\rm IR}/M_{\rm new\star})=$ 
mass of new stars formed in the current starburst.
}
\tablenotetext{}{$M_{\rm old\star} =
M_{\rm dyn}-M_{\rm gas}-M_{\rm new\star}=$  
mass of bulge stars within $R$(pc)
} 
\enddata
\end{deluxetable}
\clearpage

%TABLE 12----------------------------------------------TABLE 12 ---
\begin{deluxetable}{l cccc }
\tablewidth{0pc}
\tablecaption{PROPERTIES OF EXTREME STARBURST REGIONS}
\tablehead{
\colhead{}	 
&\colhead{Arp\,193} &\colhead{Mrk\,273} &\colhead{Arp\,220}&\colhead{Arp\,220}  \nl
\colhead{}	
&\colhead{SE core} &\colhead{core}    &\colhead{west}   &\colhead{east}  }

\startdata
{\it 1) Reference radius:}\nl
$R$(pc)						
					&150 	&120 	&68     &110	\nl
\nl
{\it 2) Gas Mass:}\nl
$M_{\rm gas}(<R)$ (10$^9$\,\Msun )	    
					&0.6    &1.0      &0.6	&1.1	\nl
\nl
{\it 3) Mean gas density:}\nl
$<N({\rm H}_2)>$ (cm$^{-3}$)    	
&$2\times 10^3$    &$5\times 10^3$  	&$2\times 10^4$	&$8\times 10^3$	\nl
\nl
{\it 4) Total Mass (Gas plus stars):}\nl
 $M_{\rm tot}(<R)$ (10$^9$\,\Msun )	 	
					&1.4 	&2.6	&1.4 	&3.2 	\nl
\nl 
{\it 5) Estimated mass in new stars:}\nl
$M_{\rm new\star}(<R)$ (10$^9$\,\Msun )	    
					&0.8    &1.6    &0.8	&2.1	\nl
\nl
{\it 6) Luminosity:}\nl
$L_{\rm FIR}(<R)$ (10$^{12}$\,\Lsun )       	
					&0.2    &0.6    &0.3    &0.2	\nl
\nl
{\it 7) Luminosity to mass ratio:}\nl
$L_{\rm FIR}/M_{\rm new\star}$ (\Lsun / \Msun)  
					&300	&360    &380	&100	\nl

\tablenotetext{}{ {\it 1)} 
Radius from measurements (Table 3 or 4).
}
\tablenotetext{}{ {\it 2)} 
$M_{\rm gas} \approx 1.0 L^\prime_{\rm CO}$ (Table 9).
}
\tablenotetext{}{ {\it 3)} 
$<N({\rm H}_2)>$(cm$^{-3} = 0.1 (M_{\rm gas}$/\Msun ) $r^3_{\rm pc}$ 
(cylinder). 
}
\tablenotetext{}{ {\it 4)} $M_{\rm tot} = R \Delta V^2/G$. 
} 
\tablenotetext{}{ {\it 5)} $M_{\rm new\star} = M_{\rm tot} - M_{\rm gas}$. 
} 
\tablenotetext{}{ {\it 6)} 
$L_{\rm IR}/$\Lsun\ $= 0.13 r^2_{\rm pc} T^4$, with $T$ from Table~6. 
} 
\tablenotetext{}{{\it 7)} 
$(L_{\rm FIR}/M_{\rm new\star})=$ luminosity-to-mass ratio
 of new stars formed in the current starburst.
}
\enddata
\end{deluxetable}
\clearpage


\begin{references}

\reference{}Allen, D.A. 1976, ApJ, 207, 367

\reference{}Armus, L., Heckman, T.M., \& Miley, G.K. 1987, AJ, 94, 831

\reference{}Armus, L., Heckman, T.M., \& Miley, G.K. 1989, ApJ, 347, 727

\reference{}Armus, L., Heckman, T.M., \& Miley, G.K. 1990, ApJ, 364, 471

\reference{}Armus, L., Mazzarella, J.M., Graham, J.R., Soifer, B.T.,
Neugebauer, G., Matthews, K., \& Gaume, R.A. 1992, BAAS, 24, 728

\reference{}Armus, L., Surace, J.A., Soifer, B.T., Matthews, K., 
Graham, J.R., \& Larkin, J.E. 1994, AJ, 108, 76

\reference{}Armus, L, Shupe, D.L., Matthews,K., Soifer, B.T., \& Neugebauer, G.
1995a, ApJ, 440, 200 

\reference{}Armus, L,  Neugebauer, G.,  Soifer, B.T., \& Matthews, K.
1995b, AJ, 110, 2610

\reference{}Asatrian, A.S., Petrosian, A.R., \& B\"orngen, F. 
1990, in Paired and Interacting Galaxies, ed. J.W.\ Sulentic, 
W.C.\ Keel, \& C.M.\ Telesco, NASA CP-3098, (Washington DC: GPO),
201

\reference{}Baan, W.A., \& Haschick, A.D. 1987, ApJ, 318, 139

\reference{}Baan, W.A., \& Haschick, A.D. 1995, ApJ, 454, 745

\reference{}Baan, W.A., van~Gorkom, J., Haschick, A.D., \& Mirabel, F.I.
1987, ApJ, 313


% \reference{}Bartel, N. 1988, in Supernova Shells and Their Birth Events,
% ed. W. Kundt, (Berlin: Springer-Verlag), 206

\reference{}Becklin, E.E., \& Wynn-Williams, C.G. 1987, in Star Formation
in Galaxies, ed. C.J. Lonsdale Persson (NASA CP-2466; 
Washington, DC:GPO), 643

\reference{}Binney, J., \& Tremaine, S.  1987,  Galactic Dynamics,
(Princeton:Princeton Univ.\ Press), 76

\reference{}Boksenberg, A., Carswell, R.F., Allen, D.A., Fosbury, R.A.E.,
Penston, M.V., \& Sargent, W.L.W. 1977, MNRAS, 178, 451

\reference{}Borgeest, U., Dietrich, M., Hopp, U., Kollatschny, W., Schramm, 
K.J. 1991, A\&A, 243, 93

\reference{}Bryant, P.M., \& Scoville, N.Z. 1996, ApJ, 457, 678

\reference{}Burbidge, G. 1996, A\&A, 309, 9

\reference{}Carico, D.P., Keene, J., Soifer, B.T., \& Neugebauer, G.
1992, PASP, 104, 1086

\reference{}Carilli, C.L., Wrobel, J.M., \& Ulvestad, J.S. 1998, AJ, 115, 928

\reference{}Chini, R., Kr\"ugel, E., Kreysa, E., \& Gem\"und, H.P. 1989, A\&A, 
216, L5

\reference{}Condon, J.J., Huang, Z.P., Yin, Q.F., \& Thuan, T.X.  1991, ApJ,
378, 65

\reference{}Crawford, T., Marr, J., Partridge, B., \& Strauss, M.A.
1996, ApJ, 460, 225

\reference{}Cutri, R.M., Rieke, G.H., \& Lebofsky, M.J. 1984, ApJ, 287, 566

\reference{} Depoy, D.L., Becklin, E.E., \& Geballe, T.R. 1987, ApJ, 316, L63

\reference{} Diamond, P.J., Norris, R.P., Baan, W.A., \& Booth, R.S.
1989, ApJ, 340, L49

\reference{} Djorgovski, S., de Carvalho, R.R., \& Thompson, D.J.
1990, AJ, 99, 1414

\reference{}Downes, D., Solomon, P.M., \& Radford, S.J.E. 1993, ApJ, 414, L13

\reference{}Doyon, R., Wells, M., Wright, G.S., Joseph, R.D., Nadeau, D.,
\& James, P.A.   1994, ApJ, 437, L23

\reference{}Dudley, C.C., \& Wynn-Williams, C.G. 1997, ApJ, 488, 720

\reference{}Dutrey, A., Guilloteau, S., \& Simon, M. 1994, A\&A, 286, 149

\reference{}Eales, S.A., Wynn-Williams, C.G., \& Duncan, W.D. 
1989, ApJ, 339, 859

\reference{}Emerson, J.P., et al.  1984, Nature, 311, 237

\reference{}Fairclough, J.H. 1986, MNRAS, 219, 1p  

\reference{}Fazio, G.G. 1978, in Infrared Astronomy, ed. G.\ Setti \& G.G.\ 
Fazio, Reidel, Dordrecht, 25

\reference{}Fischer, J., et al. 1998, in ISO to the Peaks, ed. M.\ Kessler
\& M.\ Perry, ESTEC, Noordwijk, in press

\reference{}Gao, Yu, \& Solomon, P.M. 1998, in preparation

\reference{}Gatley, I., Becklin, E.E., Werner, M.W., \& Harper, D.A.
1978, ApJ, 220, 822

\reference{}Genzel, R., et al. 1998, ApJ, 498, 579

\reference{}Goldader, J.D., Joseph, R.D., Doyon, R., \& Sanders, D.B.
1995, ApJ, 444, 97

\reference{}Goldreich, P., \& Lynden-Bell, D. 1965, MNRAS, 130, 125

\reference{}Graham, J.R., Carico, D.P., Matthews, K., Neugebauer, G., 
Soifer, B.T., \& Wilson, T.D. 1990, ApJ, 354, L5

\reference{}Guilloteau, S., et al. 1992, A\&A, 262, 624

\reference{}Hamilton, D., \& Keel, W.C. 1987, ApJ, 321, 211

\reference{}Hoyle, F., \& Burbidge, G. 1996, A\&A, 309, 335

\reference{}Hutchings, J.B., \& Neff, S.G. 1987, AJ, 92, L14

\reference{}Jennings, R.B. 1975, in H~II Regions and Related Topics,
ed. T.L.\ Wilson \& D.\ Downes (Berlin: Springer) 137

\reference{}Jog, C.J. 1996, MNRAS, 278, 209

\reference{}Kaz\`es, I., Mirabel, I.F., \& Combes, F. 1988, IAU Circ. 4629

\reference{}Kennicutt, R.C., \& Chu, Y.H. 1994, in Violent Star Formation
from 30 Doradus to QSOs, ed. G. Tenorio-Tagle, Cambridge: Cambridge Univ. 
Press, 1

\reference{}Keto, E., Ball, R., Arens, J., Jernigan, G., \& Meixner, M.
1992, ApJ, 387, L17

\reference{}Kim, D.C., Sanders, D.B., Veilleux, S., Mazzarella, J.M.,
\& Soifer, B.T. 1995, ApJS, 98, 129

\reference{}Klaas, U., \& Els\"asser, H. 1991, A\&AS, 90, 33
\reference{}Klaas, U., \& Els\"asser, H. 1993, A\&AS, 99, 71

\reference{}Knapen, J.H., Laine, S., Yates, J.A., Robinson, A., Richards, 
A.M.S., Doyon, R., \& Nadeau,~D.  1997, ApJ, in press


\reference{}Kobayashi, Y., Sato, S., Yamashita, T., Shiba, H., \& 
Takami, H. 1993, ApJ, 404, 94


\reference{}Kollatschny, W., Dietrich, M., Borgeest, U., \& 
Schramm, K.J. 1991, A\&A, 249, 57

\reference{}Koski, A.T. 1978, ApJ, 223, 56


\reference{}Krabbe, A., Colina, L., Thatte, N., \& Kroker, H.
1997, ApJ, 476, 98

\reference{}Kr\"ugel, E., \& Siebenmorgen, R. 1994, A\&A, 288, 929

\reference{}Larkin, J.E., Armus, L., Knop, R.A., Matthews, K., \& Soifer, 
B.T. 1995, ApJ, 452, 599

\reference{}Leitherer, C., \& Heckman, T.M. 1995, ApJS, 96, 9

\reference{}Lipari, S., Colina, L, \& Macchetto, F. 1994, ApJ, 427, 174

\reference{}Lonsdale, C.J., Smith, H.E., \& Lonsdale, C.J. 1993, ApJ, 405, L9

\reference{}Lonsdale, C.J., Diamond, P.J., 
Smith, H.E., \& Lonsdale, C.J. 1998, ApJ, 493, L13

\reference{}Lutz, D., et al. 1996, A\&A, 315, L137

\reference{}Majewski, S.R., Hereld, M., Koo, D.C., Illingworth, G.D., 
\& Heckman, T.M. 1993, ApJ, 402, 125

\reference{}Malkan, M.A., \& Sargent, W.L.W. 1982, ApJ, 254, 22

\reference{}Martin, J.M., Bottinelli, L., Dennefeld, M., Gouguenheim, L.,
\& Le Squeren, A.M. 1989, A\&A, 208, 39

\reference{}Matthews, K., Neugebauer, G., McGill, T., \& Soifer, B.T.
1987, AJ, 94, 297

\reference{}Melnick, J., \& Mirabel, I.F. 1990, A\&A, 231, L19

\reference{}Mestel, L. 1963, MNRAS, 126, 553

\reference{}Mezger, P.G., 1985, in Birth \& Infancy of Stars, ed.\ R.\ Lucas,
A.\ Omont, \& R.\ Stora, Elsevier, Amsterdam, 31

\reference{}Miles, J.W., Houck, J.R., Hayward, T.L., \& Ashby, M.L.N.
1996, ApJ, 465, 191 

\reference{}Murphy, T.W., Armus, L., Matthews, K., Soifer, B.T., 
Mazzarella, J.M., Shupe, D.L., Strauss, M.A., \& Neugebauer, G.
1996, AJ, 111, 1025

\reference{}Neff, S.G., \& Ulvestad, J.S. 1988, AJ, 96, 841

\reference{}Norris, R.P. 1988, MNRAS, 230, 345

\reference{}Norris, R.P., Baan, W.A., Haschick, A.D., Booth, R.S.,
\& Diamond, P.J. 1985, MNRAS, 213, 821

\reference{}Oort, J.H. 1977, ARAA, 15, 295

\reference{}Panagia, N. 1977, in Infrared \& Submillimeter Astronomy,
ed. G.G.\ Fazio, Reidel, Dordrecht, 43

\reference{}Parker, J.W. 1993, AJ, 106, 560

\reference{}Planesas, P., Mirabel, I.F.,  \& Sanders, D.B.  1991, ApJ, 370, 172

\reference{}Preuss, E., \& Fosbury, A.E. 1983, MNRAS, 204, 783

\reference{}Radford, S.J.E., Solomon, P.M., \& Downes, D. 1991a, ApJ, 368, L15

\reference{}Radford, S.J.E., et al.  1991b,  in Dynamics of Galaxies and Their 
Molecular Cloud Distributions, ed.\ F.\ Combes \& F.\ Casoli 
(Dordrecht:Kluwer), 303

\reference{}Rees, M.J., 1982, in The Galactic Center, ed.\ G.R.\ Riegler \& 
R.D.\ Blandford, Am. Inst. of Physics Conf. Proc. 83, (New York:AIP), 166 

\reference{}Rieke, G.H. 1976, ApJ, 210, L5

\reference{}Rigopoulou, D., Lawrence, A., \& Rowan-Robinson, M.
1996, MNRAS, 278, 1049

\reference{}Roche, P.F., Aitken, D.K., \& Whitmore, B. 1983, MNRAS, 205, 21P

\reference{}Safronov, V.S. 1960, Ann d'Ap, 23, 979 

\reference{}Sage, L.J.,  \& Solomon, P.M. 1987, ApJ, 321, L103

\reference{}Sanders, D.B., Young, J.S., Scoville, N.Z., Soifer, B.T., 
\& Danielson, G.E. 1987, ApJ, 312, L5

\reference{}Sanders, D.B., Phinney, S., Neugebauer, G., Soifer, B.T.,
\& Matthews, K. 1989, ApJ 347, 29

\reference{}Sargent, W.L.W. 1972, ApJ, 173, 7

\reference{}Sargent, W.L.W., \& Steidel, C.C. 1990, ApJ, 359, L37

\reference{}Schmelz, J.T., Baan, W.A., \& Haschick, A.D. 1988, ApJ, 329, 142

\reference{}Schmidt, G.D., \& Miller, J.S. 1985, ApJ, 290, 517

\reference{}Scoville, N.Z., Sanders, D.B., Sargent, A.I., Soifer, B.T.,
\& Tinney, C.G. 1989, ApJ, 345, L25 

\reference{}Scoville, N.Z., Sargent, A.I., Sanders, D.B., \& Soifer, B.T. 1991,
ApJ, 366, L5 

\reference{}Scoville, N.Z., Yun, M.S., \& Bryant, P.M. 1997, ApJ, 484, 702

\reference{}Scoville, N.Z., Evans, A.S., Dinshaw, N., Thompson, R., 
Rieke, M., Schneider, G., Low, F.J., Hines, D., Stobie, B., 
Becklin, E., \& Epps, H. 1998, ApJ, 492, L107

\reference{}Shakura, N.I., \& Sunyaev, R.A. 1973, A\&A, 24, 337

\reference{}Shaya, E.J., Dowling, D.M., Currie, D.G., Faber, S.M., \& 
Groth, E.J. 1994, AJ, 107, 1675

\reference{}Shier, L.M., Rieke, M.J., \& Rieke, G.H. 1994, ApJ, 433, L9

\reference{}Shier, L.M., Rieke, M.J., \& Rieke, G.H. 1996, ApJ, 470, 222

\reference{}Skinner, C.J., Smith, H.A., Sturm, E., Barlow, M.J., Cohen, R.J.,
\& Stacey, G.J.  1997, Nature, 386, 472 

\reference{}Smith, C.H., Aitken, D.K, \& Roche, P.F. 1989, 241, 425

\reference{}Smith, D.A., Herter, T., Haynes, M.P., Beichman, C.A., 
Gautier, T.N. 1995, ApJ, 439, 623

\reference{}Smith, D.A., Herter, T., Haynes, M.P., Beichman, C.A., 
Gautier, T.N. 1996, ApJS, 104, 217 


\reference{}Smith, H.E., Lonsdale, C.J., Lonsdale, C.J.,  1998a, ApJ, 492, 137

\reference{}Smith, H.E., Lonsdale, C.J., Lonsdale, C.J., 
\& Diamond, P.J.,  1998b, ApJ, 493, L17

\reference{}Smith, P.S., Schmidt, G.D., Allen, R.G., \& Angel, J.R.P.
1995, ApJ, 444, 146

\reference{}Solomon, P.M., Downes, D., Radford, S.J.E., 1992, ApJ, 387, L55

\reference{}Solomon, P.M., Downes, D., Radford, S.J.E., \& Barrett, J.W.
1997, ApJ, 478, 144

\reference{}Sopp, H.M., \& Alexander, P. 1991, MNRAS, 251, 112

\reference{}Staveley-Smith, L., Cohen, R.J., Chapman, J.M., Pointon, L., 
Unger, S.W. 1987, MNRAS, 226, 689

\reference{}Sturm, E., et al. 1996, A\&A, 315, L133

\reference{}Surace, J.A., Sanders, D.B., Vacca, W.D., Veilleux, S., 
\& Mazzarella, J.M. 1998,  ApJ, 492, 116

\reference{}Taylor, G.B., Vermeulen, R.C., Pearson, T.J., Readhead, A.C.S., 
Henstock, D.R., Browne, I.W.A., \& Wilkinson, P.N. 1994, ApJS, 95, 345

\reference{}Toomre, A. 1964, ApJ, 139, 1217

\reference{}Ulvestad, J.S., \& Wilson, A.S. 1984, ApJ, 278, 544 

\reference{}van der Werf, P.P. 1996, in Cold Gas at High Redshift, ed. 
M.N.\ Bremer, H.J.A.\ R\"ottgering, \& C.L. Carilli, (Kluwer: Dordrecht),
37
 
\reference{}van der Werf, P.P., \& Israel, F. 1998, in preparation.

\reference{}Veilleux, S., Kim, D.C., Sanders, D.B., Mazzarella, J.M.,
\& Soifer, B.T. 1995, ApJS, 98, 171

\reference{}Womble, D.S., Junkkarinen, V.T., Cohen, R.D., \& Burbidge, E.M.
1990, AJ, 100, 1785

\reference{}Woody, D.P., et al. 1989, ApJ, 337, L41 

\reference{}Yun, M.S., \& Scoville, N.Z. 1995, ApJ, 451, L45

\reference{}Zenner, S., \& Lenzen, R. 1993, A\&AS, 101, 363

\reference{}Zhao, J.H., Anantharamaiah, K.R., Goss, W.M., \& Viallefond, F.
1996, ApJ, 472, 54

\reference{}Zwicky, F. 1971, Catalogue of Selected Compact Galaxies and
Post-Eruptive Galaxies, Guemlingen, Switzerland

\reference{}Zylka, R., Mezger, P.G., Ward-Thompson, D., Duschl, W.J.,
\& Lesch, H. 1995, A\&A, 297, 83
\end{references}
\end{document}